\documentclass{jfm}
\pdfoutput=1
\usepackage{graphicx,psfrag,color}
\usepackage{amsmath}
\usepackage{amsfonts}
\usepackage{mathrsfs}
\usepackage{amssymb}
\usepackage{bm}
\usepackage{url}
\usepackage{color}
\usepackage{natbib}
\usepackage{mdwlist}

\newcommand{\bx}{\mathbf{x}}
\newcommand{\by}{\mathbf{y}}

\begin{document}

\def\half {\textstyle{1 \over 2}}
\def\quart{\textstyle{1 \over 4}}
\def\eighth{\textstyle{1 \over 8}}
\def\sixteenth{\textstyle{1 \over 16}}
\def\d{{\rm d}}
\def\eps{\epsilon}

\title[Spatiotemporal dynamics  in 2D Kolmogorov flow over large domains]{Spatiotemporal dynamics in 2D Kolmogorov flow over large domains} 
\author{Dan Lucas \& Rich Kerswell}
\affiliation{School of Mathematics, University of Bristol, University
  Walk, Bristol, UK.}  \maketitle

\vspace{0.25cm}
  
\begin{abstract}

Kolmogorov flow in two dimensions - the two-dimensional Navier-Stokes
equations with a sinusoidal body force - is considered over extended
periodic domains to reveal localised spatiotemporal complexity. The
flow response mimicks the forcing at small forcing amplitudes but
beyond a critical value develops a long wavelength instability.  The
ensuing state is described by a Cahn-Hilliard-type equation and as a
result coarsening dynamics are observed for random initial data.
After further bifurcations, this regime gives way to multiple
attractors, some of which possess spatially-localised time dependence.
Co-existence of such attractors in a large domain gives rise to
interesting collisional dynamics which is captured by a system of 5
(1-space and 1-time) PDEs based on a long wavelength limit.  The
coarsening regime reinstates itself at yet higher forcing
amplitudes in the sense that only longest-wavelength solutions remain 
attractors. Eventually, there is one global longest-wavelength attractor which 
possesses two localised chaotic regions - a
kink and antikink - which connect two steady one-dimensional flow regions
of essentially half the domain width each.  The wealth of spatiotemporal
complexity uncovered presents a bountiful arena in which to study the
existence of simple invariant localised solutions which presumably 
underpin all of the observed behaviour.

\end{abstract}

\section{Introduction} 

The problem of a 2-dimensional (2D) fluid on a torus $(x,y) \in [0,2\pi/\alpha]\times
[0,2\pi]$ driven by monochromatic body forcing $\sin \, ny \,\hat{\bx}$
with $n=1$ was first introduced by Kolmogorov in 1959
(\cite{Arnold:1960vx,Arnold91}) as a mathematically and experimentally
tractable flow situation in which to study instability and transition
to turbulence. The linear instability of the steady, unidirectional
base flow can be predicted analytically using continued fractions
(\cite{Meshalkin:1961ty}) and a remarkable global stability (or
`anti-turbulence') result is known for all forcing amplitudes when the
forcing wavenumber, $n$, and box aspect ratio, $\alpha$, are both 1
(\cite{Marchioro:1986tg}). The flow can be realised in the laboratory
either using electrolytic fluids (\cite{1979FizAO..15.1017B,
  Obukhov:1983wa, Sommeria:1986wm, Batchaev:1989ud, Batchaev:2012fh,
  Suri:2013td}) or driven soap films (\cite{Burgess:1999tu}). There
are many possible variations of the problem: torus aspect ratio
(e.g. \cite{Marchioro:1986tg, Okamoto:1993tk, Sarris:2007bj}), forcing
wavelength ( e.g. \cite{1988cttr.proc..374S,
  Platt:1991ed,Armbruster:1996wk}), forcing form
(e.g. \cite{Gotoh:1987ue, Kim:2003un, Rollin:2011uh, Gallet:2013gt}),
boundary conditions (\cite{Fukuta:1998uw, Thess:1992cd,
  Gallet:2013gt}), and 3-dimensionalisation (e.g. \cite{Borue:2006dt,
  Shebalin:1997jm, Sarris:2007bj, Musacchio:2014tg}). Further physics such as rotation,
bottom friction and stratification (\cite{1972JAtS...29..258L,
  Kazantsev:1998wb, Manfroi:1999, Balmforth:2002,
  Balmforth:2005tf,Tsang:2008il, Tsang:2009ef}) can also be added to
examine most commonly their effect on the inverse energy cascade in
geophysical fluid dynamics. Recently compressible
(\cite{Manela:2012by, Fortova:2013bf}), viscoelastic
(\cite{Boffetta:2005wd,Berti:2010co}) and even granular
(\cite{Roeller:2009fo}) versions of Kolmogorov flow have been treated.

Mathematically, beyond the instability result of
\cite{Meshalkin:1961ty}, past work has developed amplitude equations
to study the dynamics near the bifurcation point
(\cite{NEPOMNIASHCHII:1976,Sivashinsky:1985}) and used bifurcation
analysis coupled with branch continuation techniques to uncover a
variety of steady vortex states
(\cite{Okamoto:1993tk,Okamoto:1998wn,Kim:2003un}). These solutions are
qualitatively representative of the vortex arrays observed by numerous
authors investigating transition to chaos in these types of flows
(\cite{Obukhov:1983wa,Platt:1991ed,Chen:2004cz,Feudel:1995wd,Sarris:2007bj}). However,
with the exception of the bifurcation analysis of Okamoto
(\cite{Okamoto:1993tk,Okamoto:1996wi,Okamoto:1998wn}), all of the work
to date has avoided Marchioro's 2D global stability result by
considering a forcing wavenumber $n>1$ for $\alpha=1$, with $n=4$
being the most common choice
(\cite{Platt:1991ed,Chen:2004cz,Chandler:2013fi}). The ensuing flow
dynamics, at least at the low to intermediate forcing amplitudes
studied, is then spatially global as the domain is square. The
objective here is to extend these previous studies to spatially
extended domains with $\alpha \leq \quart$ in order to look for
spatiotemporal behaviour.

The motivation for this study is the recent work by
\cite{Chandler:2013fi} looking at 2D Kolmogorov flow over a small
$[0,2\pi]^2$ domain. These authors attempt to make sense of the
observed global chaos by extracting simple invariant sets or
`recurrent flows' (simple exact solutions of the Navier-Stokes
equations) embedded in the chaos directly from their direct numerical
simulations (DNS). The underlying idea is that the chaos as a whole
can be viewed as one phase-space trajectory transiently visiting the
neighbourhoods of simple invariants sets (equilibria, relative
equilibria, periodic orbits and relative periodic orbits) which litter
phase space. Assuming ergodicity, the properties of these simple
invariant sets can then in principle be used, in some appropriately
weighted fashion, to predict the statistical properties of the
chaos. Due to the domain size used, \cite{Chandler:2013fi} naturally
concentrated on extracting spatially-global recurrent flows from the
(temporal) chaos. However the real challenge for practical
applications is to consider larger domains where true {\em
  spatio}temporal behaviour occurs in which the flow's complexity
can differ markedly in space at any given time.  The hope in examining
2D Kolmogorov flow over an extended domain was that it would display
spatially localised chaotic flows - i.e. something approaching 2D
turbulence (spatial and temporal complexity) - and by implication
possess spatially localised recurrent flows. This would then offer an
ideal environment to develop further the recently-introduced
techniques for extracting order and coherence from the apparent
disorder of turbulence
(\cite{Kawahara:2001ft,vanVeen:2006fm,Viswanath:2007wc,Viswanath:2009vu,2010PhST..142a4007C,Kreilos:2012bd,Chandler:2013fi,Willis:2013bu}).

The structure of the paper is as follows. A short formulation section
sets the scene as considered previously in \cite{Chandler:2013fi} with
the `geometry' parameter $\alpha$ describing the aspect ratio of the
computational domain now $\leq \quart$.  Section 3 describes the
initial instability away from the steady, 1-dimensional (1D),
unidirectional flow response realised at low forcing amplitudes.  The
new preferred steady flow state, which preserves the wavelength of the
forcing but breaks its continuous translational symmetry, adopts the
longest wavelength allowed.  Beyond the initial bifurcation point,
this state quickly develops into two 1D flow regions each
over essentially half the domain joined together by a localised
2D transition regions (`kinks' and `antikinks').  A
Cahn-Hilliard-like long-wavelength amplitude equation is discussed
which explains this response and a more accurate 3-PDE long-wavelength
system introduced.  Further (primary) steady bifurcations off the
basic state introduce initially-unstable states with more and more
localised transition regions equally spaced across the domain.
Section 4 details the existence of further steady and unsteady
bifurcations and identifies the emergence of a global attractor at
sufficiently high levels of forcing which displays localised chaos.
Other disconnected states which arise in saddle node bifurcations are
also found and the bifurcation structure of one is followed in detail
from localised periodicity to chaotic attractor and then through a
boundary crisis to a chaotic repellor where the chaos is then both
spatially and temporally localised.
Section 5 discusses the interesting behaviour observed when
co-existing attractors interact in an even larger domain. A 5-PDE
extension to the long-wavelength 3-PDE system, which adds the
possibility of the kinks and antikinks developing structure, is
successful in capturing the observed behaviour in the full system.  In
this 5-PDE system, the basic building blocks - families of travelling
waves made up of various types of kink-antikink pairs with different
separations and internal structures - are isolated. A final section 6
discusses the results and implications for further work.

%
% Formulation
%
\section{Formulation}\label{sect:form}

The incompressible Navier-Stokes equations with Kolmogorov forcing are given by
\begin{align}
\frac{\partial \bm u^*}{\partial t^*} + \bm u^*\cdot\nabla^*\bm u^* +\frac{1}{\rho}\nabla^*p^* &= \nu \Delta^* \bm u^* + \chi\sin(2\pi n y^*/L_y)\bm \hat{\bx}, \\ 
\nabla^*\cdot \bm u^* &=0
\end{align}
where $\bm u = u \hat{\bx}+v \hat{\by}=(u,v)$ is the two-dimensional velocity field, $n$ is the forcing wavenumber, $\chi$ the forcing amplitude, $\nu$ kinematic viscosity, $p$ pressure and $\rho$ is the density of the fluid defined over the doubly periodic domain $(x,y) \in [0,L_x]\times[0,L_y]$. The system is then naturally non-dimensionalised with lengthscale $L_y/2\pi$ and timescale $\sqrt{L_y/2\pi\chi}$ to give
\begin{align}
\frac{\partial \bm u}{\partial t} + \bm u\cdot\nabla\bm u +\nabla p &= \frac{1}{Re} \Delta \bm u + \sin n y \,\bm \hat{x}\label{NSu}, \\ 
\nabla\cdot \bm u &=0
\end{align}
where we define the Reynolds number
\begin{equation}
Re := \frac{\sqrt{\chi}}{\nu}\left(\frac{L_y}{2\pi}\right)^{3/2}.
\end{equation}
The equations are  solved over $[0,2\pi/\alpha]\times[0,2\pi]$ where $\alpha = L_y/L_x$ defines the aspect ratio of the domain with always  the choice $n=4$ here. 
For most of the runs presented here, $\alpha=\quart$ but $\alpha=\eighth$ and exceptionally $\alpha=\sixteenth$ are considered too.
For computational efficiency and accuracy we reformulate (\ref{NSu}) so that vorticity $\omega := \nabla \times \bm u$ is our prognostic variable and solve
\begin{align}\label{NS}
\frac{\partial \omega}{\partial t} + \bm u\cdot\nabla \omega &= \frac{1}{Re} \Delta \omega -n \cos n y.
\end{align}
The system has the symmetries

\begin{align}\label{sym}
\mathcal{S}:[u,v](x,y) &\rightarrow [-u,v]\left(-x,y+\frac{\pi}{n}\right),\\
\mathcal{R}:[u,v](x,y) &\rightarrow [-u,-v]\left(-x,-y\right),\\
\mathcal{T}_l:[u,v](x,y) &\rightarrow [u,v]\left(x+l,y\right) \qquad \textrm{for } 0\leq l \leq \frac{2\pi}{\alpha},
\end{align}

\noindent
where $\mathcal{S}$ represents the discrete shift-\&-reflect symmetry, $\mathcal{R}$ rotation through $\pi$ and $\mathcal{T}_l$ is the continuous group of translations in $x$. 

\subsection{Flow measures}

In order to discuss various features of the flows considered we define here some diagnostic quantities. Total energy, dissipation and energy input are defined in the standard way as
\begin{align}
E(t) := \half \langle \bm u^2 \rangle_V, \qquad
D(t) := \frac{1}{Re} \langle|\nabla \bm u |^2\rangle_V, \qquad
I(t) :=  \langle u \sin(ny) \rangle_V 
\end{align}
where the volume average is defined as 
\[ \langle\quad\rangle_V := \frac{\alpha}{4\pi} \int_0^{2\pi}\int_0^{2\pi/\alpha} \d x \d y. \]
The base state or laminar profile and its energy and dissipation are 
\begin{align}
\bm u_{lam} := \frac{Re}{n^2}\sin ny \,\bm{\hat x}, \qquad E_{lam} := \frac{Re^2}{4n^4}, \qquad D_{lam} := \frac{Re}{2n^2}.
\label{base}
\end{align}

\subsection{Numerical methods}\label{sect:stepper}

The pseudospectral timestepping code presented in
\cite{Chandler:2013fi} was ported into CUDA to allow accelerated
calculations to be performed on GPUs. The algorithm itself remains
identical, simply the implementation is altered to take advantage of
the vast parallelism afforded by modern general purpose GPU cards (see
the supplementary material for details). Vorticity is discretised via a
Fourier-Fourier spectral expansion with resolution $N_x\times N_y$ and
dealiased by the two-thirds rule such that
\begin{align}
\omega(x,y,t) = \sum^{N_x/3-1}_{k_x=0} \sum^{N_y/3-1}_{k_y=-N_y/3} \Omega_{k_xk_y}(t)\mathrm{e}^{\mathrm{i}\left(\alpha k_x x+k_y y\right)}
\label{representation}
\end{align}
together with reality condition $\Omega_{-k_x -k_y} = \Omega^*_{k_x
  k_y}$ (a mask is also used as outlined in section 3 of
\cite{Chandler:2013fi}). Crank-Nicolson timestepping scheme is used
for the viscous terms and Heun's method for the nonlinear and forcing
terms. As pointed out in \cite{Chandler:2013fi}, this results in an
algorithm that can be restarted from a single state vector which is
very convenient for the various solution-finding algorithms employed
here. Typical numerical resolutions used were 128 Fourier modes per
$2\pi$ so $(N_x,N_y)=(512,128) $ for $\alpha=\quart$ and $(1024,128)$
for $\alpha=\eighth$ (\cite{Chandler:2013fi} used 256 Fourier modes
per $2\pi$ but tests showed they could easily have halved this as was
done here). Typical time steps were $dt=0.05$ at $Re=20$ and $0.001$
at $Re=120$ with $2\times 10^6$ time steps ($dt=0.05$, $T=10^5$) of
the $512 \times 128$ grid for $\alpha=\quart$ and $Re=20$ taking 63
minutes on a 512-core NVIDIA Tesla M2090 GPU.

In addition to the pseudospectral DNS (direct numerical simulation)
code, we also made use of the Newton-GMRES-hookstep algorithm of
\cite{Chandler:2013fi} in order to precisely converge any invariant
states (steady states, travelling waves, periodic orbits and relative
periodic orbits) that we came across. The only change from the code
used in \cite{Chandler:2013fi} was the integration of the GPU
accelerated timestepper. In contrast to \cite{Chandler:2013fi}, the
use of the technique in the current study was not to extract unstable solutions
embedded in chaos, but simply to confirm simple attractors
emerging from the simulations were actually exact solutions. We also
employed their arc-length continuation algorithm to continue solution
branches in $Re$.

%
%
%
%
%*********************************************************************************
% START OF RESULTS
%*********************************************************************************
%
%
%
%
% fig 1
%
\begin{figure}
\begin{center}
\hspace{10mm}
\begin{tabular}{cc}%
\scalebox{1.08}{\input{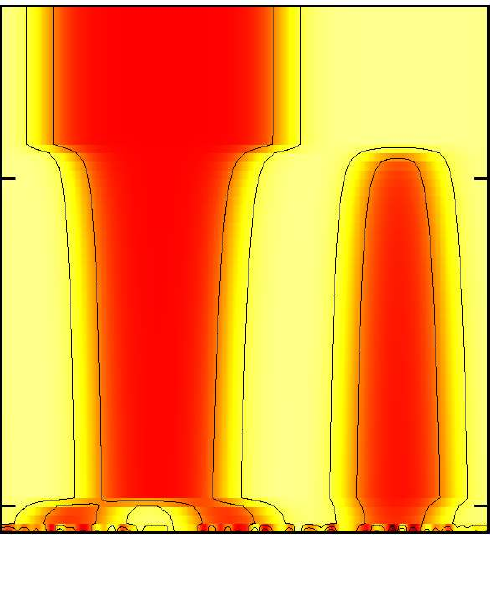}} &
\hspace{5mm}
\parbox[b]{50mm}{
\scalebox{0.9}{\input{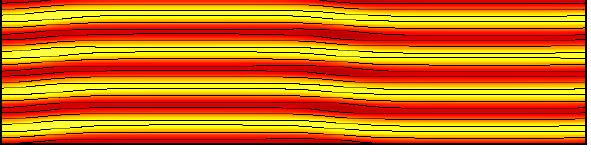}}\\
\scalebox{0.9}{\input{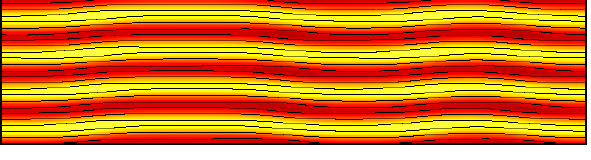}}\\
\scalebox{0.9}{\input{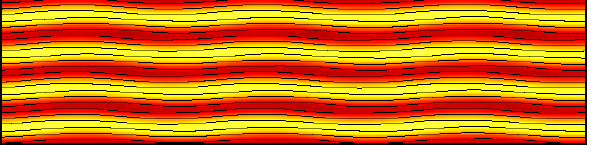}}\\
\scalebox{0.9}{\input{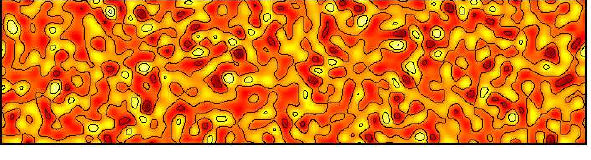}}
\vspace{1.6mm}
}
\end{tabular}
\caption{DNS calculation from a randomised initial condition at
  $Re=10$ to show how the flow very slowly converges to the longest
  wavelength state. Shown on the left is the variation in $\omega$ in
  an $(x,t)$ plane for $y=21\pi/32$ and the right snapshots of
  $\omega(x,y)$ at $t=0,5\times10^3,10^5$ and $1.5\times 10^5$. Colour
  extrema are $\omega=-5$ black, $\omega=5$ white, and 5 evenly spaced
  contours in $-4\leq\omega\leq4$ for $(x,y)$ plots and
  $-2\leq\omega\leq2$ with 4 contours for $(x,t)$.}
\label{10vort}
\end{center}
\end{figure}

%
% fig 2
%
\begin{figure}
\begin{center}
\scalebox{0.7}{\input{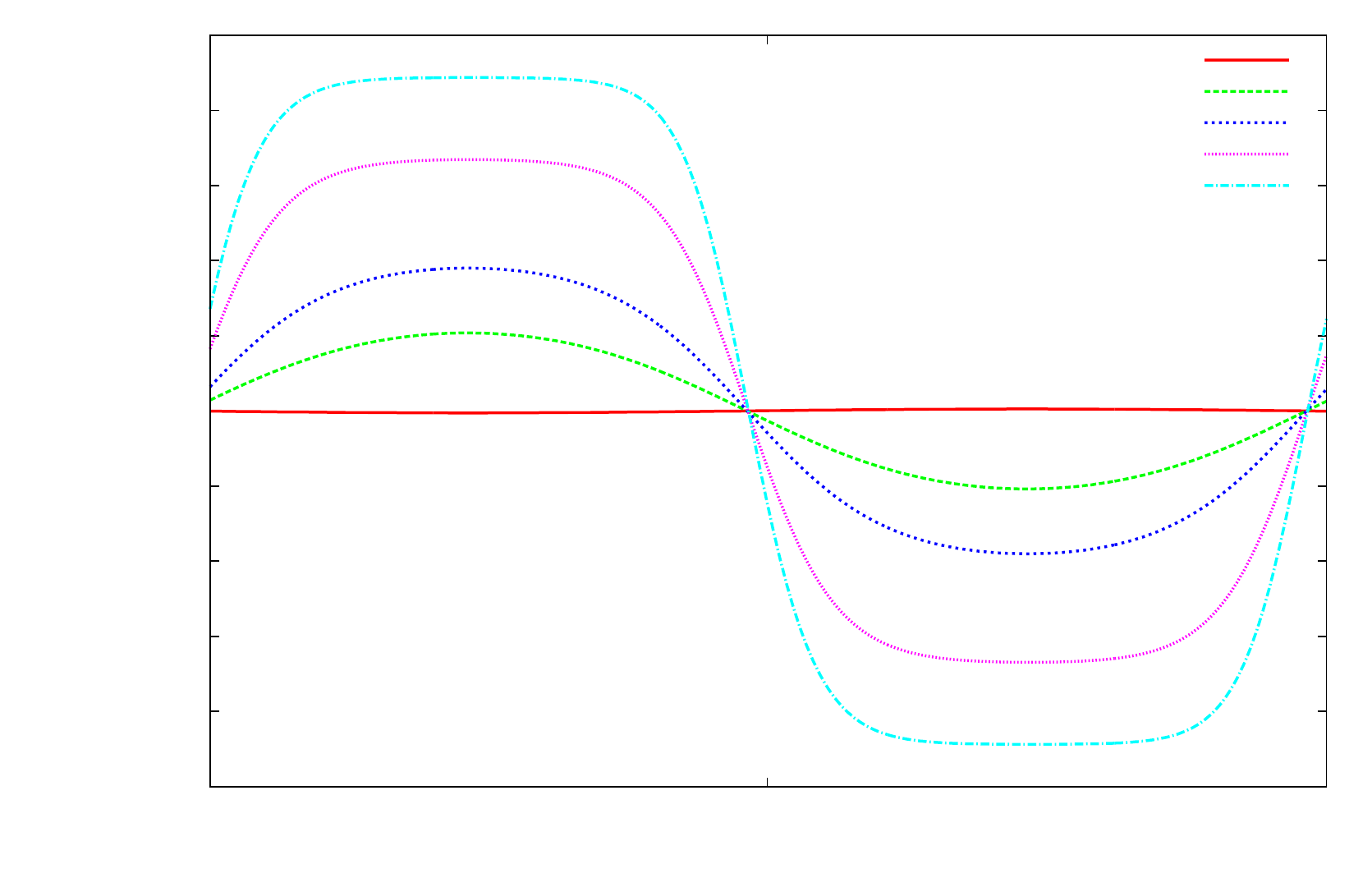}}
\caption{Profiles of vorticity at $y=\pi/8$ for $Re \gtrsim
  Re_c(\quart)=9.5416$ showing how the initial instability quickly
  generates a kink near $x=0$ and an antikink near $4\pi$.}
\label{fig:profiles}
\end{center}
\end{figure}

% S-symmetric 
%
\section{Initial instability}\label{sect:instab}

The basic flow (\ref{base}) is linearly unstable at a comparatively low $Re=Re_{c}$ with the
exact value decreasing monotonically to a limiting value of
$\sqrt[4]{2}n^{3/2}$ as the domain size $L$ becomes infinitely
large. The asymptotic approximation (derived in the Appendix),
\begin{equation}
Re_c(\alpha) = \sqrt[4]{2}n^{3/2}\biggl[ 1+\frac{3 \alpha^2}{4 n^2}+O(\alpha^4)\,\biggr],
\label{asym_linear}
\end{equation}
illustrates this behaviour as $\alpha=2\pi/L \rightarrow 0$ (this
formula is accurate to within $0.01\%$ at $\alpha=\half$).

\begin{figure}
\begin{center}
{\input{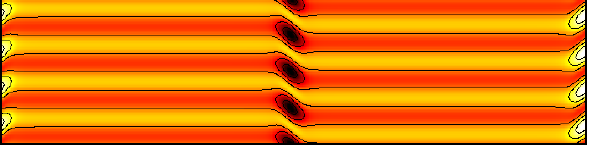}}
{\input{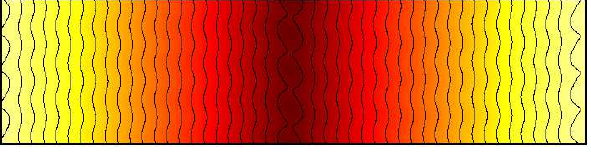}}\\
\vspace{4pt}
{\input{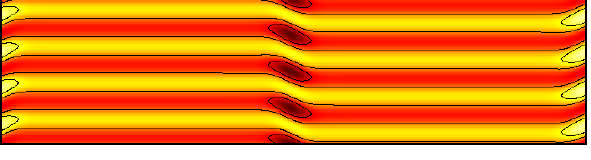}}
{\input{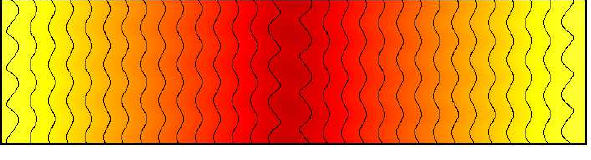}}\\
\vspace{4pt}
{\input{figs/fig3e }}
{\input{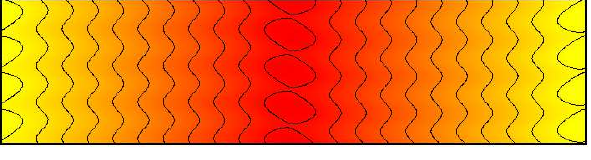}}\\
\vspace{4pt}
{\input{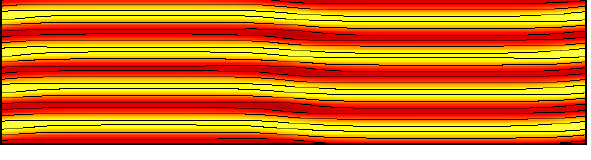}}
{\input{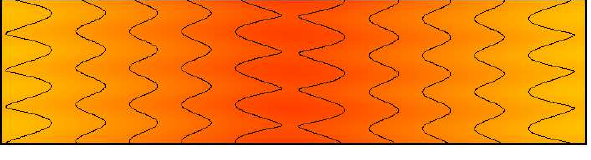}}
\caption{Vorticity field $\omega$ (left) and streamfunction $\psi=-\nabla^{-1} \omega$ (right) for the $\mathcal{S}-$symmetric kink solutions at
  $Re=10, 12, 15, 25$ in a domain $8 \pi \times 2 \pi$ to show how the
  kink and antikink intensify. Colour extrema for vorticity are $\omega=-5$ black,
  $\omega=5$ white, and 5 evenly spaced contours in
  $-4\leq\omega\leq4$, while for streamfunction we use $\psi=-8$
  black, $\psi=8$ white, and contours spaced by 0.5 in
  $-6\leq\psi\leq6$. }
\label{sym_kink}
\end{center}
\end{figure}

For a given long domain $L$ and $Re$ in the interval
$(\,Re_c(\alpha),Re_c(2\alpha)\,)$, the unique attractor is the steady
solution branch which bifurcates supercritically at $Re_c(\alpha)$ and
breaks the streamwise-independence of the basic state, i.e. the
continuous symmetry $\mathcal{T}_l$, while preserving the discrete
symmetries $\mathcal{S}$ and $\mathcal{R}$. Direct numerical
simulations (DNS) for $\alpha=\quart$ confirm this can stay a global
attractor past $Re_c(2 \alpha)=9.626$ up to 10.75. (Random initial
data were used to initialise the DNS which means here that the code is
started with uniform amplitudes but randomised phases for modes with
$2.5 < | k:=\sqrt{k_x^2+k_y^2}| < 9.5$ - see (\ref{representation}) -
normalised such that the total enstrophy is 1).  A typical such run at
$Re=10$ is shown in figure \ref{10vort} where the system reaches the
longest wavelength solution but only after a very long time.  While
the flow selects this longest overall wavelength, it also very quickly
(as $Re$ increases) separates into 2D {\em kink} and {\em antikink}
structures in the vorticity field (where length scales are $\partial_x=O(\partial_y)$)
which connect regions of essentially half the wavelength where the
flow is 1D (independent of $x$).  These 1D states take the form
\begin{equation}
\psi=\lambda x -\frac{Re}{n(Re^2 \lambda^2+n^2)} \cos ny +\frac{\lambda Re^2}{n^2 (Re^2 \lambda^2+n^2)} \sin ny
\end{equation}
where $\psi$ is the streamfunction (\,$\bm u=(\psi_y,-\psi_x)$\,) and
$\lambda$ is a parameter indicating the constant velocity in the
$-\lambda \bm{\hat y}$ direction. We use the term `kink' to refer to
the 2D flow structure connecting a leftward (decreasing $x$) 1D state
with a flow in the $-\bm{\hat{y}}$ direction to a rightward
(increasing $x$) 1D state with flow in the $+\bm{\hat{y}}$ direction:
the `antikink' does the opposite. The localising 2D structure at the kink (antikink) then corresponds to coherent negative (positive) vorticity regions. Figures \ref{fig:profiles} and
\ref{sym_kink} both show an antikink in the centre of the flow domain
and a kink at the end.  Since the 1D solutions on either side of the
kink and antikink have equal spatial extent, they must have equal in
magnitude but oppositely signed $\lambda$ values to preserve the total
linear momentum at zero in the $\bm {\hat{y}}$ direction.  The kink
and antikinks select the value for $\lambda$ as a function of $Re$ (or
the domain size) with $\lambda$ increasing to the asymptotic value of
$\lambda=\eps \sqrt{3 \sqrt{2}/n}$ as the kinks and antikinks
intensify: see the Appendix. Figure \ref{sym_kink} shows the kink and
antikink pair strengthening and localising although by $Re=25$ this
flow structure is already no longer the global attractor.

Both the system's preference for the longest wavelength instability
{\em and} the generation of these kinks and antikinks can be captured in a
long-wavelength approximation
\citep{NEPOMNIASHCHII:1976,Sivashinsky:1985} in which $\partial_x \ll
\partial_y$ is assumed close to $Re=Re_c$. Briefly (see the Appendix for details), if
\begin{equation}
\frac{1}{Re}=(1-\eps^2) \frac{1}{Re_c}
\end{equation}
with $\eps \ll 1$ and the flow varies over the scale $X:=\eps x$, a streamfunction of the form 
\begin{eqnarray}
\psi &=&-\frac{Re_c(0)}{n^3} \cos ny+A^0(X,\eps^4 t)+\eps
\biggl[ \frac{Re_c^2}{n^4} A_X^0 \sin ny +A^1_X(X,\eps^4 t) \biggr] \nonumber\\
     & &  +\eps^2 \frac{Re_c}{n}\biggl[ -\frac{1}{n^2} \cos ny+(A^0_X)^2 \frac{Re_c^2}{n^4} \cos ny +\frac{Re_c}{n^3} A_X^1(X,\eps^4 t) \sin ny \biggr] \nonumber\\
     &&  + \eps^3\biggl[ \biggl( 
\frac{3Re_c^2}{n^6} A_{XXX}^0+\frac{2 Re_c^2}{n^4} A_X^0-\frac{Re_c^4}{n^6} (A_X^0)^3+\frac{Re_c^2}{n^4} A_X^2(X,\eps^4 t)
\biggr) \biggr. \sin ny \nonumber \\
&& \hspace{4cm}+ \biggl.\frac{2 Re_c^3}{n^6} A_X^0 A_X^1 \cos ny + A^3(X,\eps^4 t) \biggr]+O(\eps^4)
\label{str}
\end{eqnarray}
emerges with a solvability condition at $O(\eps^3)$ giving the Cahn-Hilliard-type equation
\begin{equation}
A^0_{XXX}+\frac{4n^2}{3} A^0_{X}-\frac{2 \sqrt{2}n^3}{9} (A_X^0)^3=0
\label{TheEquation}
\end{equation}
for the leading unknown amplitude function $A^0$.
The appropriate solution of this is $A_X^0= \sigma $sn$(\beta X|k)$
where sn is the elliptic function and $\sigma, \beta$ and $k \in
(0,1)$ are constants (see the Appendix for details). As
$Re \rightarrow Re_c$, $k \rightarrow 0$ and $A_X^0$ tends to the
expected $sine$ function whereas as $Re \rightarrow \infty$ ($k
\rightarrow 1$) $A_X^0$ asymptotes to a $tanh$ function over each half
wavelength mimicking the behaviour seen in figure
\ref{fig:profiles}. The preference for the longest wavelength of the
system can be explained by the fact that all solutions to
(\ref{TheEquation}) are unstable to larger wavelengths disturbances
\citep{NEPOMNIASHCHII:1976,Chapman:1980} so that only the solution
already with the largest wavelength is stable.

\begin{figure}
\begin{center}
\hspace{-0.25cm}
\includegraphics[width=7cm, clip]{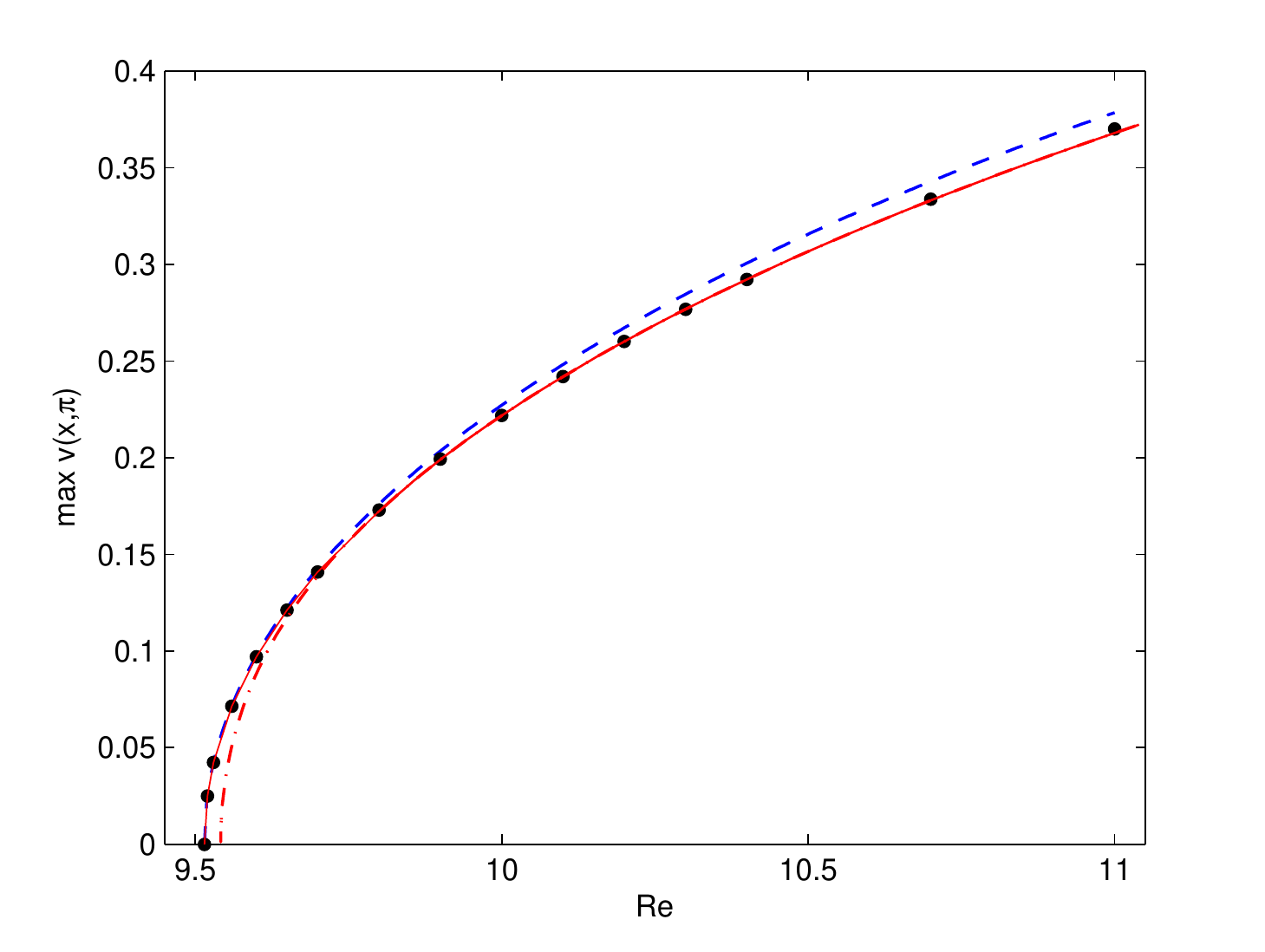}
\hspace{-0.75cm}
\includegraphics[width=7cm, clip]{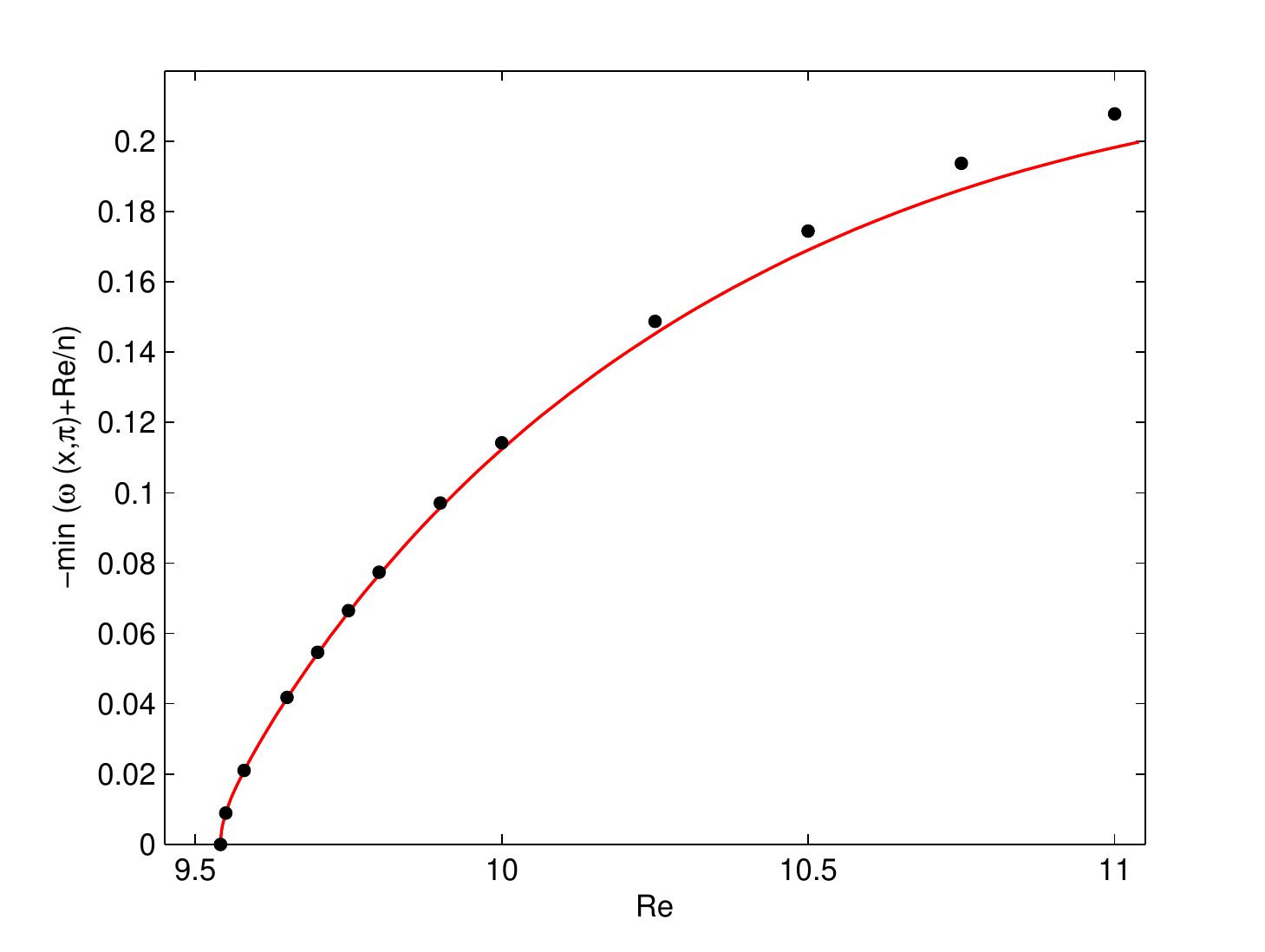}
\caption{Left, a comparison of the theoretical prediction ($\alpha
  \rightarrow \infty$; blue-dashed line) and the DNS
  ($\alpha=\sixteenth$; red solid line) for $\max_x\, v(x,\pi)$. The
  black dots are the results from the 3-PDE system. Numerical data for
  $\alpha=\quart$ is also shown as a red dash-dot line which only
  differs from the $\alpha=\sixteenth$ line near their (different)
  bifurcation points.  Right, the predictions of the 3-PDE systems
  (black dots) and the DNS data (red solid line) for $\min_x
  (\omega(x,\pi)+Re/n)$ in a domain $8 \pi \times 2 \pi$.}
\label{fig:comparison}
\end{center}
\end{figure}

Quantitative predictions for $v=-\psi_x$ are shown in figure
\ref{fig:comparison} together with the actual DNS values at
$\alpha=\quart$ and $\alpha=\sixteenth$ near the bifurcation point.
The $\alpha=\sixteenth$ DNS data is very close to the theory until $Re
\approx 11$ whereas the $\alpha=\quart$ DNS data differs near its
bifurcation point but by $Re \approx 10$ is on top of the
$\alpha=\sixteenth$ DNS data when the kinks have sufficiently
localised. No prediction can be made for the vorticity without
knowledge of $A^1(X,\eps^4 t)$ which presumably is determined by the
$O(\eps^4)$ solvability condition. Rather than systematically deriving
higher order amplitude equations in this way, we work with 3 (1-space
and 1-time) coupled PDEs instead which automatically incorporate the
first 4 amplitude functions and thereby give a much better long
wavelength approximation.

%
% fig 3
%

%
% Comparison figure  Dropbox/Kolmogorov/LK1/comparison.m and comp3D.m
%
% fig 4
%

\subsection{3-PDE long-wavelength system}\label{3_PDE}

\begin{figure}
\begin{center}
\psfrag{A}{{\color{cyan} $\omega_h$}}
\psfrag{B}{{\color{red}  $\omega_F$}}
\psfrag{C}{{\color{blue} $\omega_g$}}
\psfrag{D}{{\color{red}  $F$       }}
\psfrag{E}{{\color{cyan} $h$       }}
\psfrag{F}{{\color{blue} $g$       }}
\hspace{-0.25cm}
\includegraphics[width=7cm, height=6cm,clip]{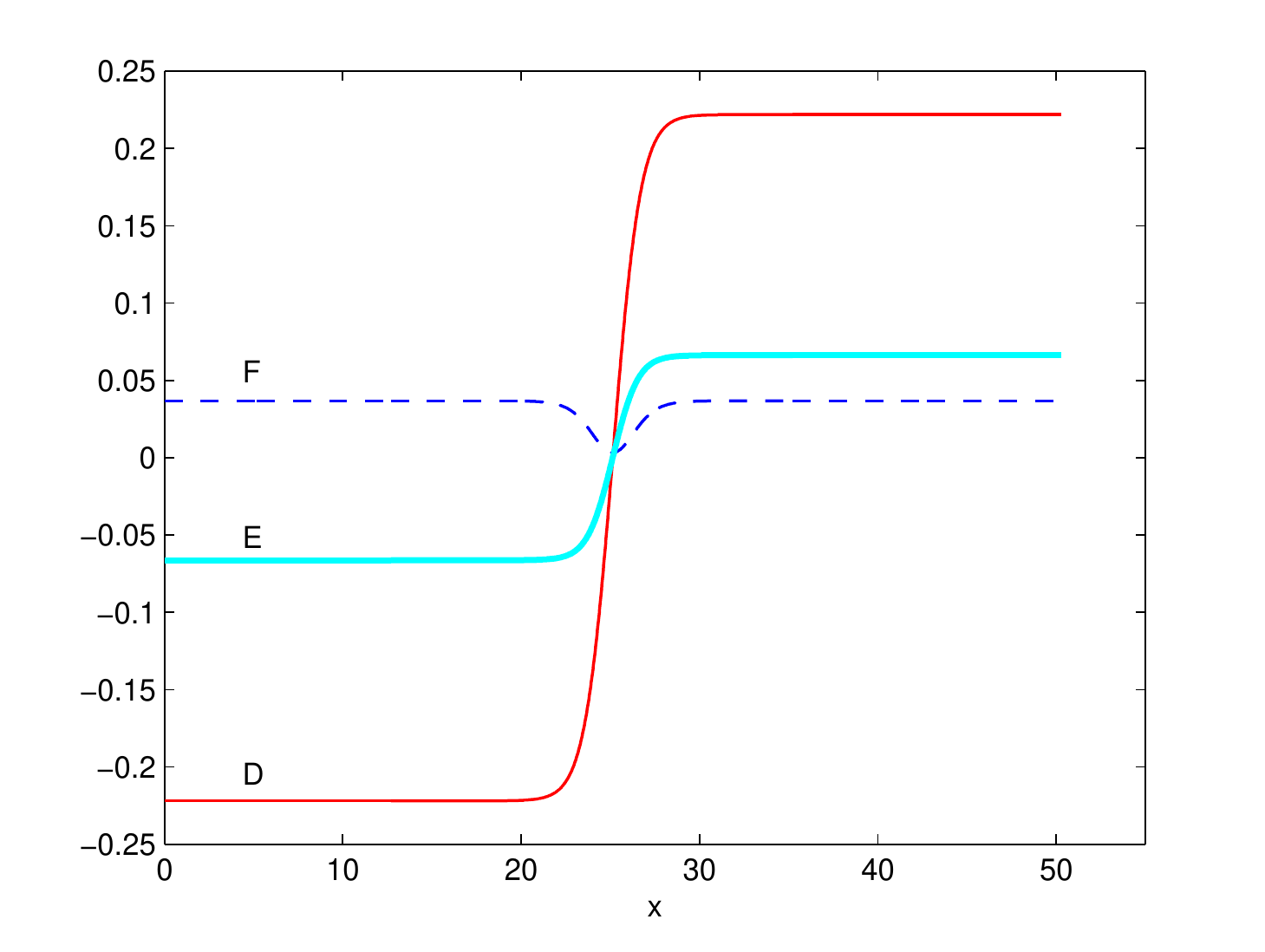}
\hspace{-0.75cm}
\includegraphics[width=7cm, height=6cm,clip]{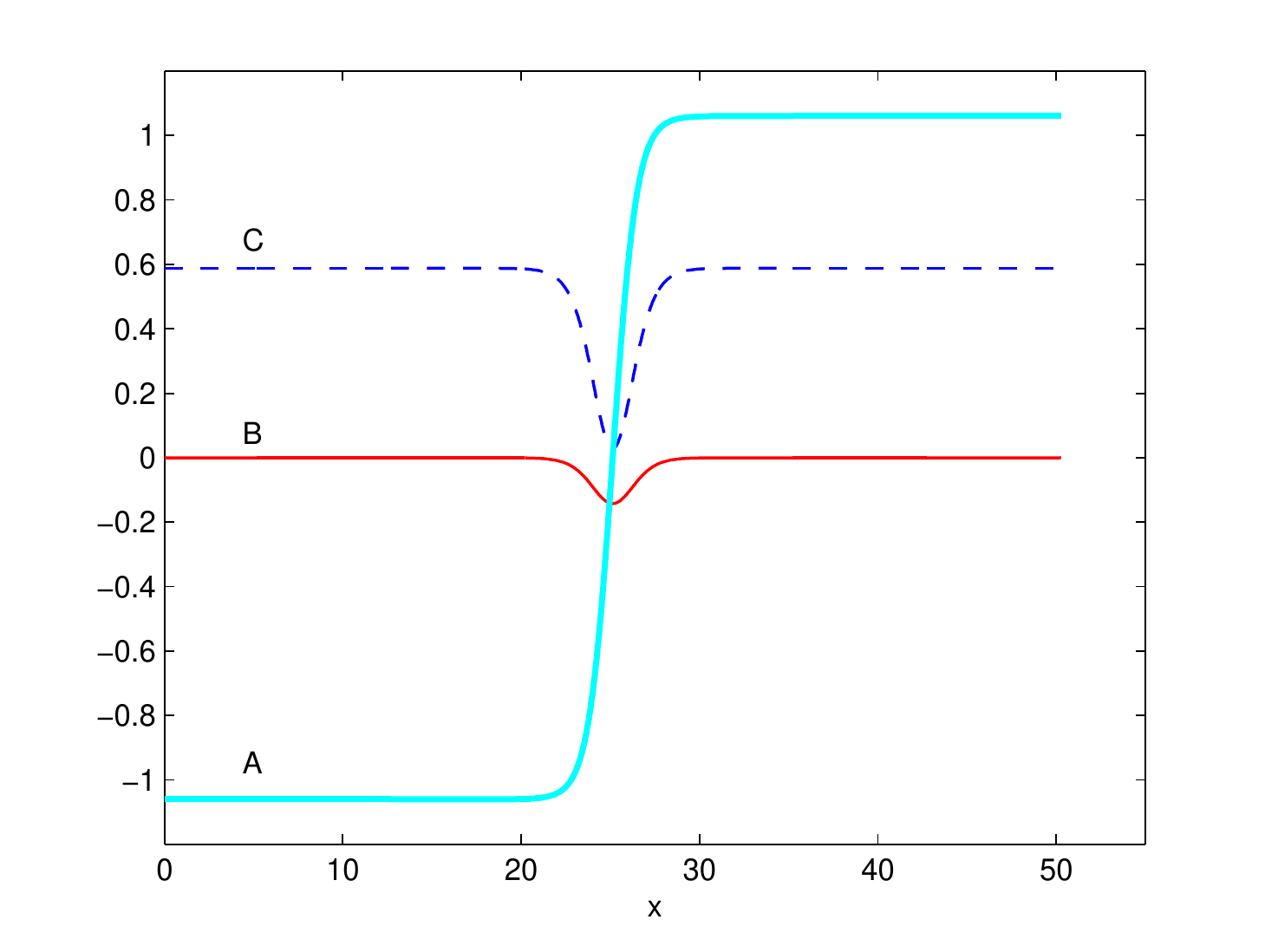}
\caption{The antikink solution at $Re=10$ in a domain $16 \pi \times 2 \pi$ ($\alpha=\eighth$) as calculated in the 3-PDE system with non-periodic boundary conditions and 4000 grid points confirming localisation. The left plot shows $F$ (red), $g$ (blue dashed) and $h$ (thick cyan), and the right plot shows $\omega_F$ (red), $\omega_g$ (blue dashed) and $\omega_h$ (thick cyan).}
\label{fig:isolate}
\end{center}
\end{figure}

The streamfunction (\ref{str}) which emerges out of the long
wavelength approximation involves just zero or first harmonics of the
forcing's periodicity in $y$ up to and including $O(\eps^3)$ (higher
harmonics $\sin 2ny$ and $\cos 2ny$ start to appear at
$O(\eps^4)$). Hence, rather than working with a sequence of amplitude
equations for $A^0$, $A^1$, $A^2$ and $A^3$, it is simpler to assume
the streamfunction takes the form
\begin{equation}
\psi(x,y,t)=\biggl(-\frac{Re}{n^3}+g(x,t) \biggr) \cos ny +h(x,t) \sin ny+ \int^x F(\xi,t) d\xi
\end{equation}
and work directly with the functions $g$, $h$ and $F$. 
Defining the associated
vorticities as $\omega_F:=-F^{'}$, $\omega_g:=-(g^{''}-n^2 g)$ and
$\omega_h:=-(h^{''}-n^2 h)$ where $(\quad)^{'}$ indicates a derivative
with respect to $x$, the higher order long wavelength system is then simply the 3 (1-space and 1-time) PDEs
\begin{eqnarray}
\frac{\partial \omega_F}{\partial t} &=& \frac{1}{Re} \omega_F^{''}+\frac{Re}{2n^2} (n^2 h-\omega_h)^{'}
+\half n (g \omega_h-h \omega_g)^{'}, \nonumber \\ 
\frac{\partial \omega_g}{\partial t} &=& \frac{1}{Re}(\omega_g^{''}-n^2 \omega_g)+n F \omega_h-n h \omega_F^{'}, \nonumber \\
\frac{\partial \omega_h}{\partial t} &=& \frac{1}{Re}(\omega_h^{''}-n^2 \omega_h) +\frac{Re}{n^2}(n^2 F-\omega_F^{'})+n(g \omega_F^{'}-F \omega_g). 
\label{3-PDE}
\end{eqnarray}
as opposed to the original 2 (2-space and 1-time) Navier-Stokes
PDEs.  As expected, this system produces a better long wavelength
prediction for $v$ than that available from (\ref{TheEquation}) {\em
  and} can yield a prediction for the vorticity too (see figure
\ref{fig:comparison}). Its real use here, however, (although see
section \ref{large_domain} later) is in studying the initial
bifurcation of the flow without the constraining influence of periodic
boundary conditions (coding up the 3 time-dependent PDEs is
straightforward using second-order finite differences - Crank-Nicolson
for the diffusive terms and Adams-Bashforth for the advective terms -
and typically 250-500 grid points per $2\pi$ length). The non-periodic
boundary conditions which do not allow energy to enter or leave the
domain are
\begin{equation}
g^{'}=h^{'}=\omega_g^{'}=\omega_h^{'}=\omega_F=\omega_g^{'}=\omega_h^{'}\biggl|_{x=0,L} \biggr.=0 \quad {\rm \&} \quad
\overline{F}:=\int^L_0 F \, dx={\rm const} 
\label{npbc}
\end{equation}
which represents conservation of momentum perpendicular to the forcing
direction (in the DNS code $\overline{F}=0$ by construction). As way
of confirmation, a linear stability analysis of the system
(\ref{3-PDE}) \& (\ref{npbc}) around
$(F,g,h,\omega_F,\omega_g,\omega_h)=0$ gives the same value of $Re_c$
as the periodic domain of twice the length (half of the wavelength of
the periodic instability fits into the non-periodic domain). In this
system, it is possible to confirm the localised nature of the
vorticity kinks and antikinks although direct time stepping to these attractors is very
inefficient due to the metastability of intermediate states (as
indicated in figure \ref{10vort}). However, a Newton-Raphson solver
converges easily given a reasonable starting guess from time stepping:
see figure \ref{fig:isolate}.

%
% Comparison figure  Dropbox/Kolmogorov/LK1/isolate.m
%
% fig 5
%

%%%%%%%%%%%%%%%%%%%%%%%%%%%%%%%%%%%%%%%%%%%%%%%%%%%%%%%%%%%%%%%%%%%%%%%%%%%%%%%%%%%%%%%%%%%%%%%%%%%%%%%%%%%%%%

%
% Fig 6
%
\begin{figure}
\begin{center}
\scalebox{0.9}{\input{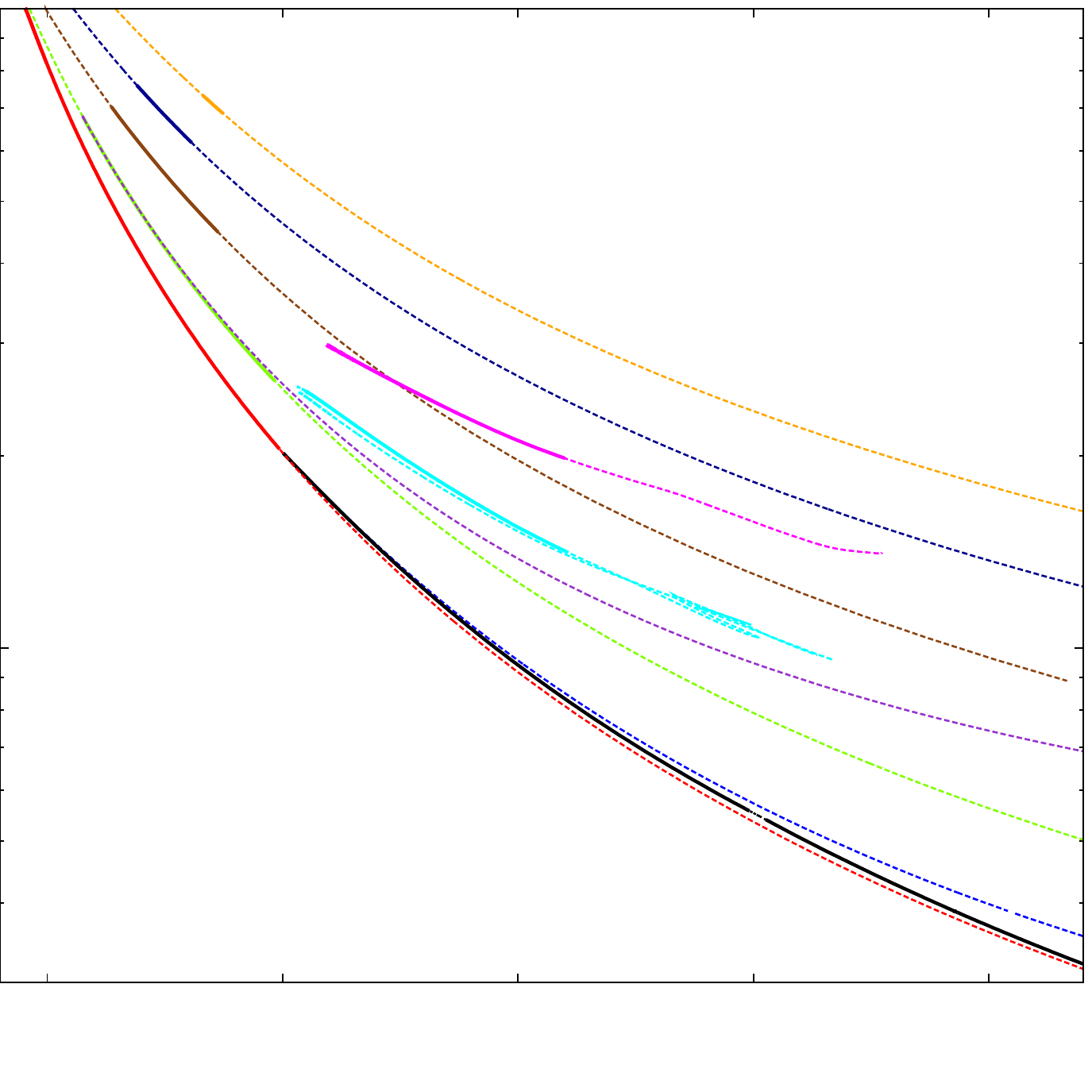}}
\caption{Dissipation against Reynolds number for $\alpha = \quart$,
  $n=4$ Kolmogorov flow. Curves indicate different $\kappa=m\alpha$ solutions together 
  with P1 and P2 solution branches. Thick solid lines
  represent stable solutions, dashed are unstable. Point (a) indicates
  the first solution, $k=2\alpha$, becomes stable at $Re=10.75$
  after bifurcating supercritically from the laminar flow. For the
  $\kappa=\alpha$ solution, the black curve represents the stable main secondary 
  branch bifurcating at (b) $Re=15$.}
\label{big_picture}
\end{center}
\end{figure}

%
% fig 7
%
\begin{figure}
\begin{center}
{\input{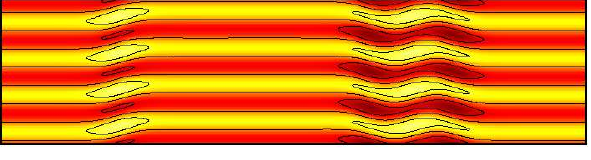}}
{\input{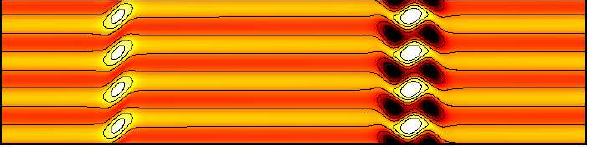}}
\caption{Vorticity fields for the `uneven' $\mathcal{R}$-asymmetric,
  $\mathcal{S}$-symmetric kink solutions ( which bifurcate off the
  primary $\kappa=2\alpha$ branch) at $Re=12$ (left) and 25
  (right). Colour extrema are $\omega=-5$ black, $\omega=5$ white, and
  5 evenly spaced contours in $-4\leq\omega\leq4$.}
\label{uneven}
\end{center}
\end{figure}

%
% fig 8
%
\begin{figure}
\begin{center}
%\scalebox{2}
{\input{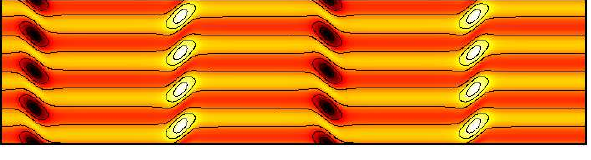}}
{\input{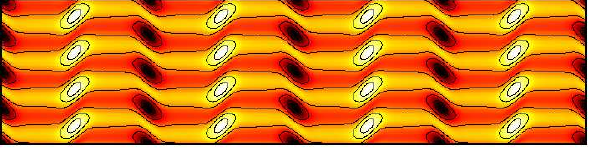}}\\
\vspace{4pt}
{\input{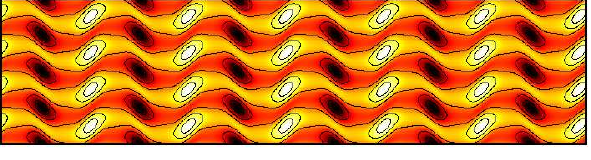}}
{\input{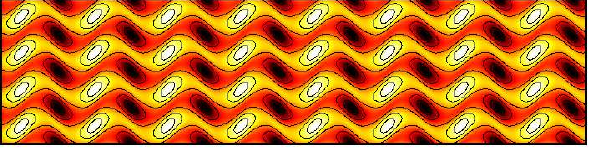}}
\caption{Vorticity fields for the primary 
  solutions $\kappa=2\alpha, 4\alpha, 6\alpha, 8\alpha$ at $Re=25$ in
  the $8 \pi \times 2 \pi$ ($\alpha=\quart$) domain. Colour extrema
  are $\omega=-5$ black, $\omega=5$ white, and 5 evenly spaced
  contours in $-4\leq\omega\leq4$.}
\label{alphavort}
\end{center}
\end{figure}

%-------------------------------------------------------------------------------------------------
%
% S-asymmetric dynamics   Re_c(0.25)= 9.5416 Re_c(0.5)=9.6258 see Documents/Kolmogorov/linear_str.m
%

\section{Bifurcations in the  $8\pi \times 2\pi$ ($\alpha=\quart$) domain }\label{sect:bifur}

We now consider bifurcations in a $8\pi \times 2\pi$ ($\alpha=\quart$)
domain.  Bifurcated solutions off the basic state (\ref{base}) are
referred to as {\em primary} solution branches, bifurcations off these
primary branches are termed {\em secondary} solution branches and so
on for {\em tertiary} solutions. Solutions can either have the
wavelength of the domain or have multiple wavelengths within the
domain. To indicate this, the smallest (base) wavenumber of the
solution, $\kappa$, is expressed as a multiple of $\alpha$: a
$\kappa=m\alpha$ solution has $m$ wavelengths across the domain so
that in the representation (\ref{representation}) $k_x=0,\pm m,\pm 2m,
\ldots$. Stability results are computed via Arnoldi iteration (using ARPACK) of the Jacobian matrix constructed during the Newton-GMRES-hookstep continuation.

\subsection{Steady bifurcations}

As stated above, the $\kappa=\alpha$ primary solution branch, which bifurcates
supercritically off the base flow (\ref{base}) at
$Re_c(\alpha)=9.542$, is steady and found to be the unique attractor
up to $Re=10.75$. At this point (point (a) in figure
\ref{big_picture}), the $\kappa=2 \alpha$ primary solution branch,
which bifurcates off the base flow at $Re(2 \alpha)=9.626$, gains
stability through a $\mathcal{R}-$breaking bifurcation.  The new
unstable secondary branch - hereafter christened the `uneven' branch -
is characterised by the uneven distribution in $x$ of the two
kink-antikink pairs across the domain. The two antikinks and one kink
quickly aggregate to form one composite antikink with much more
structure than the remaining kink: see figure \ref{uneven}. This
bifurcation is significant in that it results in coexisting, locally
attracting states for the first time and thereby provides an upper
limit on where the long-wavelength approximation - or `coarsening
regime' in Cahn-Hilliard language - is useful. The vorticity of the
$\kappa=2\alpha$ primary solution and other primary solutions with
$\kappa=4\alpha$, $\kappa=6\alpha$ and $\kappa=8\alpha$ are shown in
figure \ref{alphavort}.

%
% fig 9
%
\begin{figure}
\begin{center}
\input{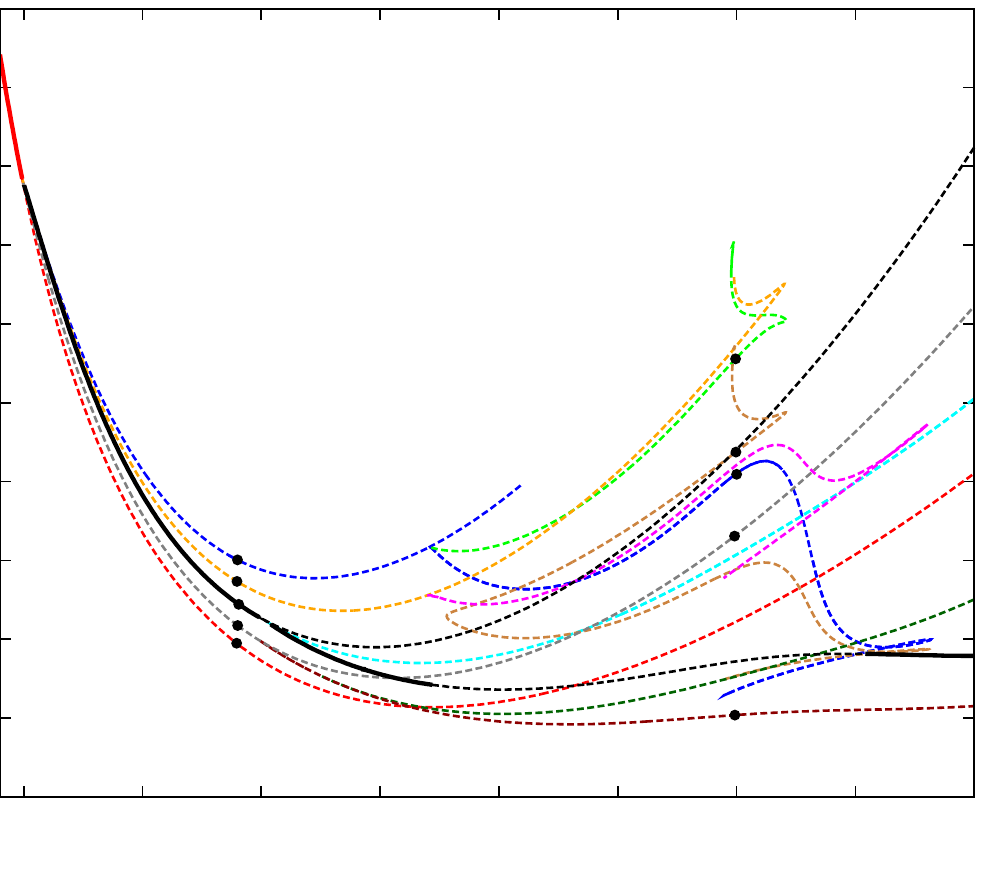}
\vspace{12pt}
\caption{Rescaled dissipation $Re^2 D/D_{lam} \propto \langle|\nabla
  \bm u |^2\rangle_V $ against $Re$ for steady primary and secondary
  solutions branches. Black dots correspond to vorticity fields
  in figure \ref{kinkvort}.}
\label{asym_plot}
\end{center}
\end{figure}

%
% fig 10
%
\begin{figure}
\begin{center}
\vspace{6pt}
{\input{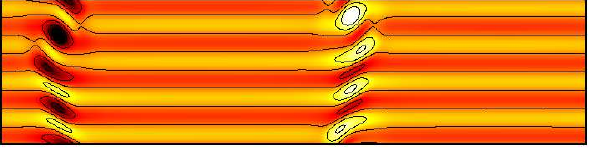}}
{\input{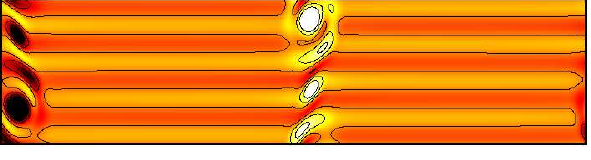}}\\
\vspace{4pt}
{\input{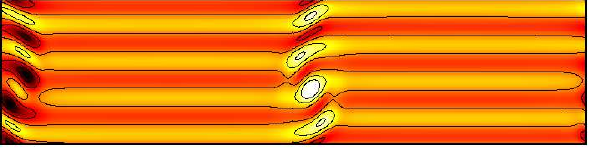}}
{\input{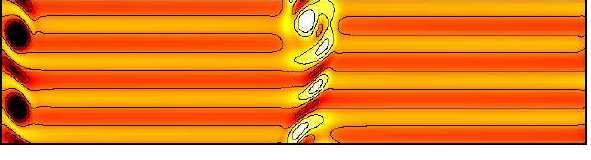}}\\
\vspace{4pt}
{\input{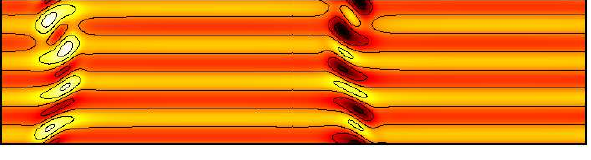}}
{\input{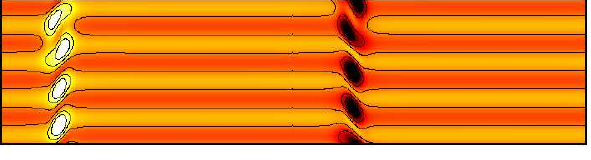}}\\
\vspace{4pt}
{\input{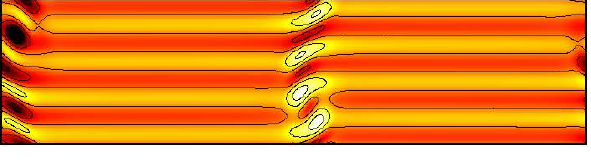}}
{\input{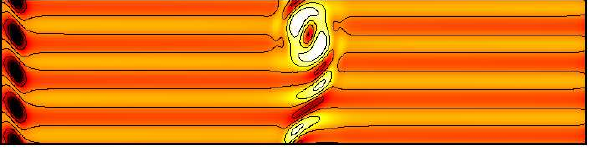}}\\
\vspace{4pt}
{\input{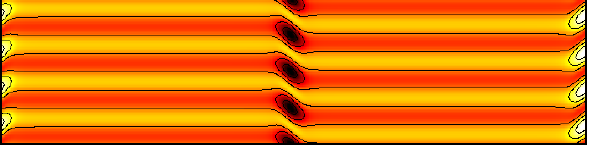}}
{\input{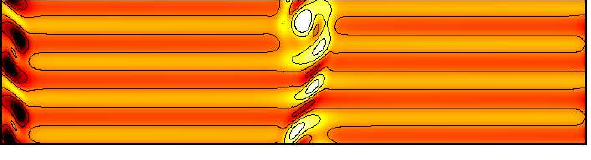}}
\caption{Vorticity fields for steady $\kappa=\alpha$ solutions at $Re=24$, left, and
  $Re=45$, right. Ordered from bottom to top in ascending dissipation
  ($D$) corresponding to black dots in figure \ref{asym_plot}. Colour
  extrema for vorticity are $\omega=-5$ black, $\omega=5$ white, and 5
  evenly spaced contours in $-4\leq\omega\leq4$. The states shown on
  the right have been chosen to illustrate (from top down to the
  bottom) both kink and antikink trying to localise, 2 localisation events in 
  antikink, little localisation, only localisation in the kink and  kink
  more localised than the antikink.}
\label{kinkvort}
\end{center}
\end{figure}

%
% fig 11
%
\begin{figure}
\begin{center}
{\input{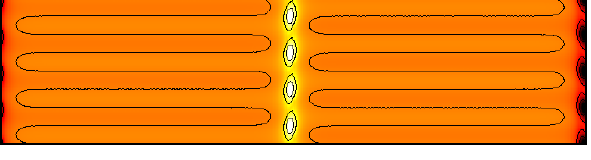}}
{\input{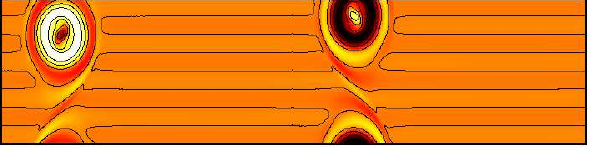}}\\

\vspace{6pt}
\caption{Vorticity fields for
  the primary $\kappa=\alpha$ branch (left) and the main secondary
  branch (right) at $Re=141$. Colour extrema are $\omega=-15$ black,
  $\omega=15$ white, and 5 evenly spaced contours in
  $-10\leq\omega\leq10$.}
\label{Re_141}
\end{center}
\end{figure}

%
% fig 12
%
\begin{figure}
\begin{center}
\scalebox{0.7}{\input{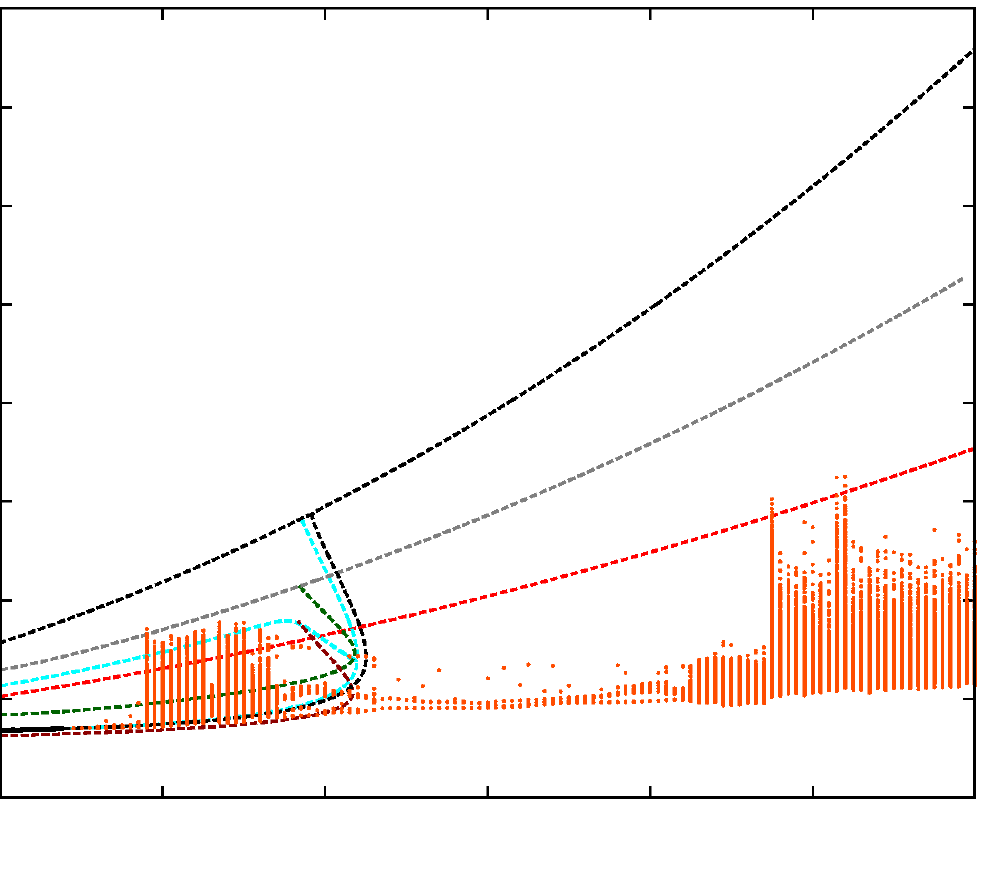}} \hspace{16mm}
\scalebox{0.75}{\input{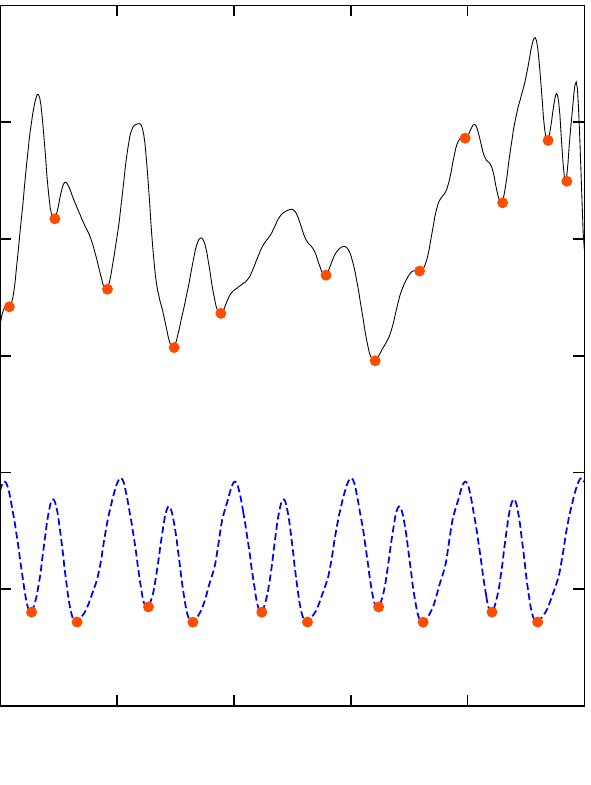}}
\vspace{12pt}
\caption{Left plot shows rescaled dissipation $Re^2 D/D_{lam}$ against
  $Re$ for the solutions shown in figure \ref{asym_plot} but now for
  $Re>60$ (red $\mathcal{S}$-symmetric primary branch and black main
  secondary $\mathcal{S}$-asymmetric branch, all line colours correspond to those in figure \ref{asym_plot} ) and time dependence
  displayed as local minima in DNS time series (orange points). Right
  shows a typical segment from the time series at $Re=72$ (black solid
  line) and $Re=90$ (blue dashed line) with circles indicating minima
  as shown in the left hand plot. At $Re=120$, (a) indicates the the
  primary $\kappa=\alpha$ branch, the middle grey line, labelled (c),  corresponds to the second state up on the right state in figure \ref{kinkvort}, and the upper black line, labelled (b) is the $\kappa=\alpha$ secondary
  branch drawn as the top black line in \ref{asym_plot} into which the
  main $\kappa=\alpha$ secondary branch terminates at $Re \approx
  80$. Note the other solution branches either terminate at other
  branches or can't be continued due to lack of numerical
  convergence.}
\label{kink_D}
\end{center}
\end{figure}

%
% fig 13
%
\begin{figure}
\begin{center}
\begin{tabular}{cc}%
\scalebox{1.15}{\input{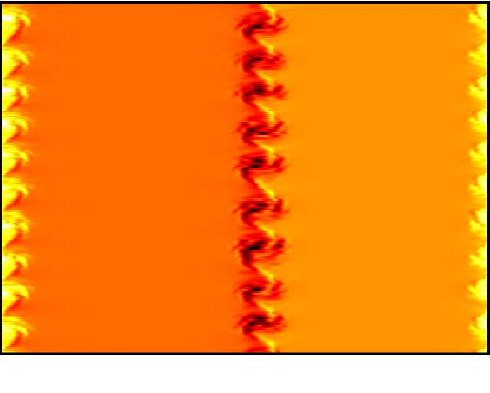}} &
\hspace{5mm}
\parbox[b]{50mm}{
\vspace{-0.5mm}
\scalebox{0.9}{\input{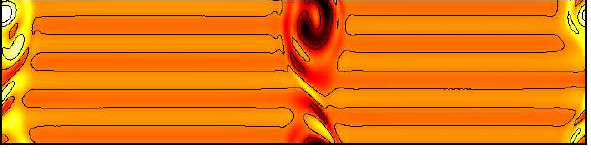}}\\
\scalebox{0.9}{\input{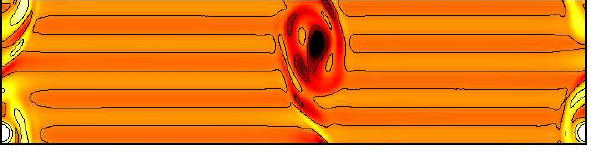}}\\
\scalebox{0.9}{\input{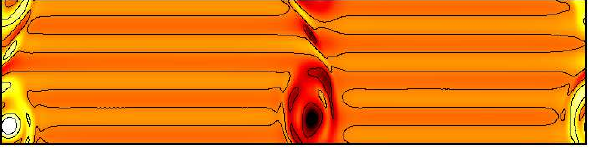}}
}
\end{tabular} 
\caption{DNS calculation at $Re=120$ which shows the chaotic
  oscillation of the kinks. Shown on the left is the variation in
  $\omega$ in an $(x,t)$ plane for $y=21\pi/32$ and the right
  corresponding snapshots of $\omega(x,y)$ at $t=50$ (bottom), $t=250$
  (middle) and $t=350$ (top). Colour extrema are $\omega=-8$ black,
  $\omega=8$ white, and 5 evenly spaced contours in
  $-8\leq\omega\leq8$ for $(x,y)$ plots.}
\label{120vort}
\end{center}
\end{figure}

%
% fig 14
%
\begin{figure}
\begin{center}
%\scalebox{2}
\scalebox{0.8}{\input{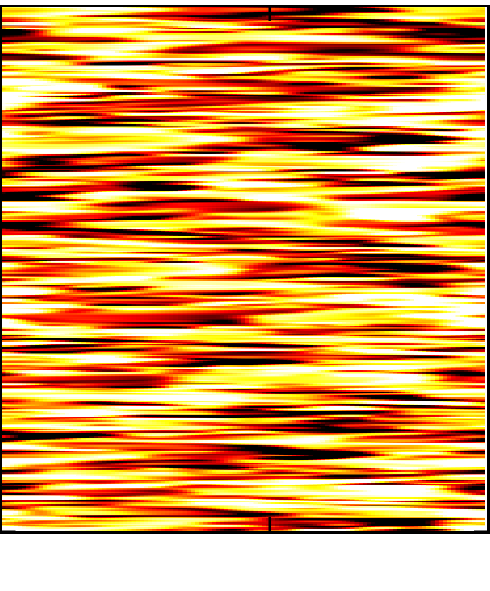}} 
\hspace{5pt}
\scalebox{0.8}{\input{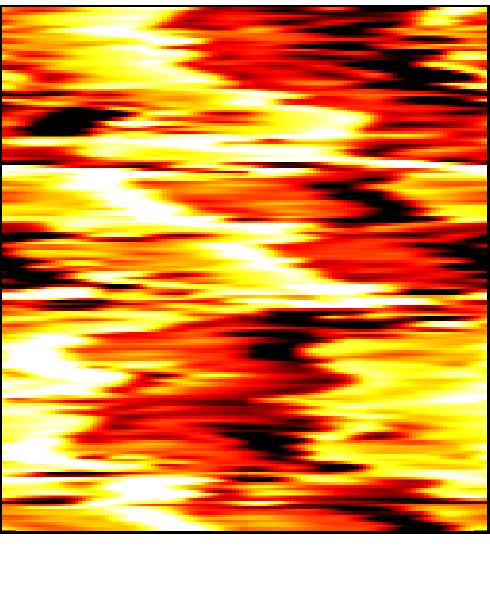}} \\
\vspace{5pt}
%\hspace{5pt}
\scalebox{0.8}{\input{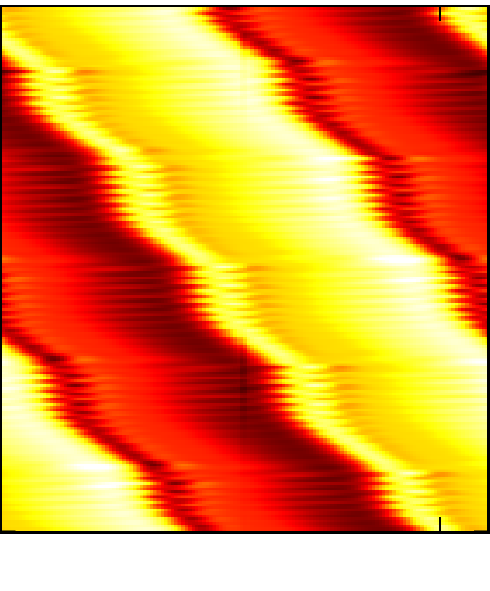}} 
\hspace{5pt}
\scalebox{0.8}{\input{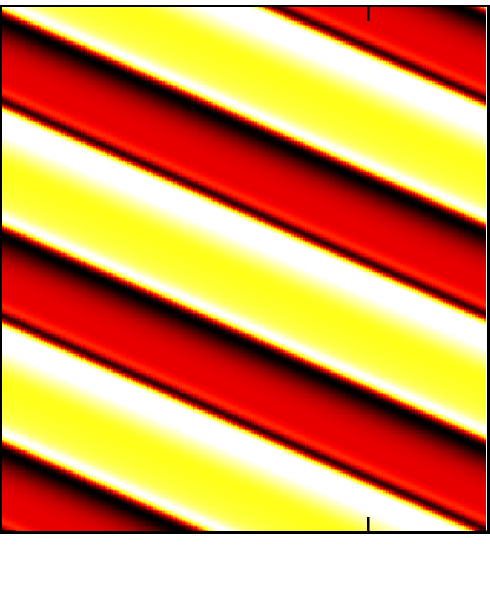}} 
\caption{Variation in $\omega$ in an $(x,t)$ plane for
  $y=21\pi/32$. Top shows chaotic attractors for $\alpha=1.1$ (left)
  and $\alpha=1$ (right), bottom left shows the translating and
  oscillating kinks for $\alpha=0.9$ and right the steady translating
  kinks for $\alpha=0.75$. All four are at $Re=40$ and are over a
  temporal window of $T=10^3$ following a longer run of $10^4$ time
  units to eliminate transients. Colours are $-3\leq\omega\leq3$ black
  to white.}
\label{XT_alpha}
\end{center}
\end{figure}

The primary $\kappa=\alpha$ solution becomes unstable at $Re \approx
15$ through two very close but separate $\mathcal{S}-$symmetry
breaking (modulational) bifurcations at $Re \approx 14.944$ and $Re
\approx 14.946$ each giving rise to two new distinct secondary
solution branches (point (b) in figure \ref{big_picture}).  Figure
\ref{asym_plot} shows all four new secondary branches and (a subset of
the) subsequent tertiary bifurcations over $15 \leq Re \leq 55$
produced via arc-length continuation (a rescaled dissipation is
plotted to draw out the variation in the solution curves which are
compressed in a $D$ vs $Re$ plot: see figure \ref{big_picture}). The
primary $\kappa=\alpha$ solution is the lowest (red) branch over $15
\lesssim Re \lesssim 30$ in figure \ref{asym_plot} and the middle
thick (black) line indicates one of the new secondary
$\mathcal{S}-$asymmetric branches - hereafter the `main secondary
branch' - which is found to be attracting from its birth near $Re=15$
up to $Re=32$ with only a small window of instability $24.97 < Re <
25.23$ (stable
solutions are shown using solid lines and unstable solutions using
dashed lines in Figure \ref{asym_plot}).  Plots of the vorticity
fields associated with the secondary branches (figure \ref{kinkvort})
indicate how the solutions vary subtly in their kink and antikink
structures.  The fact that kinks and antikinks of different `internal'
($y$) structure can be paired together will be discussed again in
section \ref{large_domain}.  A selection of tertiary solutions at
$Re=45$ are also shown in the same figure to illustrate the variety of
states which arise.  The trend is for the vorticity to accumulate in
the largest $y$-scale with the states varying in how effectively this
has occurred and whether it is happening equally in both kink and
antikink or just one.  The primary $\kappa=\alpha$ branch and its
secondary solutions can be continued to much higher $Re$ (where they are unstable) 
and then the
contrast in their vorticity structure becomes dramatic: see figure
\ref{Re_141}.

The stability of the multiple wavelength primary solutions is
summarised in figure \ref{big_picture}.  After gaining stability at
$Re=10.75$, the $\kappa=2\alpha$ primary branch becomes unstable again
at $Re=14.9$ through a modulational instability. The new solution
branches were not studied in detail and so only one secondary branch
(the `uneven' solution) is shown in figure \ref{big_picture}. The
$\kappa=4\alpha$ primary solution has a window of stability of $11.2 <
Re < 14.0$, the $\kappa=6 \alpha$ primary solution is stable over
$11.9 < Re < 13.1$ and the $\kappa=8 \alpha$ primary solution is
stable over $13.3< Re< 13.78$.

\subsection{Unsteady bifurcations}

Along the main $\kappa=\alpha$ secondary solution (thick black line in
figure \ref{asym_plot}), a Hopf bifurcation occurs at $Re \approx 32$.
A sweep of DNS is then performed from this bifurcation point, moving
in increments of $Re=0.5$ where the previous endstate (after a total
time of $T=5\times10^4$) is used to initiate the next run. Periodic
orbits are observed in the range $32\leq Re \leq 50$ after which the
main $\kappa=\alpha$ secondary solution becomes stable again over the
interval $50 \leq Re \leq 64$ (verified via Arnoldi).  For $64 < Re < 69$, we observe a
return to periodic behaviour based around the main secondary branch
until there is a window of chaos for $69 \leq Re < 77$: see figure
\ref{kink_D} (and supplementary video 1). Periodicity reappears for
$77 \leq Re <102$, a narrow window of quasiperiodicity follows before
sustained `kink-antikink' chaos sets in for $Re>103$.  Figure
\ref{120vort} shows a space-time plot of this chaos which is localised and
some snapshots of the flow illustrating this at $Re=120$. This state
appears to be the global attractor for $Re>120$; five different randomised initial conditions were simulated at $Re=200$ for $10^4$ time units and all settle upon the same $\kappa=\alpha$ kink-antikink chaotic attractor. We suspect that the transition to sustained kink-antikink chaos at $Re=103$ also indicates the threshold of this uniqueness.

\subsection{Kinks \& antikinks in smaller domains}

The ubiquity of these kink and antikink structures in sufficiently
large domains (where they can be easily recognised) poses the question
whether they are also relevant to the dynamics in smaller domains and
therefore previous work. The chaos previously studied in 2D Kolmogorov
flows has taken the form of spatially global states
(\cite{Platt:1991ed}, \cite{Chen:2004cz}, \cite{Chandler:2013fi}) in a
small $2 \pi \times 2 \pi$ domain. However, it is possible to pick out
the presence of kink-antikink pairs even here. Figure 7 from
\cite{Chandler:2013fi} shows some steady and travelling wave states at
$Re=40$ which are strikingly similar to the $\mathcal{S}$-asymmetric
branches shown in figure \ref{asym_plot}. In addition (their) figures
10 and 12 show unstable periodic orbits which seem to consist of a
coherent kink-antikink pair closely interacting (see also figure 24(b)
in \cite{Balmforth:2002}). Decreasing $\alpha$ from slightly above $1$
to below 1 has the dramatic effect of letting the kink and antikink
separate: see Figure \ref{XT_alpha} which shows $\omega(x,21\pi/32,t)$
in the $(x,t)$ plane for $Re=40$ and $\alpha=1.1$, $1$, $0.9$ and
$0.75$. For $\alpha=1.1$, the flow is organised into alternate signed
strips of vorticity (kinks and antikinks) which chaotically oscillate
and meander but are never able to spatially separate. This dynamics is repeated
at $\alpha=1$ but is noticeably less intense.  Intermittently the kink
and antikink undergo a more violent interaction which is discussed in
\cite{Chandler:2013fi} as high dissipation excursions or
`bursts'. When $\alpha$ is reduced to $0.9$, the character of the flow
changes considerably as the kink-antikink separation increases and a
coherent structure forms, oscillating and translating in a regular
fashion. A further expansion of the domain ($\alpha=0.75$), produces
an even simpler attractor where the kink-antikink pair, now quite
separated, just translates to the left with constant speed.

%%%%%%%%%%%%%%%%%%%%%%%%%%%%%%%%%%%%%%%%%%%%%%%%%%%%%%%%%%%%%%%%%%%%%%%%%%%%%%%%%%%%%%%%%%%%

%
% fig 15
%
\begin{figure}
\begin{center}
\begin{tabular}{cc}%
\scalebox{1.15}{\input{figs/fig15a }} &
\hspace{5mm}
\parbox[b]{50mm}{
\scalebox{0.9}{\input{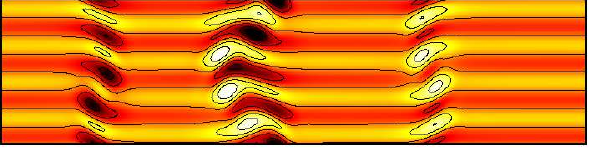}}\\
\scalebox{0.9}{\input{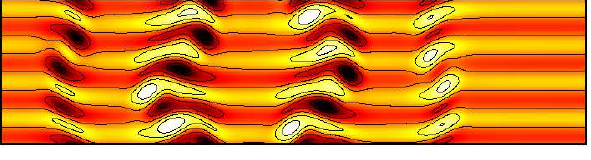}}\\
\scalebox{0.9}{\input{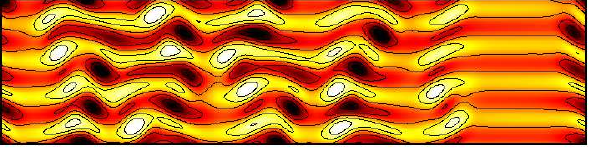}}\\
\scalebox{0.9}{\input{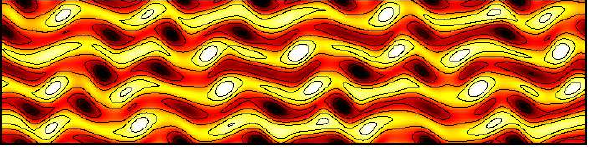}}\\
\scalebox{0.9}{\input{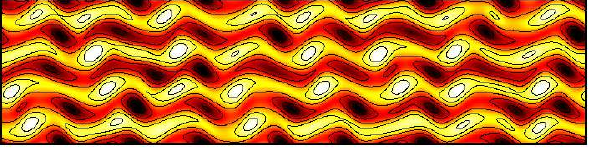}}
\vspace{1.6mm}
}
\end{tabular}
\caption{DNS calculation from a randomised initial condition at
  $Re=20$ which shows kink-antikink annihilation events. Shown on the left is  the variation in $\omega$ in an $(x,t)$ plane for $y=21\pi/32$ and
  the right snapshots of $\omega(x,y)$ at (from bottom to top)
  $t=1.25\times 10^4,5\times10^4,7.5\times10^4,8.5\times10^4$ and
  $10^5$. Colour extrema are $\omega=-5$ black, $\omega=5$ white, and
  5 evenly spaced contours in $-4\leq\omega\leq4$ for $(x,y)$ plots
  and $-3\leq\omega\leq3$ for $(x,t)$.}
\label{20vort}
\end{center}
\end{figure}

%
% fig 16
%
\begin{figure}
\begin{center}
%\scalebox{2}
\vspace{2pt}
\scalebox{0.75}{\input{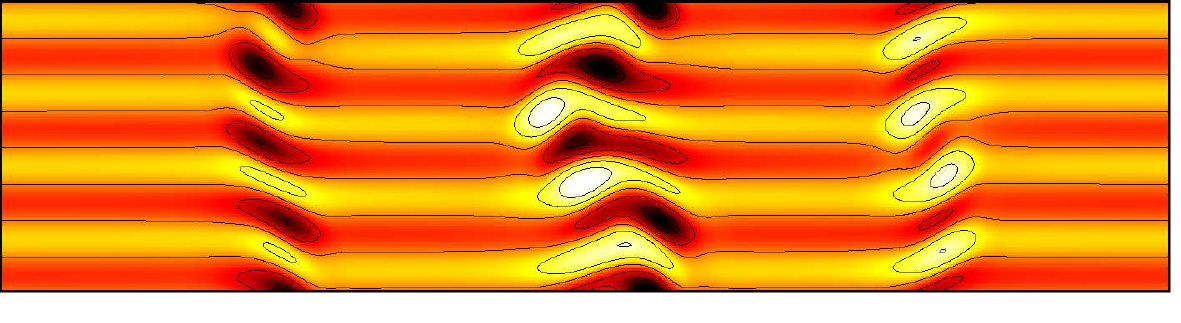}}\\
\vspace{2pt}
\scalebox{0.75}{\input{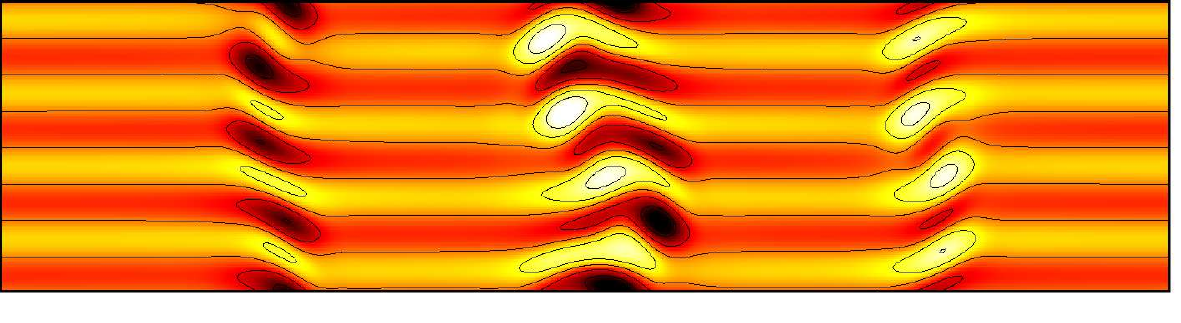}}\\
\vspace{2pt}
\scalebox{0.75}{\input{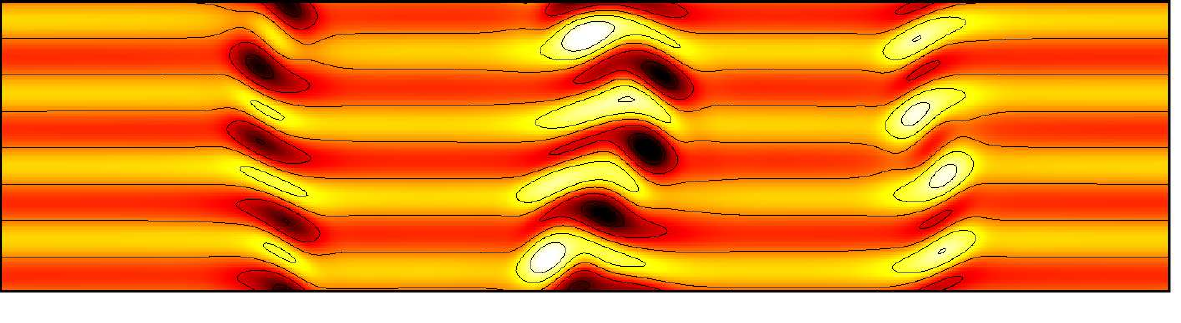}}\\
\vspace{2pt}
\scalebox{0.75}{\input{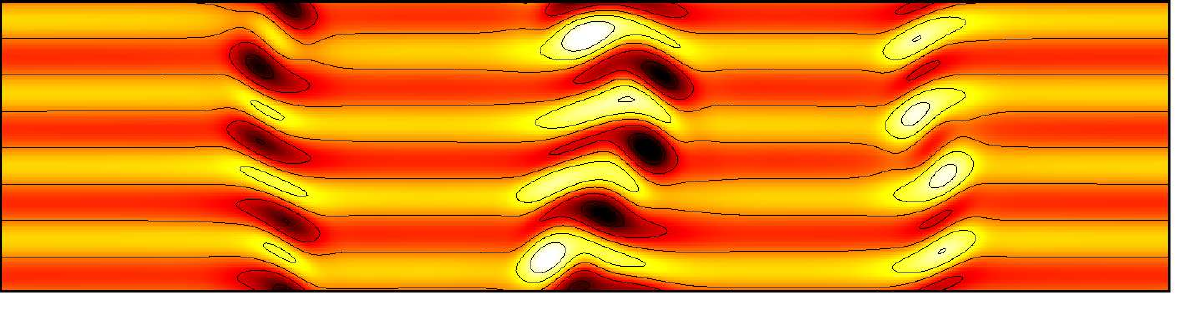}}\\
\vspace{2pt}
\scalebox{0.75}{\input{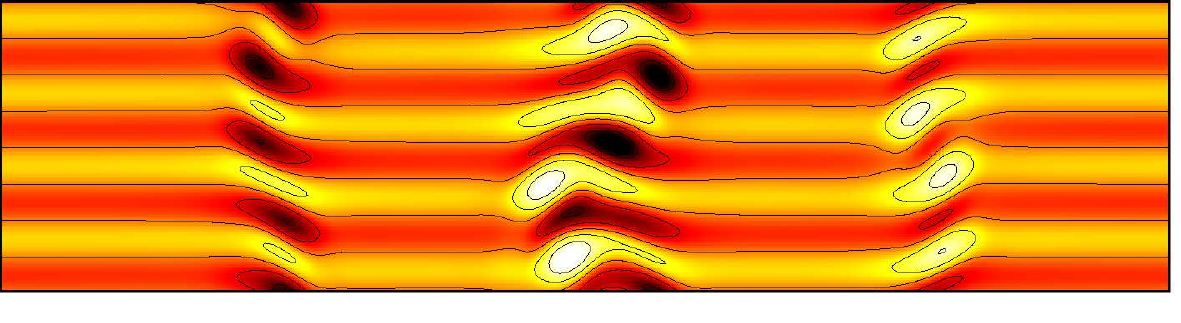}}\\
\vspace{2pt}
\scalebox{0.75}{\input{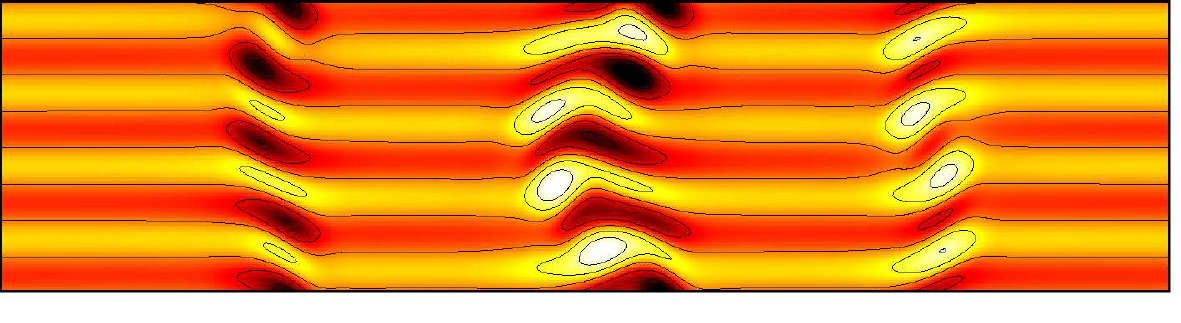}}\\
\caption{Vorticity fields for the $Re=20$ P1 orbit (period
  $T=20.61$) at $t=0,3.5,7,10.5,14,17.5$, colour extrema are
  $\omega=-5$ black, $\omega=5$ white, and 5 evenly spaced contours in
  $-4\leq\omega\leq4$ in a $8 \pi \times 2 \pi$ domain. The bordering
  kink/antikink look steady whereas the central snake oscillates
  periodically.}
\label{snakevort}
\end{center}
\end{figure}

%
% fig 17
%
\begin{figure}
\begin{center}
\scalebox{1.0}{\input{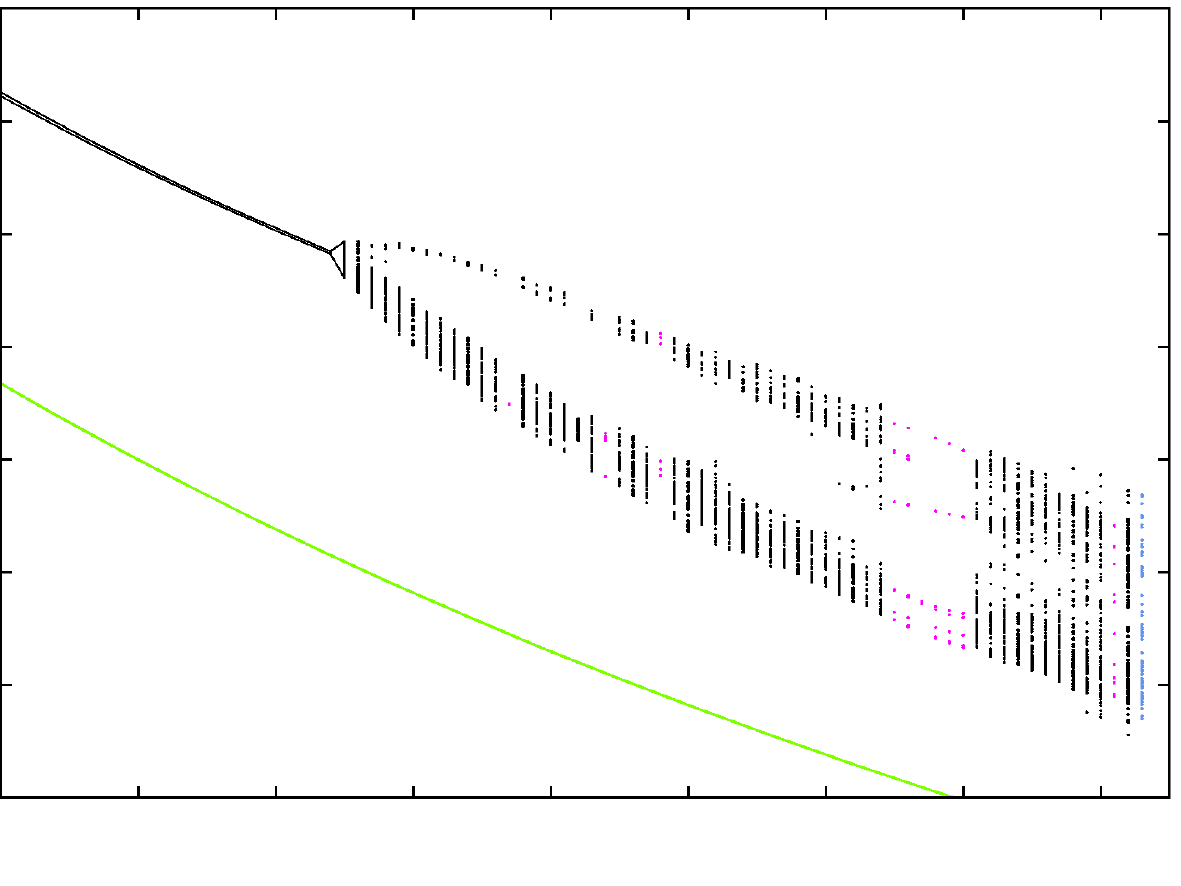}} 
\vspace{8pt}
\caption{Dissipation minima for DNS of the P1 solution branch taken
  from $T=10^5$ time series shown in black. Light green curve shows
  $\kappa=2\alpha$ primary steady state. At $Re\approx21.15$ we observe
  a torus bifurcation and quasi-periodic behaviour thereafter with
  occasional returns to periodic orbits, and full chaos at
  $Re=24$. For $T=10^5$, $Re=24.15$ is the first transient DNS along
  this sweep, however we find longer runs at $Re=24.1$ are also
  transient indicating that the boundary crisis occurs between $24.05$
  and $24.1$. Note we only plot 1\% of minima points to clearly
  indicate the structure of the various attractors. The post crisis
  transient DNS is shown in light blue (rightmost line of data) and periodic orbits are shown
  in magenta (light grey in b/w). }
\label{snk_branch}
\end{center}
\end{figure}

\subsection{P1: a disconnected solution}\label{sect:P1}
%
% introduce P1
%

Randomly-seeded DNS runs at $Re=20$ uncovered a new stable
time-periodic solution - hereafter labelled P1 - coexisting with a
steady kink-antikink solution at $Re=20$. Figure \ref{20vort}
documents one such run showing the familiar kink-antikink annihilation
or coarsening events but, rather than reaching the steady main
$\kappa=\alpha$ secondary branch, the final attractor is the new P1
state (note it took $O(10^5)$ time units to settle). This solution
appears to consist of a direction-reversing travelling wave
(\cite{LandKnob91}) flanked by a steady kink and antikink: see figure
\ref{snakevort} and video 2. \cite{LandKnob91} describe how a Hopf
bifurcation from a circle of steady states due to $O(2)$ symmetry
gives rise to an oscillatory drift along the group orbit of these
steady states (e.g. \cite{Alonso00}). There is certainly $O(2)$
symmetry present - $\mathcal{R} \times \mathcal{T}_l$ - but the
flanking kink and antikink do not appear to move: see figure
\ref{snakevort}. This suggests that P1 has arisen through a Hopf
bifurcation off a steady localised state with $O(2)$ symmetry
contained {\em between} the kink and antikink rather than the whole
global state. However this could not be confirmed because in this
domain P1 does not bifurcate off a steady state but instead is born
in a saddle node bifurcation (see below).

To confirm P1 was an exact solution and not a long-lived transient,
the Newton-GMRES-hookstep method described in \cite{Chandler:2013fi}
was used successfully to converge it as a `relative' periodic solution
of the Navier-Stokes equations. A `relative' periodic orbit is a flow
which repeats after a period $T$ {\it and} a drift in any homogeneous
direction of the system (see \cite{Chandler:2013fi} for more
discussion). At $Re=20$, the $L^2$ norm of the difference between the
velocity fields separated by a period $T=20.61$ and a shift in $x$ of
$-0.0002$ normalised by the $L^2$ norm of the starting field was
reduced to zero to machine accuracy. Once converged, P1 could be traced
using arc-length continuation over the range $15.03\leq Re \leq 26.68$
(at either end the Newton-GMRES-hookstep algorithm fails to converge).

\begin{figure}
\begin{center}
{\input{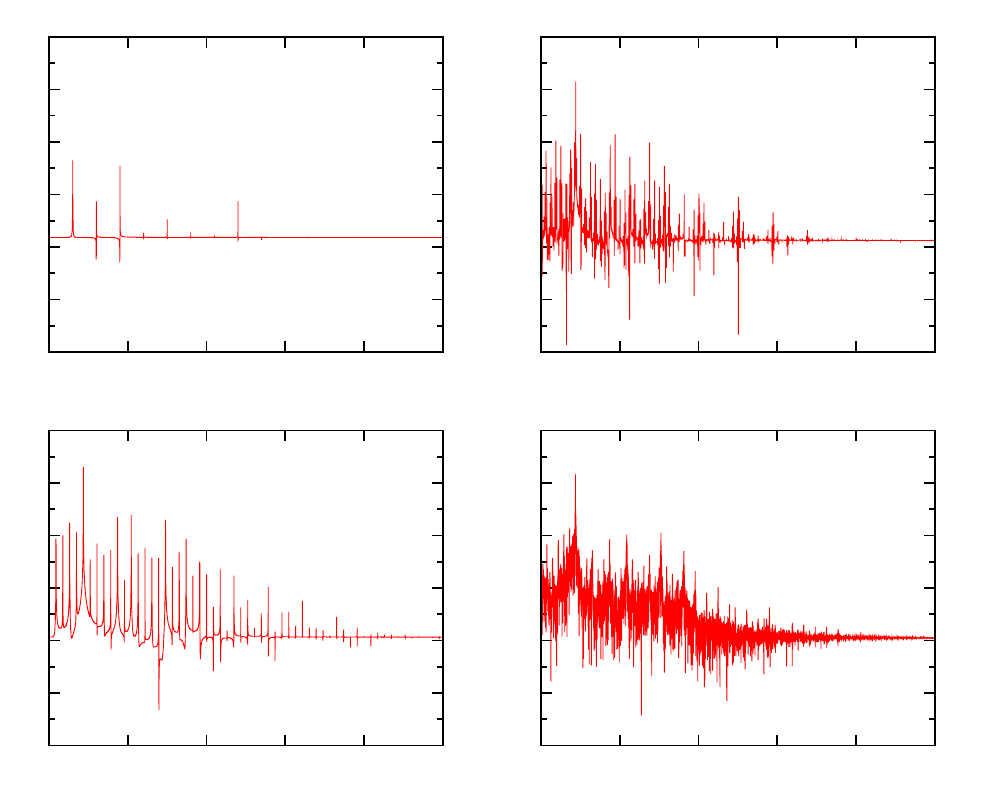}}
\vspace{24pt}
\caption{Power spectra for the time dependent signals of $D/D_{lam}$
  taken from the final $10^4$ time units of the P1 DNS sweep
  illustrated in figure \ref{snk_branch}. $Re=20$, $22.1$, $24.1$,
  $23.5$ moving clockwise from top left.}
\label{snk_pwrspc}
\end{center}
\end{figure}

\begin{figure}
\begin{center}
\fontsize{8}{10}\selectfont
\input{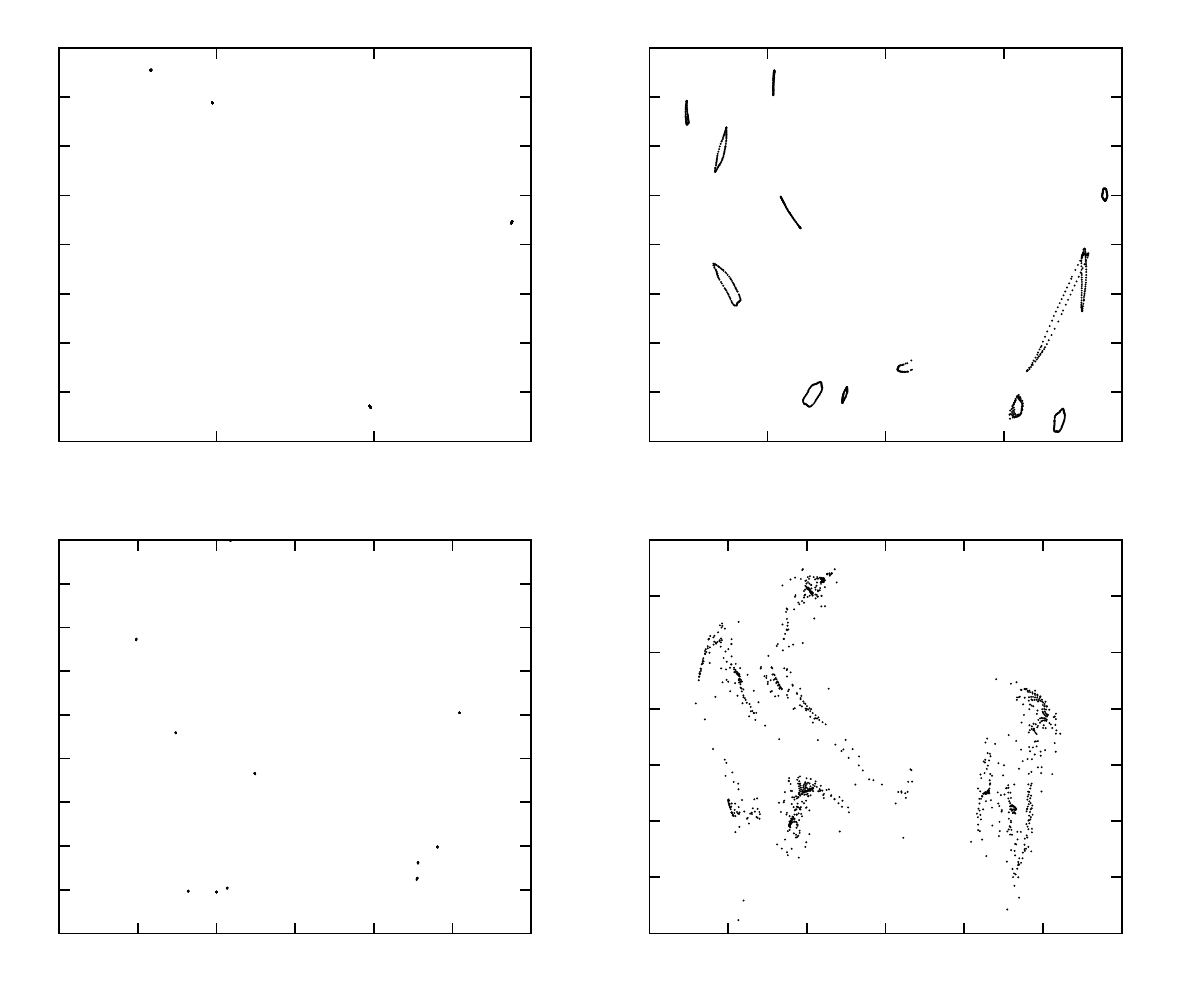}
\vspace{18pt}
\normalsize
\caption{Poincare sections for trajectories from the P1 DNS sweep
  illustrated in figure \ref{snk_branch}. The intersecting plane is
  taken to be cutting $D-I=0$ from below and we plot the variation in
  the $k_y=1, 2,$ $k_x=0$ components (chosen to negate the effect of
  $x$-translation). The final 1000 time units are taken and we show
  $Re=20$, $22.1$, $24$, $23.5$ moving clockwise from top left. These
  sections nicely highlight the periodicity, quasi-periodicity and
  chaos of the attractors at these Reynolds numbers.}
\label{snk_poinc}
\end{center}
\end{figure}

Near $15.03$, the solution branch becomes close to the primary
$\kappa=2\alpha$ solution (which has 2 kink-antikink pairs in a $8\pi
\times 2 \pi$ domain) as is perhaps to be expected since the
oscillatory centre of the P1 looks to be made up of a kink-antikink
pair. However, no connection was found and instead a partner branch of
periodic solutions is found consistent with a saddle node bifurcation.
To explore the stability of P1 and what happens to it beyond
$Re=26.68$, a sweep of DNS was undertaken starting from the $Re
\approx 17$ stable snake and incrementing $Re$ in $0.05$
steps. Initial conditions were taken to be the end state of the
previous $Re$ run and each run was integrated for $T=10^5$. The nature
of the solutions found is illustrated by plotting local minima of
dissipation $D_{min}$ from the time series against $Re$ in figure
\ref{snk_branch}. The first bifurcation experienced by P1 is
period-doubling which occurs at $Re=17.8$. After a number of further
period doublings, there is a torus bifurcation at $Re\approx 21.15$
indicated by the start of line filling in figure \ref{snk_branch} when
the dissipation minima are plotted. The difference in the power
spectra before (at $Re=20$ with one fundamental frequency of $0.30$)
and after (at $Re=22.1$ where there are two fundamental frequencies
$0.438$ and $0.064$ given dominant peaks in the spectrum at $0.438$ and $0.502$) this bifurcation is clear from figure
\ref{snk_pwrspc} as well as from the Poincar\'{e} sections in figure
\ref{snk_poinc}. Thereafter trajectories appear to be embedded in such
tori with sporadic returns to periodic behaviour with the most notable
being the interval $23.25 \leq Re \leq 23.5$ (see figure
\ref{snk_branch}).  For $Re>23.5$, the quasiperiodicity returns
followed by a chaotic regime at $Re=24$ (see figures \ref{snk_pwrspc}
and \ref{snk_poinc}), a periodic orbit at $Re=24.05$ and then an
apparent boundary crisis at $Re=24.1$ whereupon the state becomes
transient.

%
% fig 18
% old fig 20
%

%
% fig19
% old fig 21
%

%%%%%%%%%%%%%%%%%%%%%%%%%%%%%%%%%%%%%%%%%%%%%%%%%%%%%%%%%%%%%%%%%%%%%%%%%%%%%%%%%%%%%%%%%%%%%%%%%%%%%

\subsection{Boundary crisis for P1}\label{sect:BC}

\begin{figure}
\begin{center}
{\input{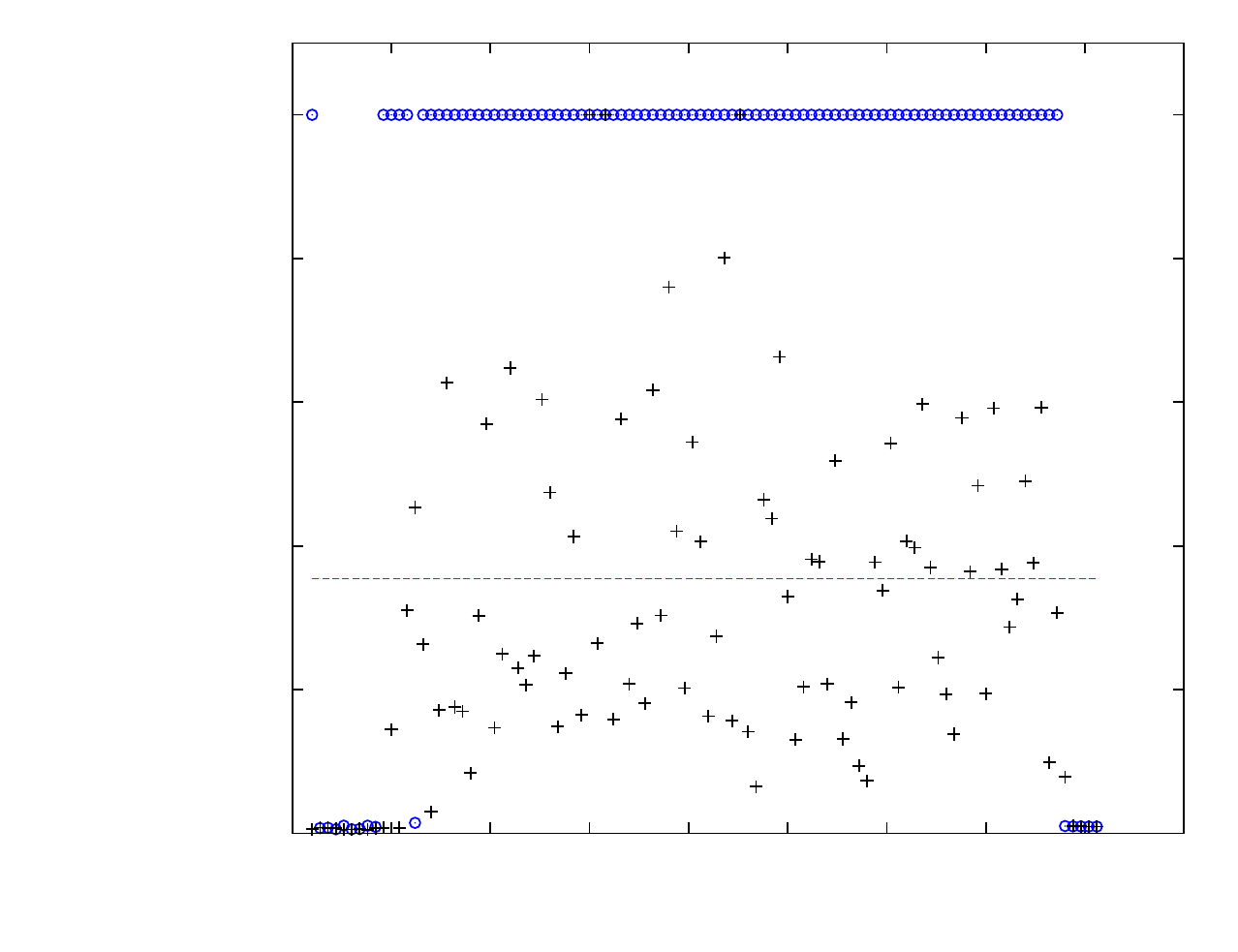}}
\vspace{6pt}
\caption{Lifetimes for the P1 chaotic attractor at $Re=24$ (blue
  circles) and chaotic transient at $Re=24.2$ (black $+$) against
  $\lambda$, the initial condition scaling. Also plotted as a dashed
  red line is the mean lifetime $\bar{\tau}=35,488.7$ for transient
  trajectories at $Re=24.2$.}
\label{lifetime}
\end{center}
\end{figure}

\begin{figure}
\begin{center}
%\scalebox{2}
{\input{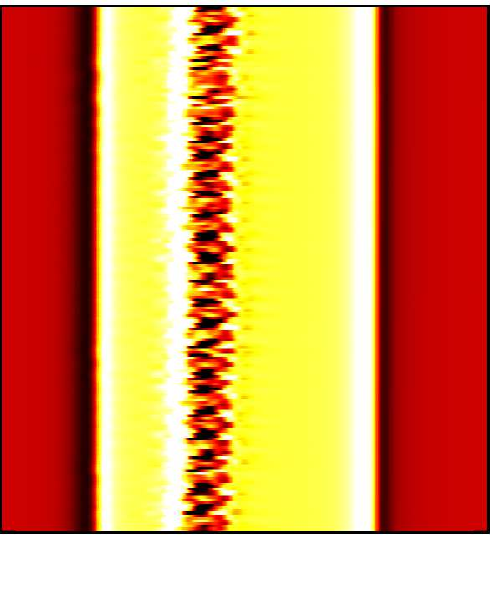}} 
\hspace{20pt}
{\input{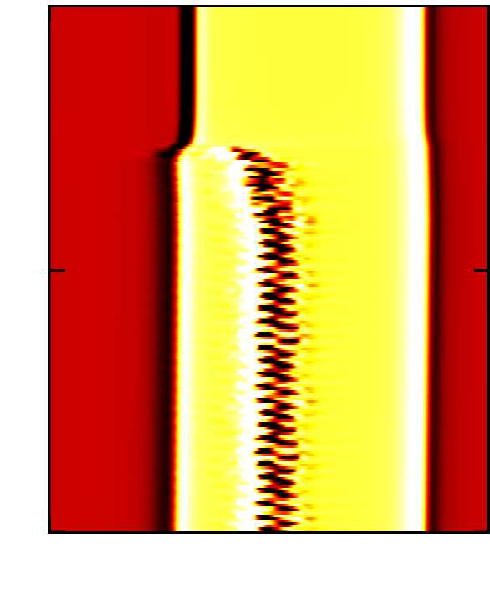}} 
\caption{Variation in $\omega$ in an $(x,t)$ plane for
  $y=21\pi/32$. Left shows the P1 chaotic attractor at $Re=24$ to
  $T=10^5$ and right shows a shorter time integration
  ($T=2\times10^4$) for the chaotic saddle at $Re=24.1$. Colours are
  $-2\leq\omega\leq2$ black to white.}
\label{XT_snk}
\end{center}
\end{figure}

\begin{figure}
\begin{center}
{\input{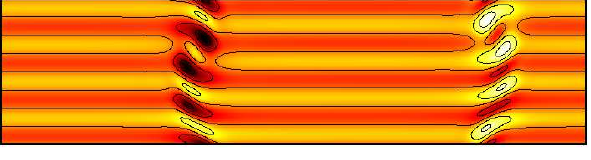}}
{\input{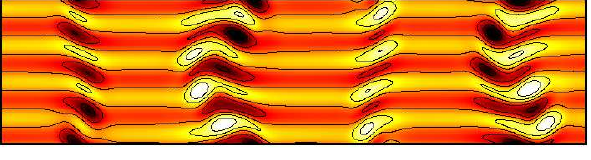}}\\ 
{\input{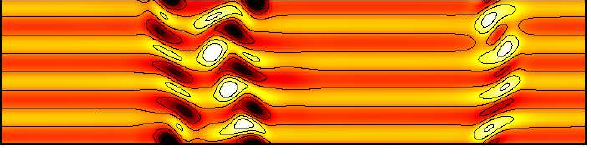}}
{\input{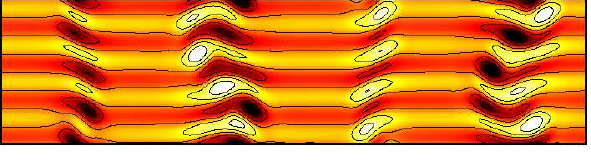}}\\
{\input{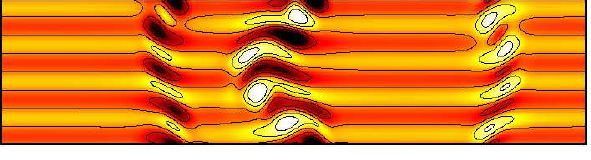}}
{\input{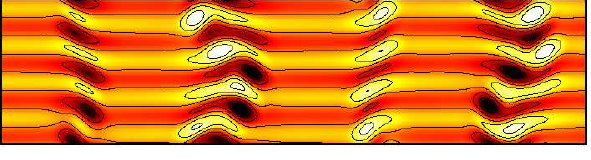}}\\
{\input{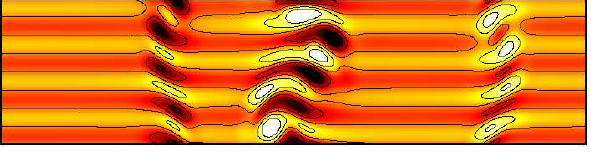}}
{\input{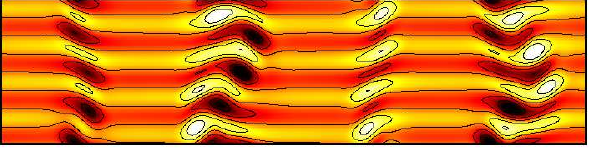}}
\caption{Left shows vorticity fields for the $Re=24.1$ P1 chaotic
  transient at $t=12000,14000,14500,15000$ (bottom to top), showing
  the departure from the chaotic saddle by $t=15000$. Right shows
  vorticity fields for the $Re=20$ P2 periodic orbit (period
  $T=21.42$) at $t=0,5,10,15$ (bottom to top). Colour extrema are
  $\omega=-5$ black, $\omega=5$ white, and 5 evenly spaced contours in
  $-4\leq\omega\leq4$.}
\label{sadvort}
\end{center}
\end{figure}

Given the recent interest in transient turbulence in wall-bounded
shear flows (e.g. \cite{Avila11} and references therein) and the fact
that for $Re \approx 24.1$ P1 appears to offer a simpler 2D
spatially-localised version, this transient state was examined
further. The growing range of the dissipation minima shown in figure
\ref{snk_branch} as $Re$ approaches the chaotic attractor-repellor
transition points towards a boundary crisis in which the chaotic
dynamics have grown with $Re$ to collide with its basin boundary. To
corroborate this hypothesis, lifetimes of trajectories initiated in
the former attractor were computed. A 1-D cross-section of initial
conditions across the former attractor was generated by taking the
periodic orbit converged at $Re=23.5$, $\omega_1(\bm x,t)$, and
systematically rescaling it as follows
\[ \omega_\lambda(\bm x, t=0) = \lambda \omega_{1}(\bm x,t=0)\]
where $\lambda$ is some real number such that $\lambda=1$ should be
in the attractor pre-crisis and for sufficiently larger and smaller
values, trajectories of $\omega_\lambda$ will exit (leading to quick
convergence to the kink attractor).  The lifetime within the
attractor, $\tau$, is defined as the time until the normalised
dissipation $D(t)/D_{lam}$ falls below some threshold deemed to be
outside the basin of attraction. Given at $Re\geq24.1$ we observe
trajectories attracted to the main $\kappa=\alpha$ secondary solution,
which has significantly lower dissipation, the lifetime threshold is
set to be $D_{tresh}=0.07$ (see figure \ref{snk_branch}). The total
time integration is capped at $T=10^5$ and 100 steps are taken in
$\lambda$ across the interval $[0.76,1.16]$.  Pre-crisis at $Re=24$,
all initial conditions $0.8 \lesssim \lambda \lesssim 1.15$ stay in
the attractor (over the $10^5$ time interval) whereas post-crisis at
$Re=24.2$, the lifetimes vary enormously across the small steps taken
in $\lambda$ consistent with a chaotic saddle: see figure
\ref{lifetime}. As a further check, a chaotic saddle should exhibit
lifetimes which are exponentially distributed indicating that the
probability of leaving the saddle in a given time period depends only
on the length of the period. Checking this property, however, requires
extensive DNS to build up enough data and as a result was not pursued
further. In figures \ref{XT_snk} (see video 3) and
\ref{sadvort} (left), we show a typical ending for the chaotic saddle
at $Re=24.1$. Efforts to analyse this chaotic saddle and attractor
using the recurrent-flow analysis performed in \cite{Chandler:2013fi}
will be reported elsewhere.

\subsection{P2: another disconnected solution}

In addition to the P1 solution discovered as an attractor at $Re=20$,
further DNS runs lead to another related solution stable for $Re
\lesssim 21$.  This orbit - labelled P2 - has period $T=21.42$ and
$x$-shift $s=-0.0015$ at $Re=20$ and has two regions of
spatially-localised oscillations: see figure \ref{sadvort} (right) and
video 4. It is distinct from P1 which only has its oscillatory part in the
region where the fluid is moving in the $-\bm{\hat{y}}$ direction. P2
has an oscillatory region in both the $-\bm{\hat{y}}$ {\em and}
$+\bm{\hat{y}}$ moving regions (see also figure
\ref{big_picture}). Exploratory computations indicate that P2
undergoes a similar bifurcation sequence to (attractive) chaos before
becoming a chaotic repeller. For $25 \lesssim Re$, the only attractors found were longest-wavelength solutions in which a single kink-antikink pair exist and have a variety of time-dependent behaviours but with no mean motion.

\begin{figure}
\begin{center}
\begin{tabular}{cc}%
\scalebox{1.08}{\input{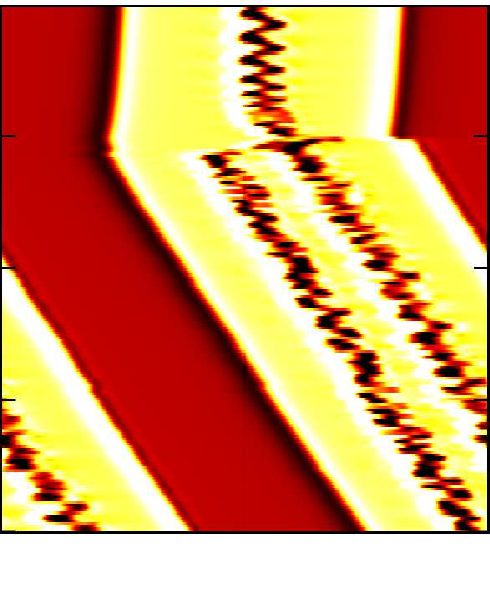}} &
\parbox[b]{50mm}{
\scalebox{0.9}{\input{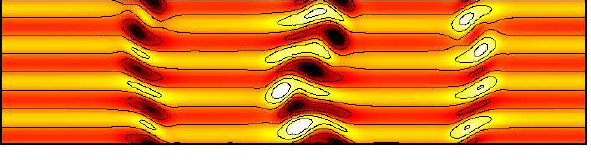}}\\
\scalebox{0.9}{\input{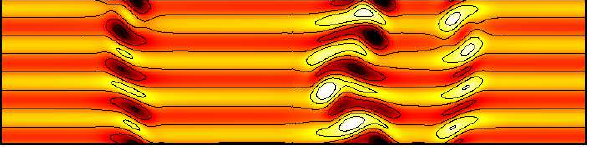}}\\
\scalebox{0.9}{\input{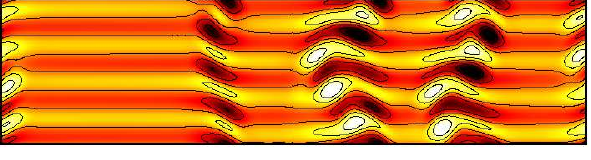}}\\
\scalebox{0.9}{\input{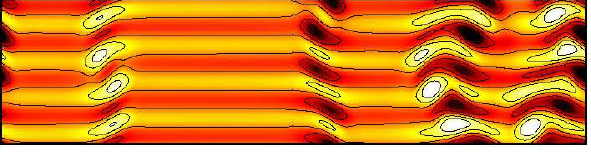}}
\vspace{1.6mm}
}
\end{tabular}
\caption{DNS calculation at $Re=20.75$ showing a double oscillation
  long-lived transient. Shown on the left is the variation in $\omega$
  in an $(x,t)$ plane for $y=21\pi/32$ and the right snapshots of
  $\omega(x,y)$ at (from bottom to top) $t=2.5\times
  10^4,5\times10^4,7.5\times10^4$ and $10^5$. Colour extrema are
  $\omega=-5$ black, $\omega=5$ white, and 5 evenly spaced contours in
  $-4\leq\omega\leq4$ for $(x,y)$ plots and $-2\leq\omega\leq2$ for
  colours in $(x,t)$.}
\label{dblevort}
\end{center}
\end{figure}

Another long-lived but ultimately transient state was also found at
$Re=20.75$ and is shown in figure \ref{dblevort} for $t<7.5\times10^4$
(see video 5). This state has two spatially localised time-dependent
patches \textit{within} the same flanking kink-antikink pair and
appears chaotic before ultimately settling down to P1 (no attempt was
made to identify a stable version of this state by reducing $Re$).
The evolution shown in figure \ref{dblevort} (left) highlights the
differing translational speeds of this chaotic transient and P1. The
co-existence of apparently-localised flow structures with different
translational speeds begs the question: in a large domain where they
can coexist spatially, what will happen when they collide?

%%%%%%%%%%%%%%%%%%%%%%%%%%%%%%%%%%%%%%%%%%%%%%%%%%%%%%%%%%%%%%%%%%%%%%%%%%%%%%%%%%%%%%%%%%%%%%%%%%%%%%%%%%%%%%%%%%%%%%%%%%%%%%

\section{Behaviour in a $16\pi \times 2 \pi$ ($\alpha=\eighth$) domain} \label{large_domain}

\begin{figure}
\begin{center}
\begin{tabular}{cc}%
\scalebox{0.95}{\input{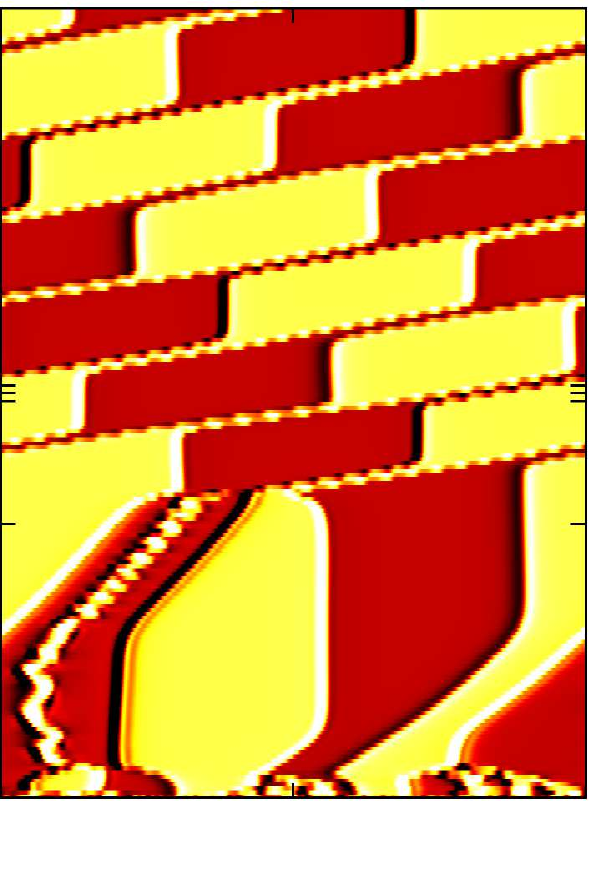}} &
\parbox[b]{80mm}{
\scalebox{0.8}{\input{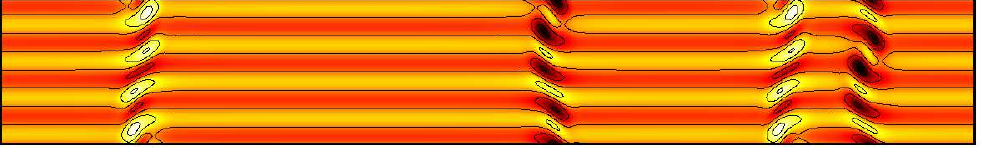}}\\
\scalebox{0.8}{\input{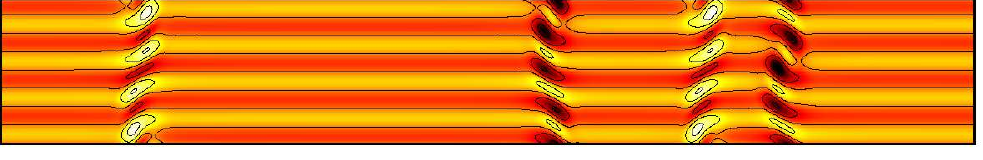}}\\
\scalebox{0.8}{\input{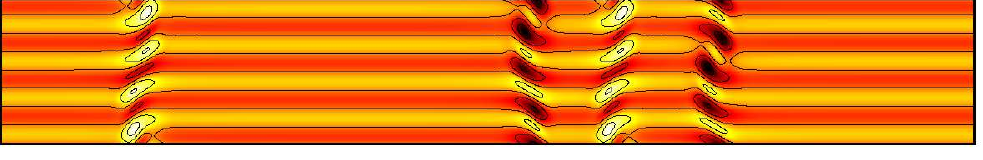}}\\
\scalebox{0.8}{\input{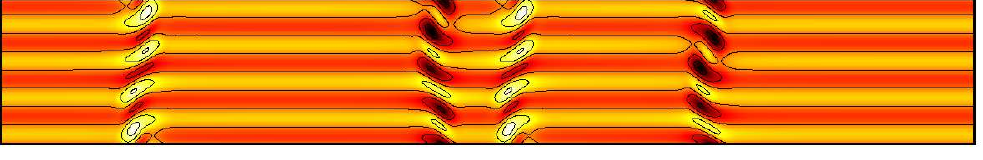}}\\
\scalebox{0.8}{\input{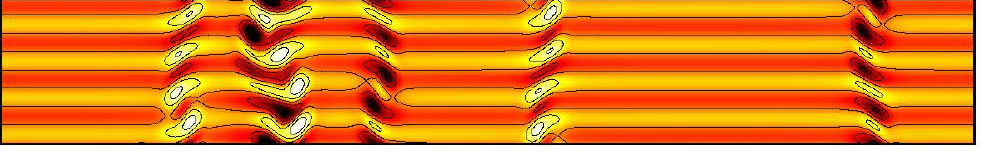}}\\
%\scalebox{0.9}{\input{figs/kol_100}}\\
\scalebox{0.8}{\input{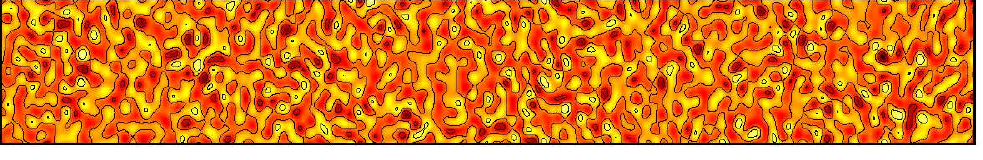}}
\vspace{1.6mm}
}
\end{tabular}
\caption{DNS calculation for randomised initial condition in
  $\alpha=1/8$, $Re=22$. Shown on the left is the variation in
  $\omega$ in an $(x,t)$ plane for $y=21\pi/32$ and the right
  snapshots of $\omega(x,y)$ at (from bottom to top)
  $t=0,3.4\times10^4,4.85\times 10^4,4.95\times10^4, 5.05\times10^4$
  and $5.15\times10^4$ (times indicated as black tics on the left
  figure). Colour extrema are $\omega=-5$ black, $\omega=5$ white, and
  5 evenly spaced contours in $-4\leq\omega\leq4$ for $(x,y)$ plots
  and $-2\leq\omega\leq2$ for colours in $(x,t)$.}
\label{kolmo}
\end{center}
\end{figure}

To investigate the possibility of different flow structures
interacting, some exploratory DNS calculations were performed with
randomised initial data in an extended $16 \pi \times 2 \pi$ domain
($\alpha=\eighth$). At $Re=20$ and $25$, the initial data gradually
evolved into the familiar secondary $\kappa=\alpha$ solution but at
$Re=22$, differentially-propagating localised states quickly emerged
out of the initial data. These interacted and ultimately formed a
periodic state in which kinks and antikinks repeatedly collide. Figure
\ref{kolmo} shows the total time history on the left and some selected
snapshots of the vorticity field on the right (see video 6).  At
$t=3.4 \times 10^4$, there is a state resembling a compressed version
of P1 propagating slowly in the positive $x$-direction and a kink and
antikink some distance apart to the right of it. At $4.85 \times
10^{4}$, the P1-like state has broken down into a kink-antikink pair
making a total of two pairs in the domain. One pair has become close
and propagates as a coherent unit with constant speed. When this
antikink-kink pair collides with the stationary antikink (e.g. $t=4.95
\times 10^4$) from the left, the leftmost antikink becomes stationary
and a new kink-antikink pairing is created which then moves off at the
same speed after the `collision' (e.g. $t=5.05 \times 10^4$). This
swapping phenomenon repeats periodically in time, due to the periodic
 boundary conditions. The period was too large to converge
using the Newton-GMRES-hookstep method.

\begin{figure}
\begin{center}
\begin{tabular}{cc}%
\scalebox{0.85}{\input{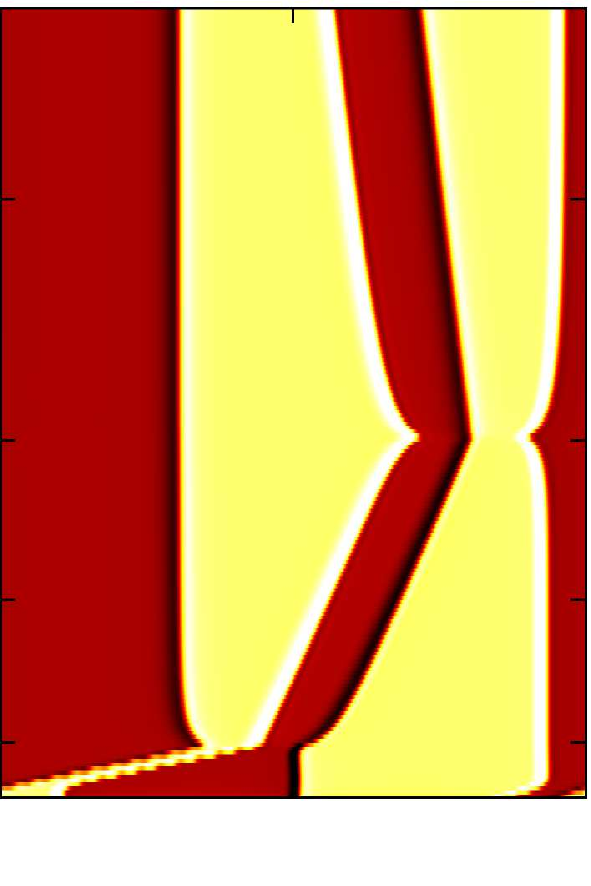}} &
\parbox[b]{80mm}{
\scalebox{0.8}{\input{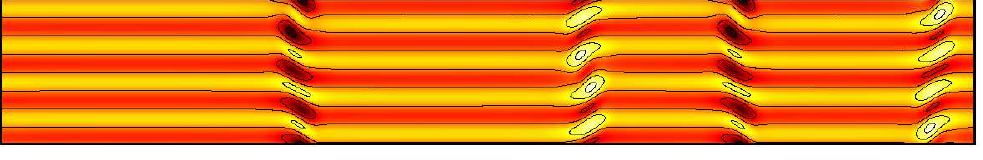}}\\
\scalebox{0.8}{\input{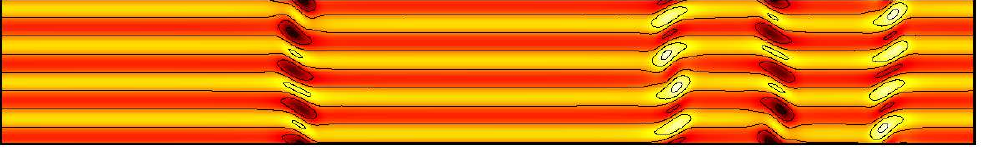}}\\
\scalebox{0.8}{\input{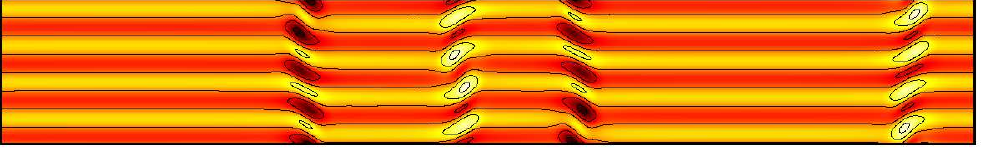}}\\
\scalebox{0.8}{\input{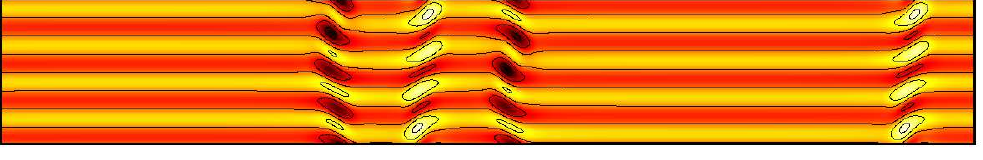}}\\
\scalebox{0.8}{\input{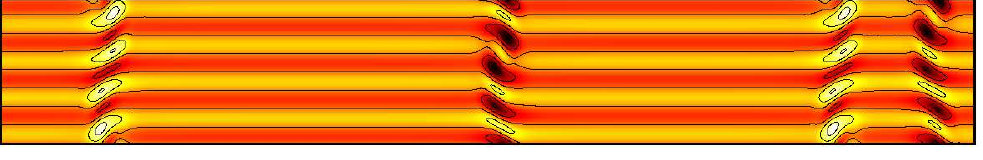}}
\vspace{5mm}
}
\end{tabular}
\caption{DNS calculation for in $\alpha=\eighth$, $Re=18$ initialised with the
  the end state from the $Re=22$ run. Shown on the left is the
  variation in $\omega$ in an $(x,t)$ plane for $y=21\pi/32$ and the
  right snapshots of $\omega(x,y)$ at (from bottom to top)
  $t=0,6.5\times10^3,1.2\times 10^4,4.45\times10^4$ and
  $7.5\times10^4$ (times indicated as black tics on the left
  figure). Colour extrema are $\omega=-5$ black, $\omega=5$ white, and
  5 evenly spaced contours in $-4\leq\omega\leq4$ for $(x,y)$ plots
  and $-2\leq\omega\leq2$ for colours in $(x,t)$.}
\label{kolmo18}
\end{center}
\end{figure}

The interaction of these kinks and antikinks can be more
complicated. Using the final state calculated at $Re=22$ as an initial
condition for a DNS at $Re=18$, we find that the kinks and antikinks
can rebound rather than colliding: see figure
\ref{kolmo18}. Interestingly, there is some adjustment in the internal
structure of the kink-antikink pair after this episode presumably to
produce the reversal in propagation speed. There is also more separation
between the kink and antikink in this pairing compared to that found
at $Re=22$ and the propagation speed is correspondingly smaller.

\begin{figure}
\begin{center}
\begin{tabular}{cc}%
\scalebox{0.85}{\input{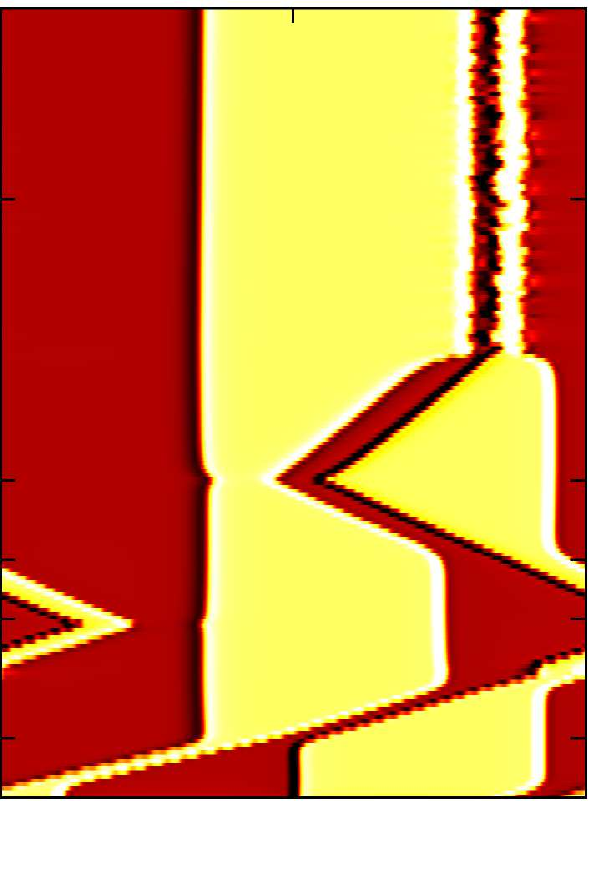}} &
\parbox[b]{80mm}{
\scalebox{0.8}{\input{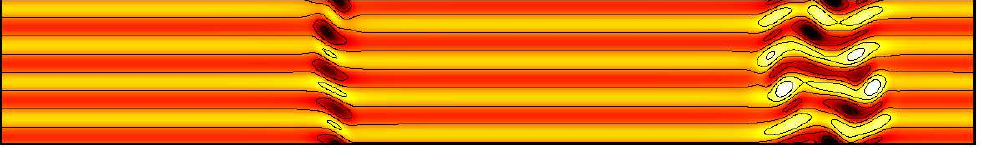}}\\
\scalebox{0.8}{\input{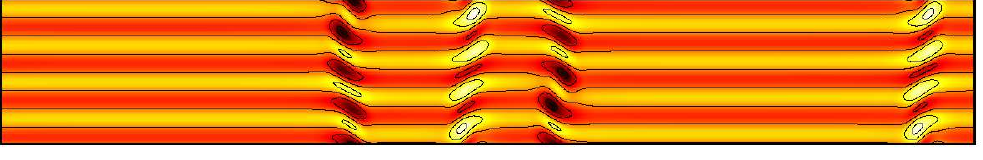}}\\
\scalebox{0.8}{\input{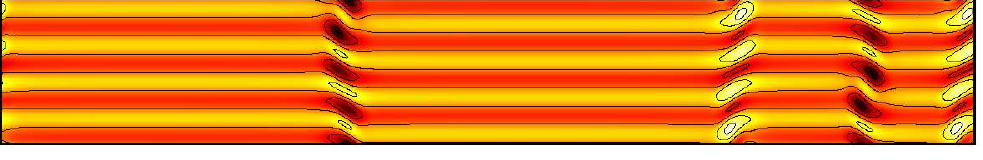}}\\
\scalebox{0.8}{\input{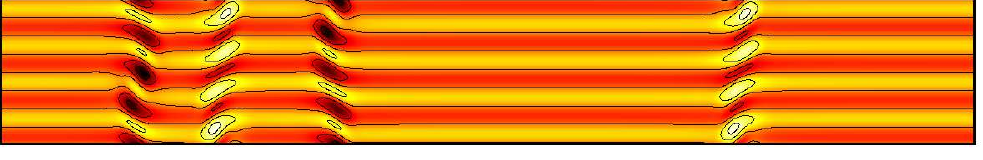}}\\
\scalebox{0.8}{\input{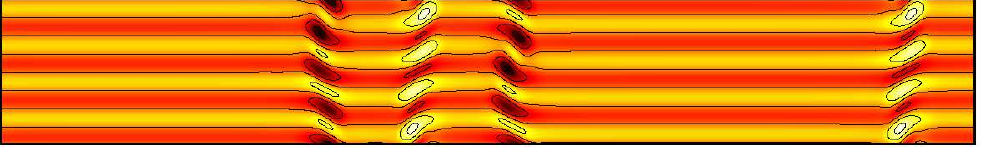}}
\vspace{5mm}
}
\end{tabular}
\caption{DNS calculation for in $\alpha=\eighth$, $Re=19$ initialised
  with the end state from $Re=22$. Shown on the left is the
  variation in $\omega$ in an $(x,t)$ plane for $y=21\pi/32$ and the
  right snapshots of $\omega(x,y)$ at (from bottom to top)
  $t=7\times10^3,2.2\times 10^4,2.95\times10^4,3.95\times10^4$ and
  $7.5\times10^4$ (times indicated as black tics on the left
  figure). Colour extrema are $\omega=-5$ black, $\omega=5$ white, and
  5 evenly spaced contours in $-4\leq\omega\leq4$ for $(x,y)$ plots
  and $-2\leq\omega\leq2$ for colours in $(x,t)$.}
\label{kolmo19}
\end{center}
\end{figure}

Repeating the exercise at $Re=19$, figure \ref{kolmo19} (see also
video 7) shows three different types of behaviour in one run. The
first is a kink-antikink partner swap at $t \approx 7 \times 10^3$ (as
per figure \ref{kolmo}), then two rebounds at $2.2 \times 10^4$ and
$3.95 \times 10^4$ (as per figure \ref{kolmo18}) and then finally, a
fusion of a kink-antikink pair with a stationary kink which leads to a
complicated oscillating structure reminiscent of the central part of
P1. The endstate is then a stationary antikink separated from a non-propagating
kink-antikink-kink chaotic structure. 

For $25 \lesssim Re$, DNS initiated with random initial conditions were found to invariably settle to a longest-wavelength ($\kappa=\alpha$) kink-antikink solution. Therefore, as in the $8 \pi \times 2 \pi$ domain, the coarsening regime re-establishes itself albeit now with a variety of local longest-wavelength attractors.

%
% fig 24
%

%
% fig 25
%

%
% fig 26
%

\subsection{5-PDE system}\label{5-PDE}

To gain some understanding of these DNS results, we looked for the
simplest system which retains this rich behaviour and then tried to
isolate the fundamental unit of the dynamics which appear to be kink
and antikinks travelling together as bound states (or TWs).  The
long-wavelength 3-PDE system (see (\ref{3-PDE})) is not complicated
enough to display any time-dependent attracting states but adding
subharmonics (to capture the secondary instabilities which are
precursors to the phenomena) changes this: the smallest such extension
leads to a 5-PDE system. The presence of the subharmonics breaks the
$\mathcal{S}^2$-symmetry of the kinks and antikinks (invariance under
a shift of $2\pi/n$ in $y$) and is crucial in creating translational
motion since the kink and antikink can then have different internal
structure.  Then if the kinks and antikinks are close enough their
interacting vorticity fields can induce motion.  This minimal
extension of the 3-PDE system requires adopting a slightly more
general streamfunction
\begin{eqnarray}
\psi(x,y,t) &=&\int^x F(\xi,t) d\xi+\biggl(-\frac{Re}{n^3}+g(x,t) \biggr) \cos ny +h(x,t) \sin ny \nonumber \\
&&\hspace{2cm} +G(x,t) \cos \half ny +H(x,t) \sin \half ny.
\end{eqnarray}
Defining the subharmonic vorticities 
as $\omega_G:=-(G^{''}-\quart G)$ and $\omega_H:=-(H^{''}-\quart H)$, the new truncated set of equations is then

\begin{eqnarray}
\frac{\partial \omega_F}{\partial t} &=& \frac{1}{Re} \omega_F^{''}+\frac{Re}{2n^2} (n^2 h-\omega_h)^{'}
+n \biggl[ \half (g \omega_h-h \omega_g)^{'} +\quart(G \omega_H-H \omega_G)^{'} \biggr], \nonumber \\
\frac{\partial \omega_g}{\partial t} &=& \frac{1}{Re}(\omega_g^{''}-n^2 \omega_g)+n F \omega_h-n h \omega_F^{'} 
-\quart nG \omega_H^{'}-\quart nH \omega_G^{'}+\quart n G^{'} \omega_H+\quart n H^{'} \omega_G,
\nonumber \\
\frac{\partial \omega_h}{\partial t} &=& \frac{1}{Re}(\omega_h^{''}-n^2 \omega_h) +\frac{Re}{n^2}(n^2 F-\omega_F^{'})
+n(g \omega_F^{'}-F \omega_g) +\quart nG \omega_G^{'}-\quart n H \omega_H^{'} \nonumber\\
&& \hspace{5cm} -\quart n G^{'} \omega_G+\quart n H^{'} \omega_H, \nonumber \\
%
% subharmonics
%
\frac{\partial \omega_G}{\partial t} &=& \frac{1}{Re}(\omega_G^{''}-\quart n^2 \omega_G)
+\frac{Re}{2n^2}(n^2 H^{'}-\omega_H^{'})
+\half n g \omega_H^{'} -\half n h \omega_G^{'}+\quart nG \omega_h^{'}-\half nH \omega_F^{'} \nonumber \\
&& \hspace{1cm}-\quart nH \omega_g^{'} +\half n F \omega_H+\quart ng^{'} \omega_H-\quart nh^{'} 
\omega_G+\half nG^{'}\omega_h-\half nH^{'} \omega_g,
\nonumber \\
\frac{\partial \omega_H}{\partial t} &=& \frac{1}{Re}(\omega_H^{''}-\quart n^2 \omega_H)
+{Re \over 2n^2}(n^2 G^{'}-\omega_G^{'})
+\half n g \omega_G^{'} +\half n h \omega_H^{'}-\quart nG \omega_g^{'}+\half nG \omega_F^{'} \nonumber \\
&& \hspace{1cm}
-\quart nH \omega_h^{'}
-\half n F \omega_G+\quart n g^{'} \omega_G+\quart nh^{'} \omega_H-\half n G^{'}\omega_g-\half nH^{'} \omega_h.
\end{eqnarray}

To truly isolate flow structures, the same non-periodic boundary conditions were used as for the 3-PDE system (see (\ref{npbc})) 
augmented by 
\begin{equation}
G^{'}=H^{'}=\omega_G^{'}=\omega_H^{'}\biggl|_{x=0,L} \biggr.=0. 
\label{npbc_augment}
\end{equation}
%
%
% n=2 symmetric bifurcation Re=3.40316 alpha=1/4   3.37 for alpha=1/8
%    asymmetric bifurcation   =8.75371
% 
Crucially, if the forcing wavenumber is changed to $n=2$, the 5-PDE
system is {\em identical} (up to its boundary conditions) to a DNS run
in which only the $k_y$ wavenumbers $(-2,-1,0,1,2)$ are included.
This is important as the more efficient pseudospectral DNS code (with
periodic boundary conditions) could then be used to hunt for
kink-antikink TWs. Once found, parts of the flow domain could
subsequently be used to generate initial guesses for TWs in the 5-PDE
system shifted to a Galilean frame $(X=x-ct,y)$ moving at some constant
but {\em a priori} unknown speed $c$.

\begin{figure}
\begin{center}
\begin{tabular}{cc}%
\scalebox{1.1}{\input{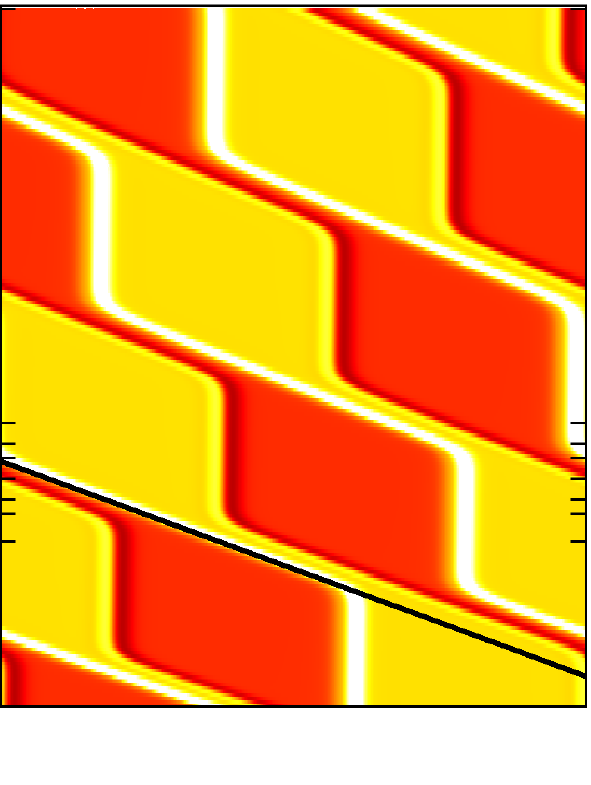}} &
\hspace{5mm}
\parbox[b]{80mm}{
\scalebox{0.65}{\input{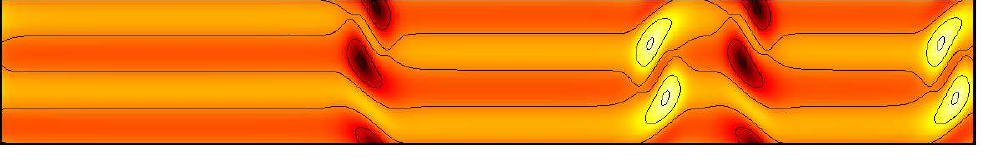}}\\
\scalebox{0.65}{\input{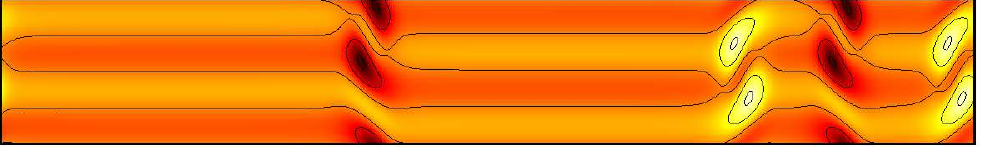}}\\
\scalebox{0.65}{\input{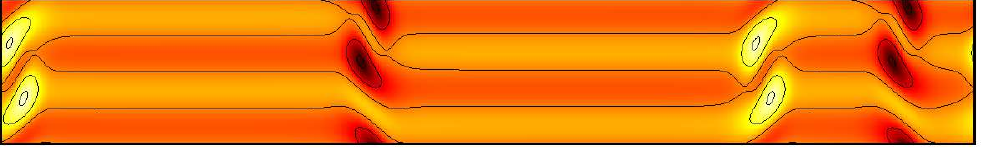}}\\
\scalebox{0.65}{\input{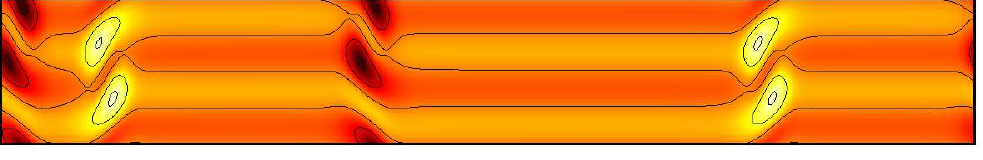}}\\
\scalebox{0.65}{\input{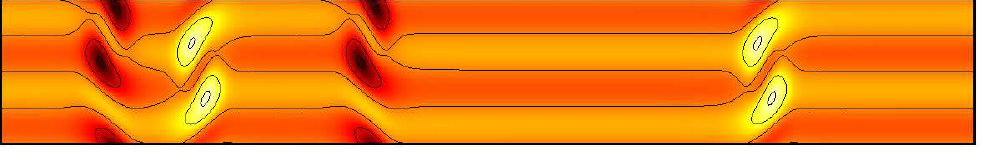}}\\
\scalebox{0.65}{\input{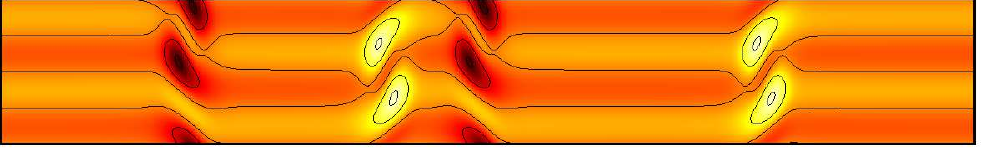}}\\
\scalebox{0.65}{\input{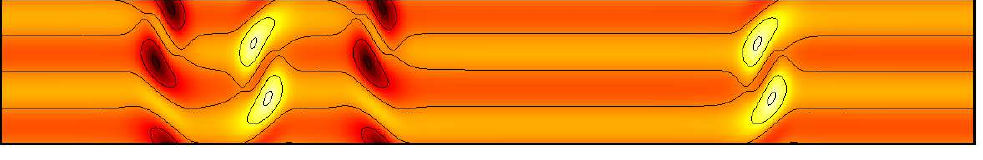}}
\vspace{8mm}
}
\end{tabular}
\caption{A DNS calculation of TWs in the
  $k_y$-truncated system with $n=2$, $\alpha=\eighth$, $Re=13$. Shown
  on the left is the variation in $\omega$ in an $(x,t)$ plane for
  $y=21\pi/32$. The black line overlaying the plot shows an estimate
  of phase speed $c=3.27\times10^{-3}$. The right shows snapshots of
  $\omega(x,y)$ at (from bottom to top) $t=1.15\times10^4,1.25\times
  10^4,1.35\times10^4, 1.45\times10^4, 1.5\times10^4, 1.85\times10^4$
  and $ 2\times10^4$, indicated by tick marks on the $(x,t)$
  plot. Colour extrema are $\omega=-5$ black, $\omega=5$ white, and 5
  evenly spaced contours in $-4\leq\omega\leq4$ for $(x,y)$ plots and
  $-2\leq\omega\leq2$ for colours in $(x,t)$.}
\label{kolmo_1}
\end{center}
\end{figure}

\begin{figure}
\begin{center}
\begin{tabular}{cc}%
\scalebox{1.0}{\input{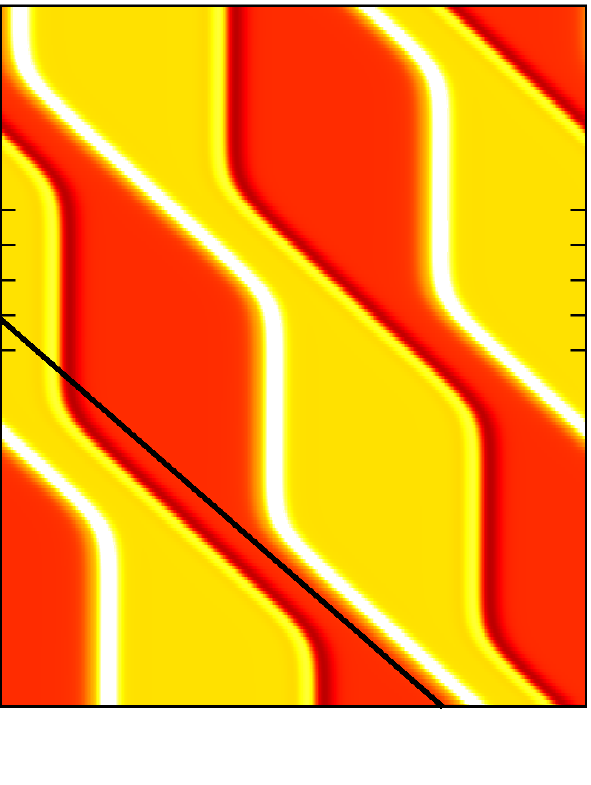}} &
\hspace{5mm}
\parbox[b]{80mm}{
\scalebox{0.78}{\input{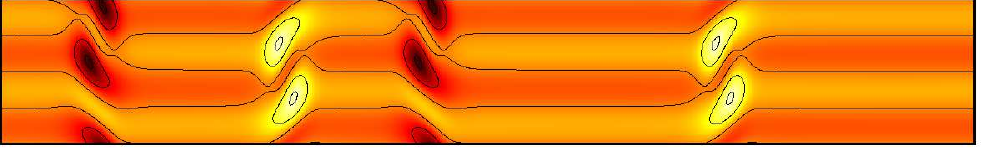}}\\
\scalebox{0.78}{\input{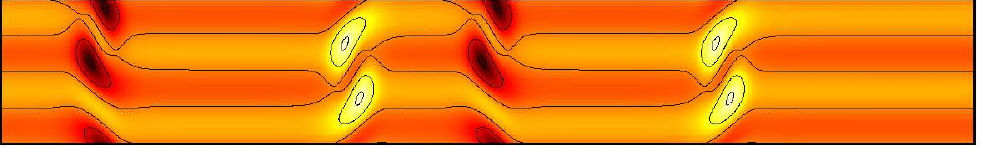}}\\
\scalebox{0.78}{\input{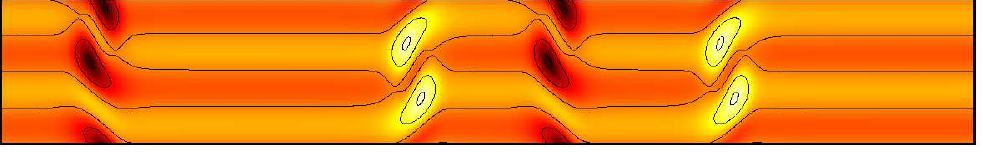}}\\
\scalebox{0.78}{\input{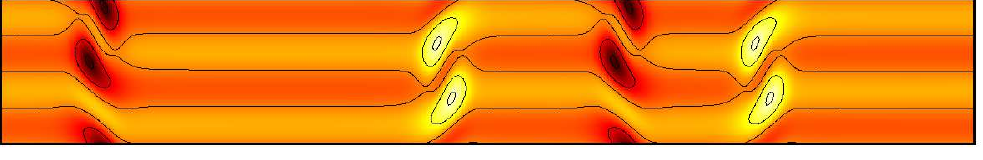}}\\
\scalebox{0.78}{\input{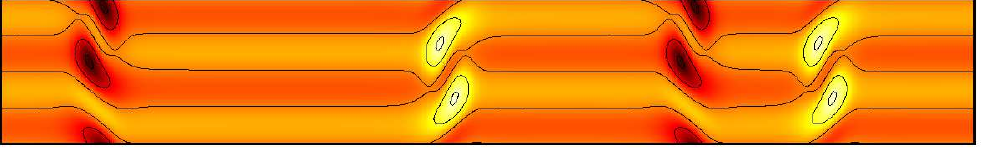}}
\vspace{8mm}
}
\end{tabular}
\caption{A DNS calculation of TWs in the
  $k_y$-truncated system with $n=2$, $\alpha=\eighth$, $Re=13$. Shown
  on the left is the variation in $\omega$ in an $(x,t)$ plane for
  $y=21\pi/32$. The black line overlaying the plot shows an estimate
  of phase speed $c=3.34\times10^{-4}$. The right shows snapshots of
  $\omega(x,y)$ at (from bottom to top) $t=1.1\times10^5,1.2\times
  10^5,1.3\times10^5, 1.4\times10^5$ and $1.5\times10^5$, indicated by
  tick marks on the $(x,t)$ plot. Colour extrema are $\omega=-5$
  black, $\omega=5$ white, and 5 evenly spaced contours in
  $-4\leq\omega\leq4$ for $(x,y)$ plots and $-2\leq\omega\leq2$ for
  colours in $(x,t)$.}
\label{kolmo_2}
\end{center}
\end{figure}

\begin{figure}
\begin{center}
\psfrag{A}{{\color{cyan}  $\omega_h$}}
\psfrag{B}{{\color{red}   $F$}}
\psfrag{C}{{\color{blue}  $\omega_g-Re/n$}}
\psfrag{E}{{\color{green} $\omega_G$   }}
\psfrag{F}{{\color{black} $\omega_H$   }}
\includegraphics[width=12cm, height=9cm,clip]{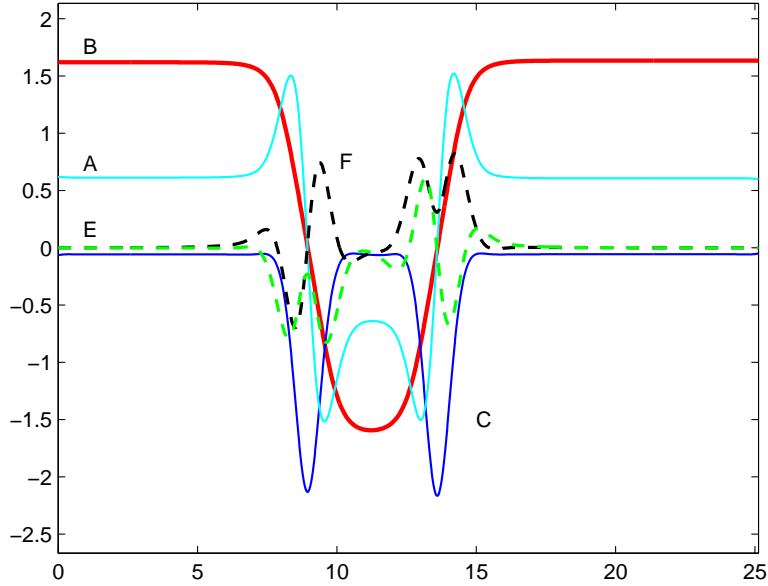}
\caption{The vorticity field of the converged TW with $c=-3.22 \times
  10^{-3}$ and $\overline{F}=1.0241$ found from the DNS solution of
  figure \ref{kolmo_1}. $F$ is the thickest (red) line,
  $\omega_g-Re/n$ is the thin (dark) blue line, $\omega_h$ the thin
  (light) cyan line, $\omega_G$ the dashed (light) green line and
  $\omega_H$ the dashed (dark) black line.}
\label{vnewTW}
\end{center}
\end{figure}

\begin{figure}
\begin{center}
\hspace{-0.5cm}
% a Symm_state_Re8-22
\includegraphics[width=7cm, height=5cm,clip]{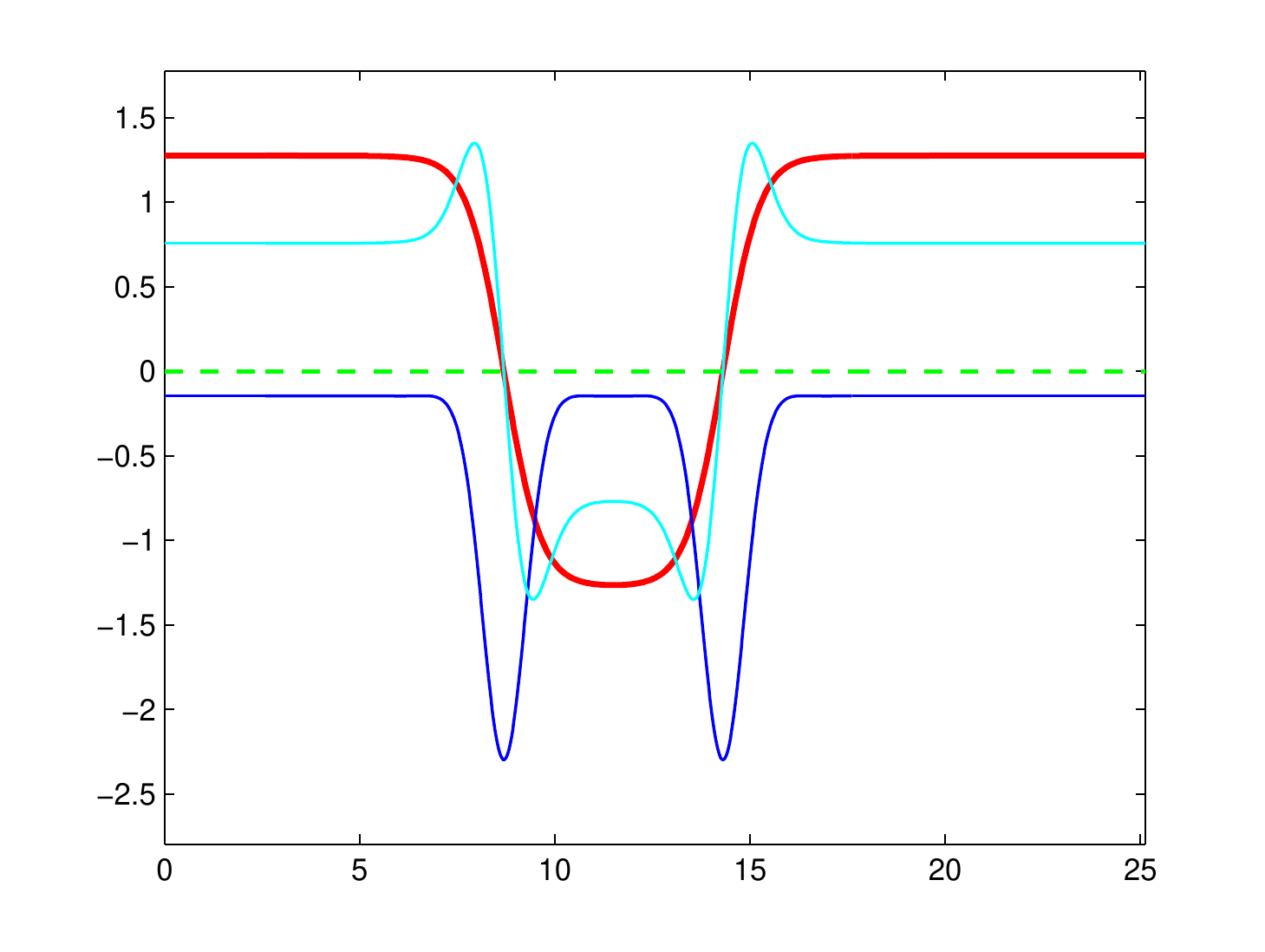}
\hspace{-0.5cm}
% b cRe_5_state_c=0
\includegraphics[width=7cm, height=5cm,clip]{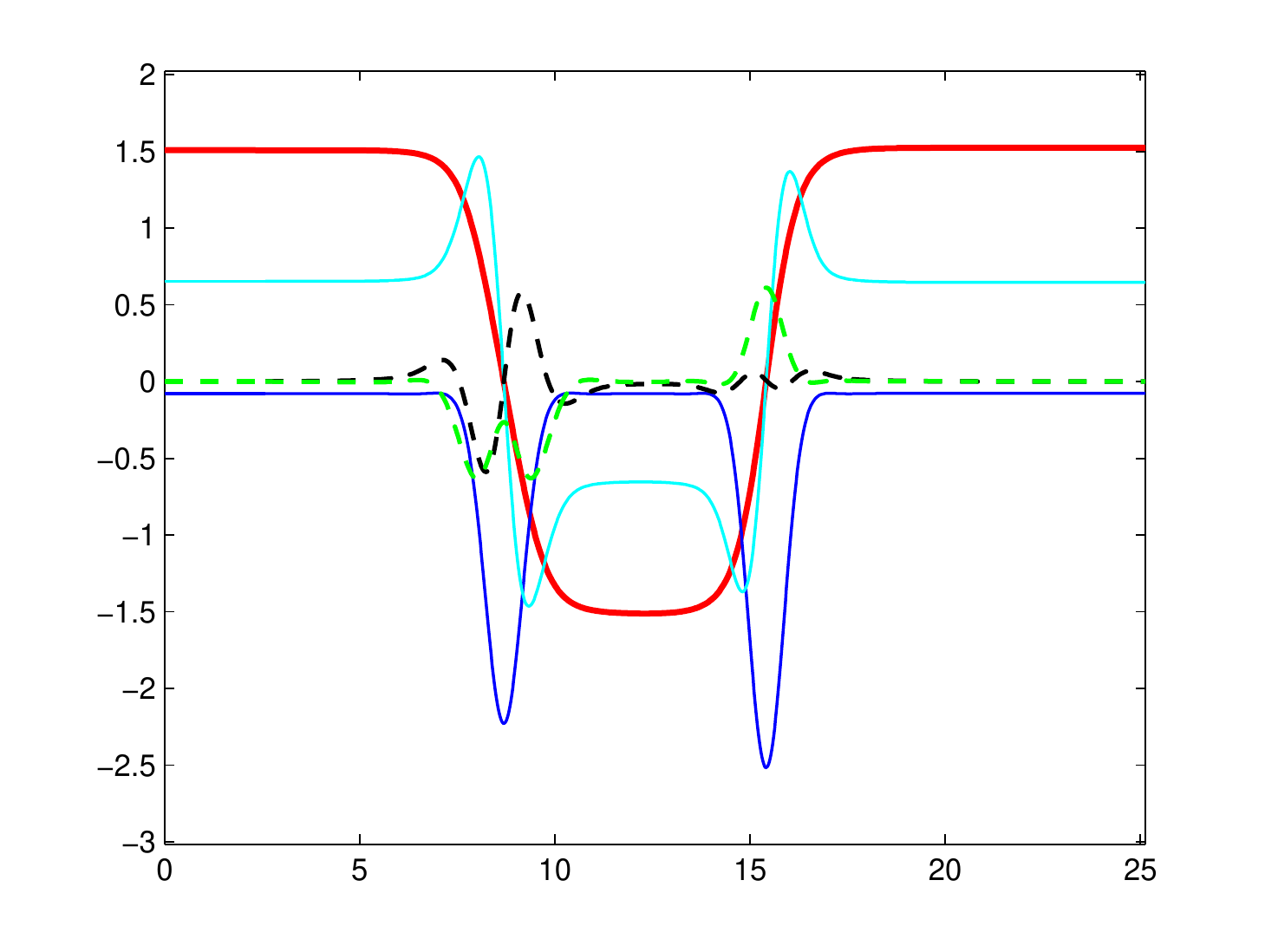}\\
\hspace{-0.5cm}
% c cRe_5_state_Re13
\includegraphics[width=7cm, height=5cm,clip]{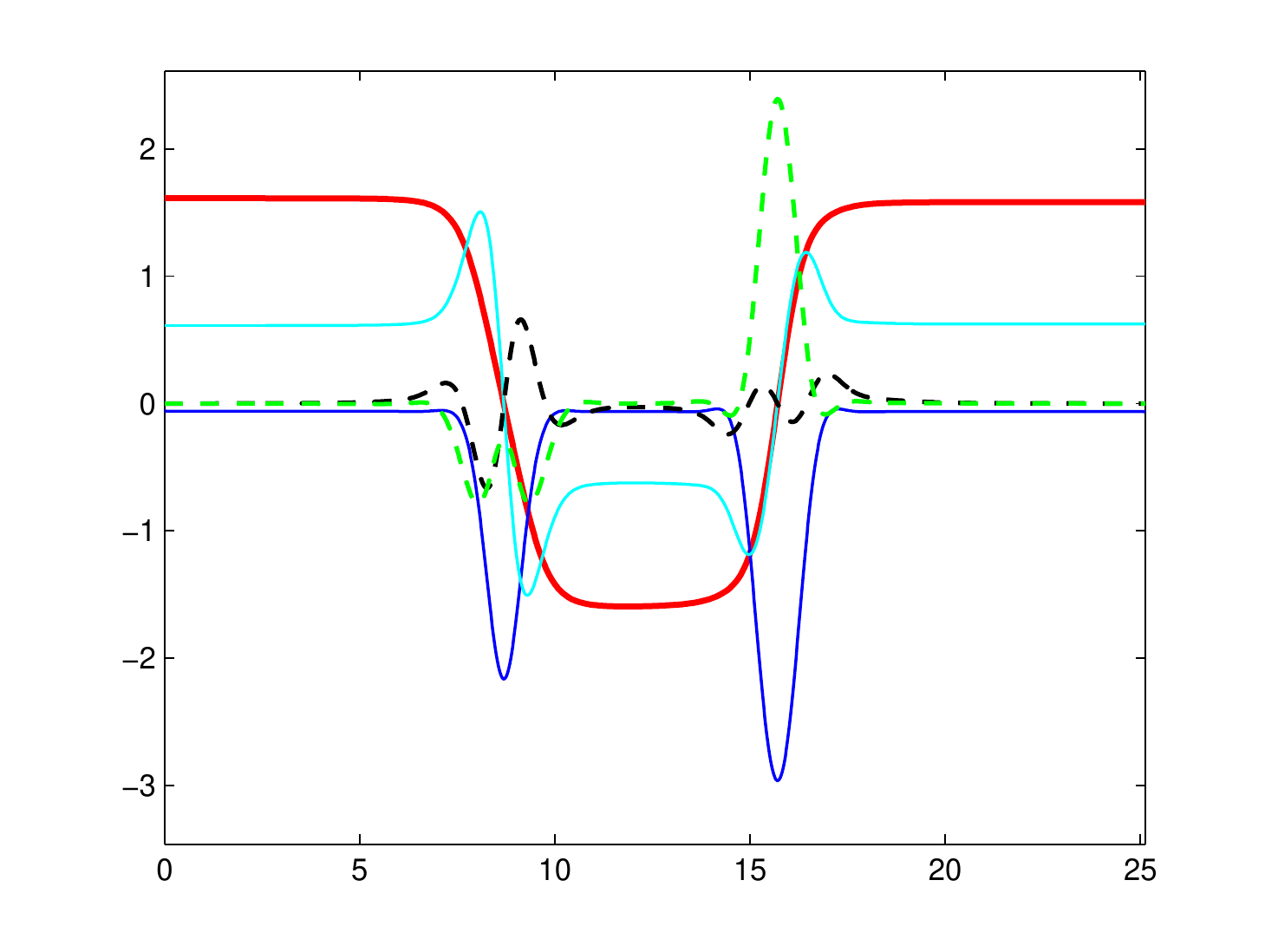}
\hspace{-0.5cm}
% d cRe_1_state_Re13a
\includegraphics[width=7cm, height=5cm,clip]{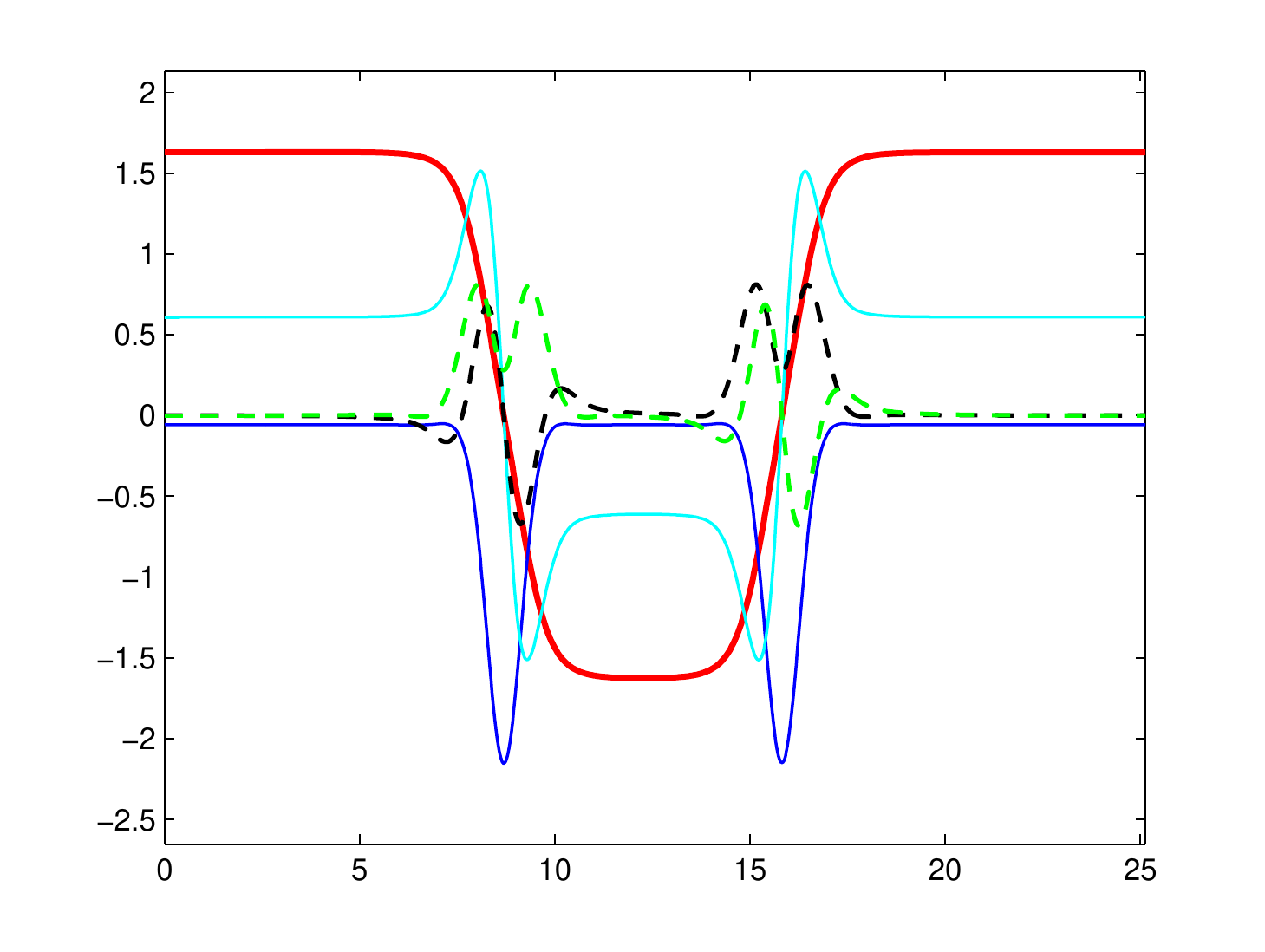}\\
\hspace{-0.5cm}
% e cRe_1_state
\includegraphics[width=7cm, height=5cm,clip]{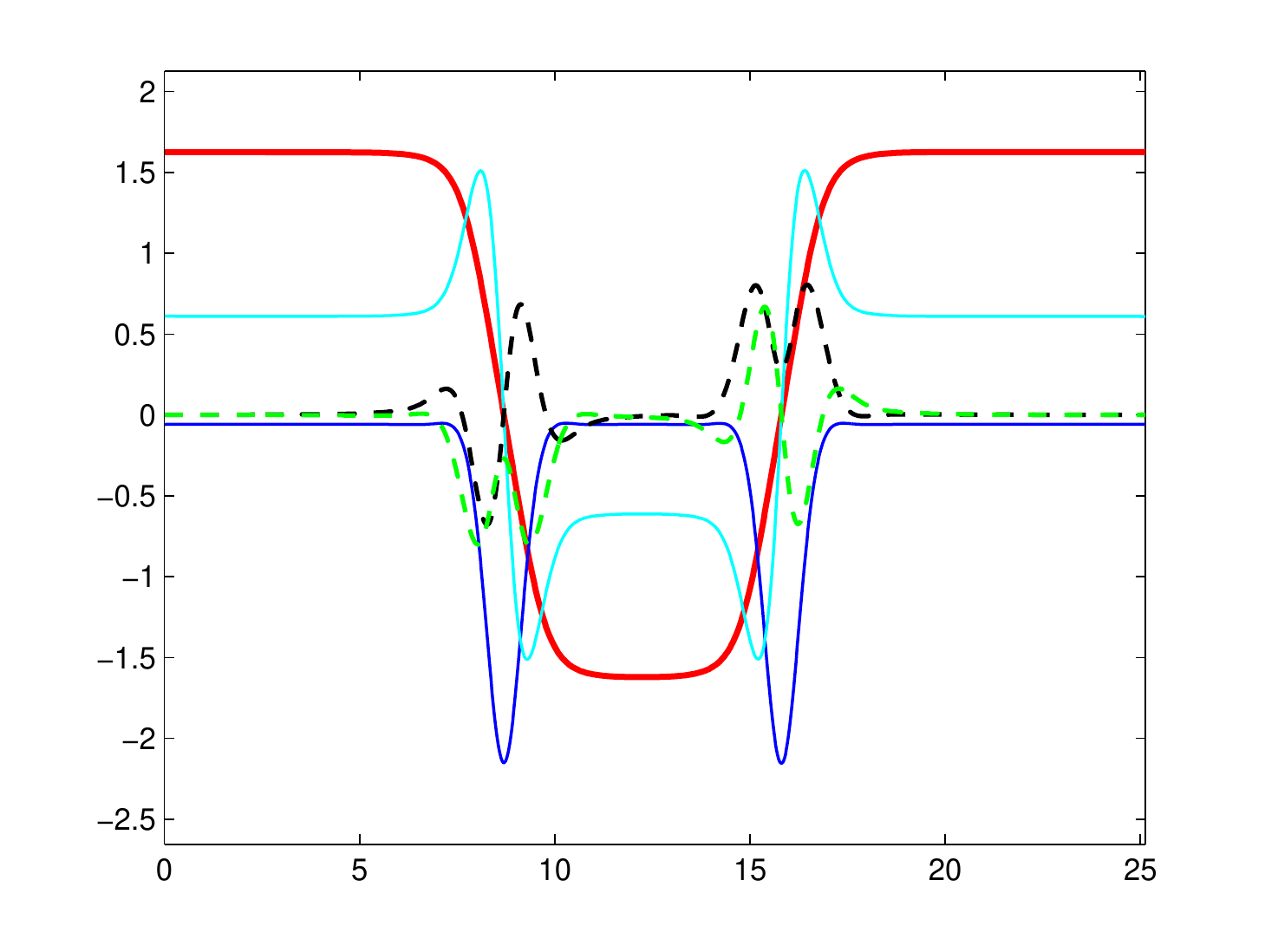}
\hspace{-0.5cm}
% f cRe_3_state
\includegraphics[width=7cm, height=5cm,clip]{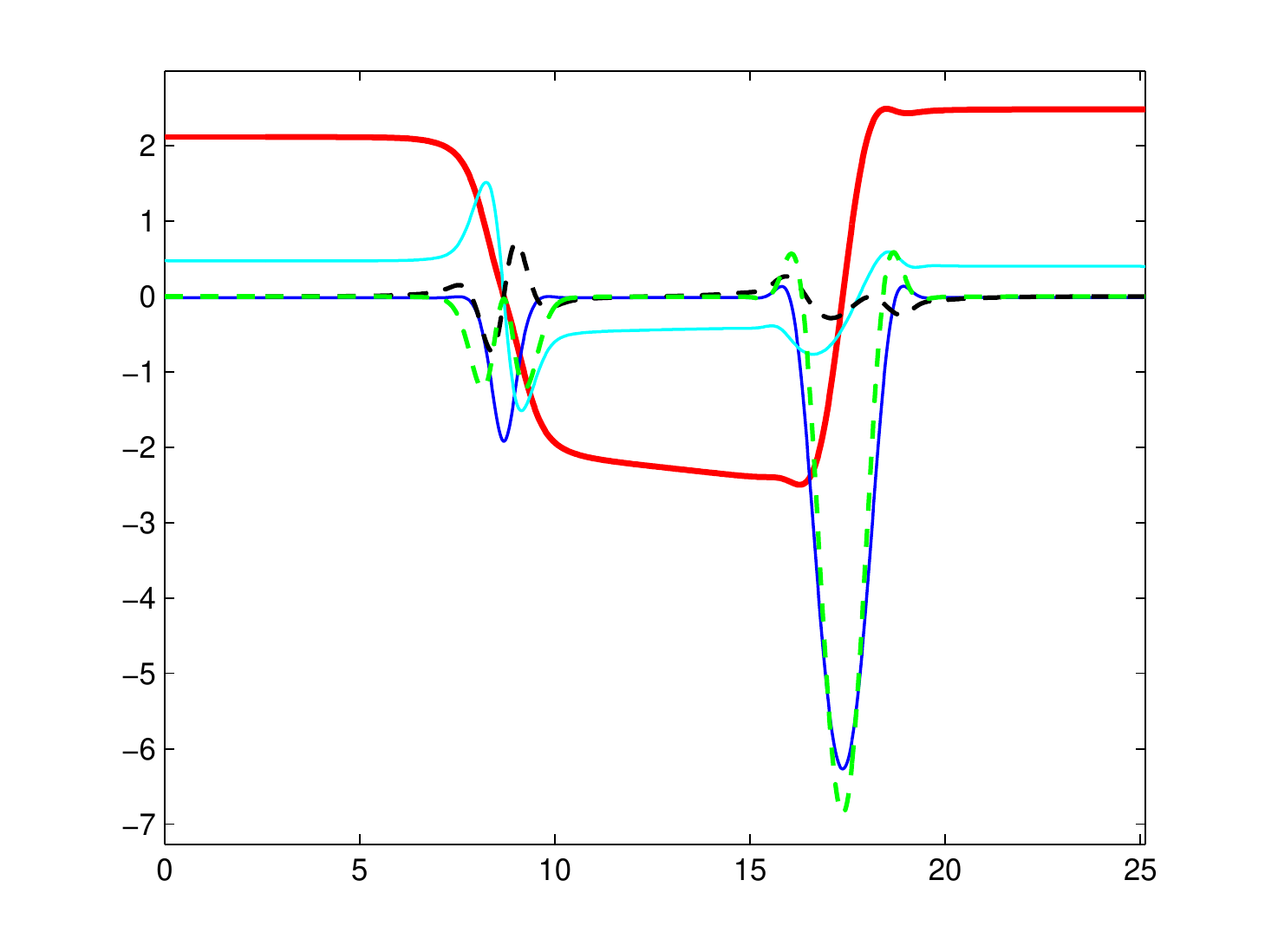}
\caption{The vorticity fields of the solutions labelled $a$ (top
  left), $b$ (top right), $c$ (middle left), $d$ (middle right), $e$
  (bottom left), $f$ (bottom right) in figure \ref{cRe}. In each, $F$
  is the thickest (red) line, $\omega_g-Re/n$ is the thin (dark) blue
  line, $\omega_h$ the thin (light) cyan line, $\omega_G$ the dashed
  (light) green line and $\omega_H$ the dashed (dark) black line (as
  in figure \ref{vnewTW}). $\overline{F}=0.7055$ for all solutions.}
\label{fields}
\end{center}
\end{figure}

\begin{figure}
\begin{center}
\includegraphics[width=12cm, height=9cm,clip]{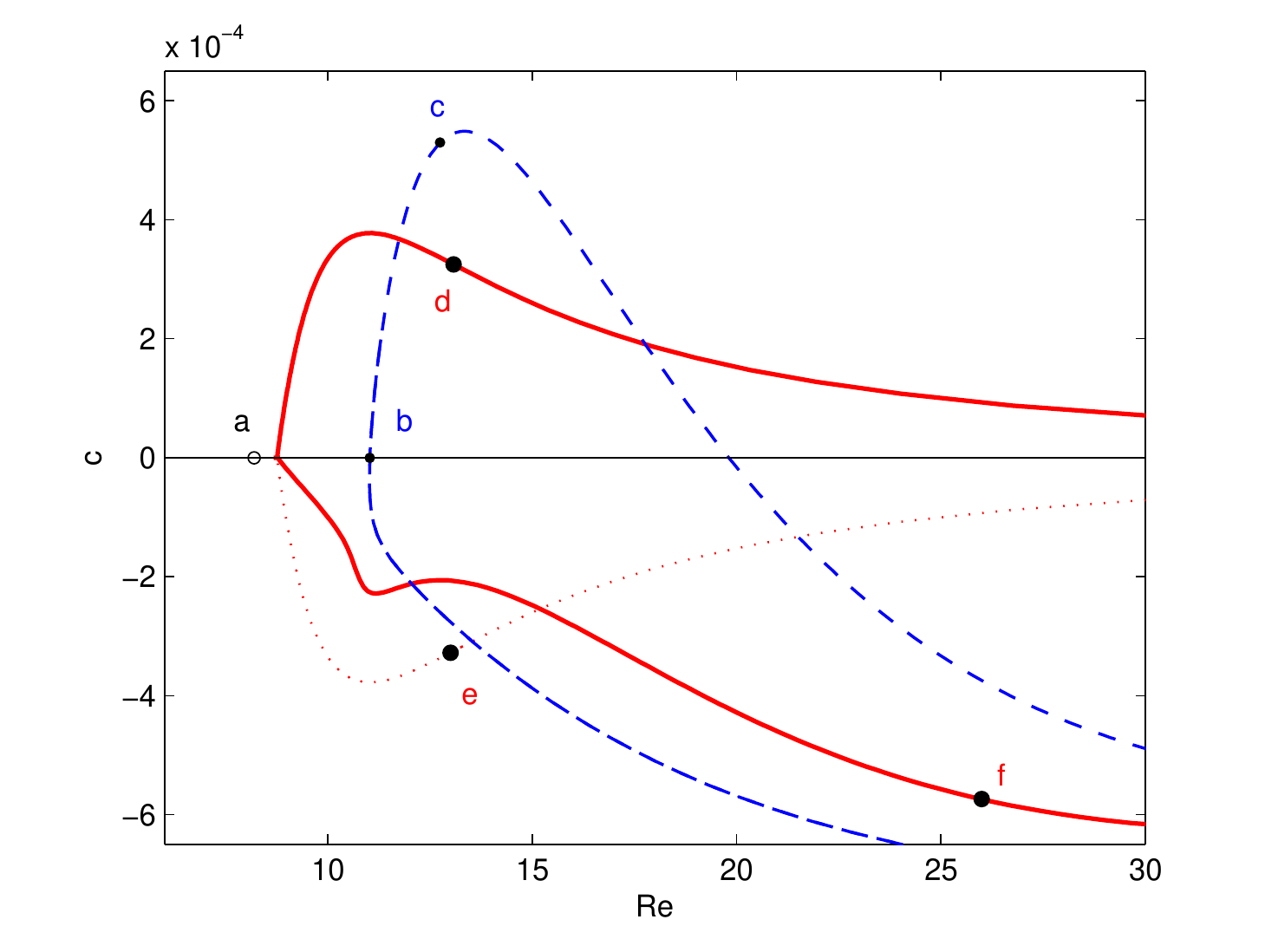}
\caption{The travelling wave solution branches found in the 5-PDE
  system: similar branches with $c \rightarrow -c$ exist via symmetry
  transformations as, for example, has been shown for the upper solid
  red branch to produce the lower dotted branch. Plots of the
  vorticity fields at points labelled $a,b,c,d,e,f$ are given in
  figure \ref{fields}. Solution $a$ is $\mathcal{S}-$ symmetric and so
  $c=0$. The thick red $\mathcal{S}-$ asymmetric branch crosses the
  $\mathcal{S}-$ symmetric branch at the asymmetric bifurcation point
  $Re \approx 8.75$ . Solution $b$ has $c=0$ but is {\em not}
  $\mathcal{S}-$ symmetric. Solutions $d$ and $e$ are complements of
  each other (same branch but opposite $c$). Solution $f$ shows the
  tendency for the kink pairs to separate and become
  unequal. $\overline{F}=0.7055$ for all solutions.}
\label{cRe}
\end{center}
\end{figure}

Calculations confirm that kink-antikink TWs do exist still in this
reduced system. Figures \ref{kolmo_1} and \ref{kolmo_2} show two time
sequences from DNS with $n=2$ and $\alpha=\eighth$ {\em both} at $Re=13$ (note $Re_c(\eighth)=3.37$ for
$n=2$). In the former, the kink-antikink pairs are noticeably tighter
and correspondingly the propagation speed $c$ (estimated as $-3.27
\times 10^{-3}$) is higher that the latter's value (estimated as
$-3.34 \times 10^{-4}$). In both cases, the vorticity field at a
chosen time can be used to extract half the spatially periodic DNS
domain containing only the kink-antikink pair. This can then be used
as an initial guess for a Newton-Raphson search for a
uniformly-propagating TW in the 5-PDE system. In both cases,
convergence is surprisingly straightforward (i.e. just 4 or 5 steps to
reduce an initial residual of $10^{-2}$ down to $10^{-12}$) with phase
speeds of $c=-3.22 \times 10^{-3}$ and $c=-3.28 \times 10^{-4}$
emerging respectively (the Newton-Raphson assumes the existence of $c$
but its value is found as part of the convergence). The $c=-3.22
\times 10^{-3}$ TW is shown in figure \ref{vnewTW} and the $c=-3.27
\times 10^{-4}$ TW as solution $e$ in figures \ref{fields} and
\ref{cRe}: note the difference in kink-antikink separations.

The TWs which emerge from this process of taking half the DNS domain
have $\overline{F} \neq 0$ (although the full domain has
$\overline{F}=0$) and depend on $Re$ so $c=c(Re,\overline{F})$ (they
are sufficiently localised not to depend on $\alpha \leq
\quart$). Branch continuing this solution by varying $Re$ while
$\overline{F}$ is held fixed gives the red solid line in figure
\ref{cRe}. These are primary TW solutions since they bifurcate off an
$\mathcal{S}-$symmetric branch of a stationary kink pair (solution
$a$ in figure \ref{cRe}). By symmetry, this branch can be reflected in the $Re$-axis as
partly done in figure \ref{cRe}.  A secondary branch of
$\mathcal{S}-$asymmetric states (the blue dashed line in figure
\ref{cRe}) was found serendipitously while branch-continuing the
primary branch using too big a step in $Re$ (a useful `quick-\&-dirty'
technique to branch switch). This secondary branch also crosses the
$c=0$ line but remains $\mathcal{S}-$asymmetric (i.e. the subharmonics
do not vanish): see solution $b$ in figure \ref{fields}. Presumably
further secondary branches exist with different separations between
the kinks and antikinks and $\mathcal{S}-$asymmetries of each (a
systematic search was not carried out).  All the solution branches
move continuously with $\overline{F}$, a good example of this being
solution $e$. This has $\overline{F}=0.7055$ but branch continuing it
(with $Re=13$ fixed) to $\overline{F}=1.0241$ leads to precisely the
TW shown in figure \ref{vnewTW}, that is, the kink and antikink come
together and the phase speed increases tenfold.
 
All the TWs found consist of a kink and antikink separated by a
distance which varies continuously with either $\overline{F}$ or/and
$Re$. Individually, a given kink or antikink has symmetry-group-orbit
counterparts obtained by applying the transformations $\mathcal{S}^m
$, $m=1,2,\ldots,2n-1$ and/or $\mathcal{R}$ to it (note $\mathcal{S}$
and $\mathcal{R}$ do not commute). For $n=2$, the effect of the
transformations are as follows (with only fields which change
indicated for brevity)
\begin{equation}
\begin{array}{lcl}
\mathcal{S}  :[\omega_F,G,\omega_G, H, \omega_H,c](x)  & \rightarrow & [-\omega_F,-H,-\omega_H, G, \omega_G,-c](-x) \\
                                                       &             &                                              \\
\mathcal{R}  :[F,h,\omega_h,H, \omega_H,c](x)          & \rightarrow & [-F,-h,-\omega_h,-H, -\omega_H,-c](-x) 
\end{array}
\end{equation} 
with $\mathcal{S}^4=\mathcal{R}^2=I$, the identity.  This observation
raises the interesting possibility of generating new TWs by bringing
different versions of a kink and antikink together 
(note not all versions of kinks are compatible with a given antikink and vice versa). Some of the
various possibilities for the kinks and antikinks in the full system
are well illustrated in figure \ref{kinkvort}.  The various
different vorticity distributions a kink and antikink can possess individually,
coupled with the kink-antikink separation determines the propagation
speed of the pair as a whole.

Finally, it's worth remarking that applying a transformation to a
given TW as a whole can reverse its phase speed and
suggests arranging collisions between two TWs to see what
results. Trying this for solution $e$ and $\mathcal{S}$(solution $e$)
produced a steady endstate of 4 unequally spaced kinks and antikinks. 
Firing the roughly 10 times faster travelling wave
$\mathcal{S}$(TW in figure \ref{vnewTW}) towards solution $e$ also
produced the same type of stationary endstate. It's
not inconceivable that more exotic combinations (in which different
types of kinks and antikinks are involved) may give rise to yet more states 
but this was not pursued further here.

%
% fig 27
%
%
% fig 28
%

%
% Comparison figure  Dropbox/Kolmogorov/LK1/plotter
%
% fig 29
%

%
% fig 30
%

%
% Comparison figure  Dropbox/Kolmogorov/LK1/isolate.m
%
% fig 31
%

\section{Discussion}

%
% summary of what has been found
%

The original motivation for this study was to look for an accessible
arena in which to study spatiotemporal behaviour in the Navier-Stokes
equation. Kolmogorov flow is by design a minimalist model of a viscous
fluid flow being 2-dimensional, driven by a simple body force and
subject to periodic boundary conditions in both directions. The
results presented here indicate that despite this, it possesses rich
spatiotemporal behaviour once larger domains are considered. Key to
the observed dynamics is the existence of localised flow structures -
kinks and antikinks - which, directly or indirectly through their
bifurcated derivatives, appear to underpin all of the flow dynamics
observed.  The kinks and antikinks first arise during the initial
long-wavelength instability away from the basic flow response which
mirrors the forcing at low amplitudes. The structure of the ensuing
state as a function of the forcing amplitude is described by a
Cahn-Hilliard-type equation and quickly separates into two regions of
roughly half the domain where a constant flow exists in either of the
two directions perpendicular to the forcing.  These regions are
connected via kink and antikink transition regions and coarsening
dynamics leading to this state are observed for random initial data.

After further bifurcations (typically modulational instabilities in
$y$) as the forcing strengthens, this regime gives way to multiple
attractors, some of which possess spatially-localised time dependence
(P1 and P2).  Co-existence of such attractors in a large domain gives
rise to a variety of interesting collisional dynamics. A minimal 5-PDE
(1-space and 1-time) system has been built by taking a 3-PDE
improved long-wavelength approximation and incorporating the possibility of
subharmonic instability.  This extended system captures the behaviour
seen and allows the basic building blocks of the behaviour -
travelling waves - to be easily isolated. These consist of
kink-antikink pairs whose phase speed and collisional behaviour depend
on the separation between the kink and antikink and their internal
vortical structures. At larger forcing amplitudes ($25 \lesssim Re$ for
a $8 \pi \times 2 \pi$ domain), the coarsening regime reinstates itself when various longest-wavelength solutions consisting of
just one time-varying kink-antikink pair again become the only attractors.
Beyond a yet higher forcing threshold ($120 \lesssim Re$ for
a $8 \pi \times 2 \pi$ domain), one longest-wavelength solution
emerges as a unique attractor. In this, the kink and antikink are chaotic yet the two one-dimensional regions of essentially half the domain width remain steady.

The Cahn-Hilliard coarsening behaviour, demonstrated here in Kolmogorov
flow for the first time, is not new but the exploration beyond this
regime is for the 2D Navier-Stokes equations.
Complementary work has been performed in simpler modelling
equations (e.g \cite{Gelens11} in the complex Swift-Hohenberg equation
and \cite{vanHecke99} for coupled complex Ginzburg-Landau equations)
and found similarly, if not more, complicated behaviour (e.g. see
figure 11 of \cite{Gelens11}). Here we find no spontaneous sources or
sinks of kinks and antikinks (\cite{vanHecke99,Gelens11}) but instead
conservation of both after collisions.  Presumably the dynamics are
smoother in 2D Kolmogorov flow because of the extra spatial dimension present rather
than other differences like the character of the underlying
instability (e.g. finite wavenumber and oscillatory in
\cite{Gelens11} as opposed to vanishing wavenumber and steady
here). Certainly, the presence of this extra dimension allows the
kinks and antikinks to have various different internal distributions
of vorticity which, along with the kink-antikink separation,
determines how they move as a bound state.  \cite{Gelens11} also
report the existence of stable `breathing sinks' (their figure 12(j))
which resemble the state P1 found here and other work by
\cite{Beaume11} focussing on the phenomenon of homoclinic snaking in
2D doubly diffusive convection has found localised chaos (their figure
13).  While a complete mathematical rationale for the spatiotemporal
behaviour found here is beyond the scope of the current work, we have
at least identified a simpler 5-PDE system based upon a long
wavelength limit and started to map out some of the kink-antikink
travelling waves it possesses. Hopefully, this system can form the
basis of further analysis.  Perhaps the most important observation
made here is that the coarsening regime seems to reinstate itself at
large forcing amplitudes and eventually a global attractor emerges again 
dominated by the longest wavelength allowed by the system albeit with
localised chaotic kink and antikink regions.

In terms of the original aim of this paper, 2D Kolmogorov flow has
been shown to provide a rich environment to explore the existence of
simple exact localised solutions to the 2D Navier-Stokes equations and
to probe their relevance to the complicated flow dynamics seen using
recurrent-flow analysis in the spirit of \cite{Chandler:2013fi}.  The
disparity between the large domains used here and the spatial extent
of the localised chaos which exists, however, highlights a key
challenge: to develop efficient (practical) recurrent-flow analysis
strategies which look for solutions only over sub-domains of the full
simulated system. We hope to be able to report on progress soon.

\vspace{2cm}
\noindent
{\em Acknowledgements}.  We would like to thank Gary Chandler for advice on
the numerical codes used here, to Basile Gallet for very helpful
comments on an earlier version of the manuscript and to an anonymous
referee who alerted us to relevant literature on direction-reversing
travelling waves and helped us improve the presentation in many
ways. We are also very grateful for numerous free days of GPU time on
`Emerald' (the e-Infrastructure South GPU supercomputer:
\url{http://www.einfrastructuresouth.ac.uk/cfi/emerald}) and the
support of EPSRC through grant EP/H010017/1.

\appendix

%----------------------------------------------------------------------------------------------------------------
%----------------------------------------------------------------------------------------------------------------
%
% Appendix 
%
\section*{Appendix: The long wavelength expansion}

Providing $Re$ is close to $Re_c(\alpha)$, the  unstable wavenumbers are 
$O(\sqrt{Re-Re_c(\alpha)})$ from the neutral curve opening the way up to a small wavenumber/long wavelength expansion in which $\partial_x \ll \partial_y$ (Nepomniashchii 1976, Chapman \& Proctor 1980, Sivashinsky 1985).  
We just summarise this calculation here since the details (albeit with a different non-dimensionalisation
) are in Sivashinsky (1985) (Chapman \& Proctor 1980 derived the same
long wave equation but for convective cells in a nearly insulated
liquid layer). We work with the streamfunction version of the
governing equations (\ref{NS})
\begin{equation}
\nabla^2 \psi_t + \psi_y \nabla^2 \psi_x-\psi_x \nabla^2\psi_y= \frac{1}{Re} \nabla^4 \psi +n \cos (n y)
\label{stream_eqn}
\end{equation}
where subscripts indicate derivatives, $\omega=-\nabla^2 \psi$, 
and $\eps \ll 1$ is defined by
\begin{equation}
\frac{1}{Re}=(1-\eps^2) \frac{1}{Re_c}.
\end{equation}
(here $Re_c$ without an argument means $Re_c(0)$). The appropriately rescaled space and
time variables are $\tau=\eps^4 t$, $X=\eps x$ and $Y=ny$. The
streamfunction $\psi$ is expanded as $\psi=\psi^0+\eps \psi^1+\eps^2
\psi^2+\eps^3 \psi^3+\ldots$ and is crucially assumed to share the
same spanwise periodicity as the forcing function, that is, $\psi$ is
$2\pi$ periodic in $Y$ so it is $\mathcal{S}-$symmetric. 
This means that $\int^{2\pi}_0 \,dY$ of
\ref{stream_eqn} gives simply
\begin{equation}
\eps\int^{2\pi}_0 \psi_{XX\tau} \, dY
+n \int^{2\pi}_0(\psi_Y \psi_{XX})_X \,dY
=\frac{\eps(1-\eps^2)}{Re_c} \int^{2\pi}_0 \psi_{XXXX} \, dY.
\label{mean_eqn}
\end{equation}
At $O(1)$, (\ref{stream_eqn}) becomes simply
\begin{equation}
\frac{n^4}{Re_c} \psi^0_{YYYY}+n \cos Y=0
\end{equation}
with the solution 
\begin{equation} 
\psi^0=-Re_c/n^3 \cos Y +A^0(X,\tau)
\end{equation}
(the zeroth
approximation of (\ref{mean_eqn}) is automatically satisfied).  At
$O(\eps)$, (\ref{stream_eqn}) requires the solution
\begin{equation}
\psi^1=\frac{Re_c^2}{n^4}A_X^0 \sin Y +A^1(X,\tau)
\end{equation}
with (\ref{mean_eqn}) giving $Re_c=\sqrt[4]{2} n^{3/2}$ at $O(\eps)$. 
At $O(\eps^2)$, (\ref{stream_eqn}) requires
\begin{equation}
\psi^2=\frac{Re_c}{n}\biggl(-\frac{1}{n^2} \cos Y +(A_X^0)^2 \frac{Re_c^2}{n^4} \cos Y +\frac{Re_c}{n^3} A_X^1 \sin Y
\biggr)
+A^2(X,\tau)
\end{equation}
((\ref{mean_eqn}) at $O(\eps^2)$ is automatically satisfied) and at
$O(\eps^3)$
\begin{eqnarray}
\psi^3 &=&\biggl[ 
\frac{3Re_c^2}{n^6} A_{XXX}^0+\frac{2 Re_c^2}{n^4} A_X^0-\frac{Re_c^4}{n^6} (A_X^0)^3+\frac{Re_c^2}{n^4} A_X^2
\biggr] \sin Y \nonumber \\
&& \hspace{3cm}+\frac{2 Re_c^3}{n^6} A_X^0 A_X^1 \cos Y + A^3(X,\tau).
\end{eqnarray}
The $O(\eps^3)$
approximation to (\ref{mean_eqn}) then gives (after integrating immediately twice in $X$) the evolution equation
for $A^0$:
\begin{equation}
A_\tau^0+\frac{3Re_c^3}{2 n^8} A^0_{XXXX}+\frac{2 Re_c^3}{n^6} A^0_{XX}-\frac{Re_c^5}{3n^8} (A_X^0)_X^3=0.
\end{equation}
where $Re_c^2=\sqrt{2}n^3$. This equation has a Lyapunov functional
\begin{equation}
{\cal F}[A^0]:= \frac{Re^3_c}{12 n^8} \int^{\eps L}_0 \, \biggl[
  9(A^0_{XX})^2-12n^2(A^0_X)^2+\sqrt{2}n^3 (A^0_X)^4 \biggr] \, dX
\end{equation}
such that
\begin{equation}
A^0_{\tau}=-\frac{\delta {\cal F}}{\delta A}.
\end{equation}
As a result $d{\cal F}/{d\tau}=-\int^{\eps L}_0 (A^0_\tau)^2 dX \leq 0$ and
the dynamics is a monotonic approach to a local or global minimum  of
${\cal F}$. Such a minimum satisfies
\begin{equation}
A^0_{XXX}+\frac{4n^2}{3} A^0_{X}-\frac{2 \sqrt{2}n^3}{9} (A_X^0)^3=\gamma.
\label{ELeqns}
\end{equation}
where $\gamma$ is an integration constant. Multiplying by $A^0_{XX}$
and integrating again gives
\begin{equation}
\half (A^0_{XX})^2+\frac{2n^2}{3} (A^0_{X})^2-\frac{\sqrt{2}n^3}{18}
(A_X^0)^4-\gamma A^0_X=E.
\label{particle-eqn}
\end{equation}
with $E$ a further constant. As Chapman \& Proctor (1980) point out,
this equation represents a `particle' with position $A_X^0$
oscillating {\em spatially} in a potential well  
\begin{equation}
V(A^0_X):= \frac{2n^2}{3} (A^0_{X})^2-\frac{\sqrt{2}n^3}{18}
(A_X^0)^4-\gamma A^0_X
\end{equation}
with total constant energy $E$. Since $A^0_X$ represents the (leading
$O(\eps)$) spanwise velocity, it has zero mean,
\begin{equation}
\int^{\eps L}_0 \, A^0_X \, dX \,=\,0, 
\label{F_bar}
\end{equation} 
which implies $\gamma=0$ otherwise
the oscillation is not (spatially) centred on $A_X^0=0$. For a given
domain length, there are then a countably infinite number of
(spatially-) periodic solutions which fit into the domain although the
wavelengths will eventually approach the width of the domain
where the long-wavelength assumption breaks down. For solutions of  the Euler-Lagrange equations (\ref{ELeqns}) with $\gamma=0$, the integral relationship
\begin{equation}
\int^{\eps L}_0 \, A^0_X (\ref{ELeqns})\,dX=\int^{\eps L}_0
\frac{3}{2}(A^0_{XX})^2-2n^2(A^0_X)^2+\frac{\sqrt{2}}{6}(A^0_X)^4 \, dX=0
\end{equation}
can be used to simplify the stationary value of ${\cal F}$ to just
\begin{equation}
{\cal F}_{stat}= -\frac{\sqrt{2} Re_c^3}{12 n^5} \int^{\eps L}_0 (A_X^0)^4 \,dX.
\end{equation}
Unfortunately, this does not immediately indicate that the steady solution with longest wavelength is the global minimiser but stability analysis does indicate this (Nepomniashchii 1976, Chapman \& Proctor 1980). All solutions of (\ref{ELeqns}) are found to be unstable to perturbations of longer wavelength implying that the solution with longest possible wavelength will be the global attractor. This is borne out by simulations (e.g. \cite{She:1987}).

The solution to (\ref{ELeqns}) (with $\gamma=0$) can be written down in terms of elliptic functions as
\begin{equation}
A_X^0 =\sigma {\rm sn}(\beta X|k)
\end{equation}
where, without loss of generality, $A_X^0(0)=0$ has been imposed,
\begin{equation}
{\rm sn}(\beta X |k):=\sin \phi \quad {\rm where} \quad \beta X=\int^{\phi}_0 
\frac{d \theta}{\sqrt{1-k^2 \sin^2 \theta}},
\end{equation}
and $\sigma, \beta$ and $k \in [0,1]$ are defined by the relations
\begin{equation}
(\beta \sigma)^2=2E, \quad \beta^2(1+k^2)=\frac{4n^2}{3},\quad \frac{\beta^2}{\sigma^2}k^2=\frac{\sqrt{2}n^3}{9}. 
\end{equation}
There is one further relation which is the second boundary condition on $A_X^0$. Since $A_X^0$ is periodic over $\eps L=2 \pi \eps/\alpha$, we can look for one cell over half the (spatial) period by setting $A^0_X=0$ at $X=\pi \eps/\alpha$ or equivalently 
$A^0_{XX}(\pi \eps/(2 \alpha))=0$,
\begin{equation}
\frac{n \pi}{\sqrt{3(1+k^2)}} \biggl(\frac{\eps}{\alpha} \biggr)=
K(k):= \int^{\pi/2}_0 \frac{d \theta}{\sqrt{1-k^2 \sin^2 \theta}}
\end{equation}
where $K(k)$ is the complete elliptic integral of the first kind. This last condition defines $k \in [0,1]$  as a function of $\eps/\alpha$. The bifurcation point is approached as $k \rightarrow 0$ (which minimises $K(k)$) and yields the critical $\eps$ value as a function of the geometry
\begin{equation}
\eps_{crit}(\alpha):= \frac{\sqrt{3} \alpha}{2n}
\end{equation}
which, when converted into a critical $Re$, is the result quoted in (\ref{asym_linear}).
In the other limit of $k \rightarrow 1$ corresponding to increasing $Re$ or a lengthening domain ($2\pi/\alpha$) or both,
\begin{equation}
\lim_{k \rightarrow 1} E = n\sqrt{2} = \max_{A^0_X} \, V(A_X^0)
\end{equation}
which is the maximum possible energy for a spatially oscillatory solution to (\ref{particle-eqn}) (with $\gamma=0$). In other words, there is no upper limit on $Re$ or $1/\alpha$ where such a solution no longer is available in this long-wave approximation.
The solution simplifies in this $k \rightarrow 1$ limit  to the localised solution
\begin{equation}
A_X^0=\sqrt{ \frac{3 \sqrt{2}}{n} } \tanh \biggl( n \sqrt{\frac{2}{3}} X \biggr).
\label{kink}
\end{equation}

As $X=\pm \infty$, $A^0_X \rightarrow \pm \sqrt{3 \sqrt{2}/n}$ so that the solution (\ref{kink}) acts to connect exact solutions of the form
\begin{equation}
\psi=\lambda x -\frac{Re}{n(Re^2 \lambda^2+n^2)} \cos(ny)+\frac{\lambda Re^2}{n^2 (Re^2 \lambda^2+n^2)} \sin (ny)
\label{gs_lambda}
\end{equation}
where $\lambda=\eps \sqrt{3 \sqrt{2}/n}$ represents the constant velocity component perpendicular to the forcing direction. In fact solutions (\ref{gs_lambda}) exist for {\it all} $\lambda$ but clearly only two states are viable asymptotic limits for $A_X^0$ (no net flow in the $y$ direction forces the two values to be equal in magnitude but opposite in sign).

\bibliographystyle{jfm}
\bibliography{papers}
\end{document}